\documentclass[runningheads]{svmult}
\usepackage{makeidx}   
\usepackage{graphicx}  
\usepackage{subeqnar}  
\usepackage{multicol}  
\usepackage{cropmark} 
\usepackage{physprbb}  
\newcommand{\ch}[2]{$\rm ^{#1}#2 $}
\newcommand{\bea}{\begin{eqnarray}}
\newcommand{\eea}{\end{eqnarray}}

\newcommand{\nn}{\nonumber}

\def \beq {\begin{equation}}
\def \eeq {\end{equation}}

\def \onbb {$0\nu\beta\beta$ }

\begin{document}
%
\title*{Status of Evidence for 
\protect\newline Neutrinoless Double Beta Decay}
\toctitle{Status of Evidence for Neutrinoless Double Beta Decay}
%
%
\titlerunning{Status of Evidence for Neutrinoless Double Beta Decay}
%
\author{H.V. Klapdor-Kleingrothaus\inst{1}
\thanks{Spokesman of the HEIDELBERG-MOSCOW and GENIUS Collaborations, 
	\protect\newline e-mail: klapdor@gustav.mpi-hd.mpg.de, 
	\protect\newline home page: 
	http://www.mpi-hd.mpg.de/non$\_$acc/}
\and A. Dietz\inst{1}
\and I.V. Krivosheina\inst{1,2}}
\authorrunning{H.V. Klapdor-Kleingrothaus}
%
%
%
\institute{Max-Planck-Institut f\"ur Kernphysik, 
 	Postfach 10 39 80, D-69029 Heidelberg, Germany,
\and Radiophysical-Research Institute, Nishnii-Novgorod, Russia}

\maketitle              

\begin{abstract}
	The present experimental status in the search 
	for neutrinoless double beta decay is reviewed, 
	with emphasis on the first 
	indication for neutrinoless double beta decay 
	found in the HEI\-DELBERG-MOSCOW experiment, 
	giving first evidence for lepton number violation 
	and a Majorana nature of the neutrinos. 
	Future perspectives of the field 
	are briefly outlined.
%
\end{abstract}


\section{Introduction}
	The neutrino oscillation
	interpretation of the atmospheric and solar neutrino data,  
	deliver a strong indication for a non-vanishing neutrino mass.
	While such kind of experiments yields information on the difference of 
	squared neutrino mass eigenvalues and on mixing angles 
	(for the present status see, e.g. 
\cite{Bahc01,Barg02-Sol}), the 
	absolute scale of the neutrino mass is still unknown.
	Information from double beta decay experiments is indispensable to
	solve these questions 
\cite{KKPS1-2,KK60Y}.
	Another important problem is that of the fundamental
	character of the neutrino, whether it is a Dirac or a Majorana
	particle 
\cite{Majorana37,Rac37}.
	Neutrinoless double beta decay could answer also this question. 
	Perhaps the main question, which can be investigated 
	by double beta decay with high sensitivity, is  
	that of lepton number conservation or non-conservation.

	Double beta decay, the rarest known nuclear decay process, can occur
	in different modes:

\begin{tabbing}
\hspace*{4.5cm} \= \hspace*{7.9cm} \= \kill
\qquad $ 2\nu\beta\beta- \rm decay:$ \> $ A(Z,N) \rightarrow
A(Z\!+\!2,N\!-\!2) + 2e^- + 2\bar{\nu}_e$ \hspace*{1.05cm}(1)\\[-0ex]
\qquad $ 0\nu\beta\beta- \rm decay:$ \> $ A(Z,N) \rightarrow
A(Z\!+\!2,N\!-\!2) + 2e^-$ \hspace*{2cm}(2) \\[-0ex]
\qquad $ 0\nu(2)\chi\beta\beta- \rm decay:$ \> $ A(Z,N) \rightarrow
A(Z\!+\!2,N\!-\!2) + 2e^- + (2)\chi$ \hspace*{0.9cm}(3) \\[-0ex]
\end{tabbing}

\begin{figure}[h]
\begin{center}
\includegraphics[height=5.cm,width=9cm]{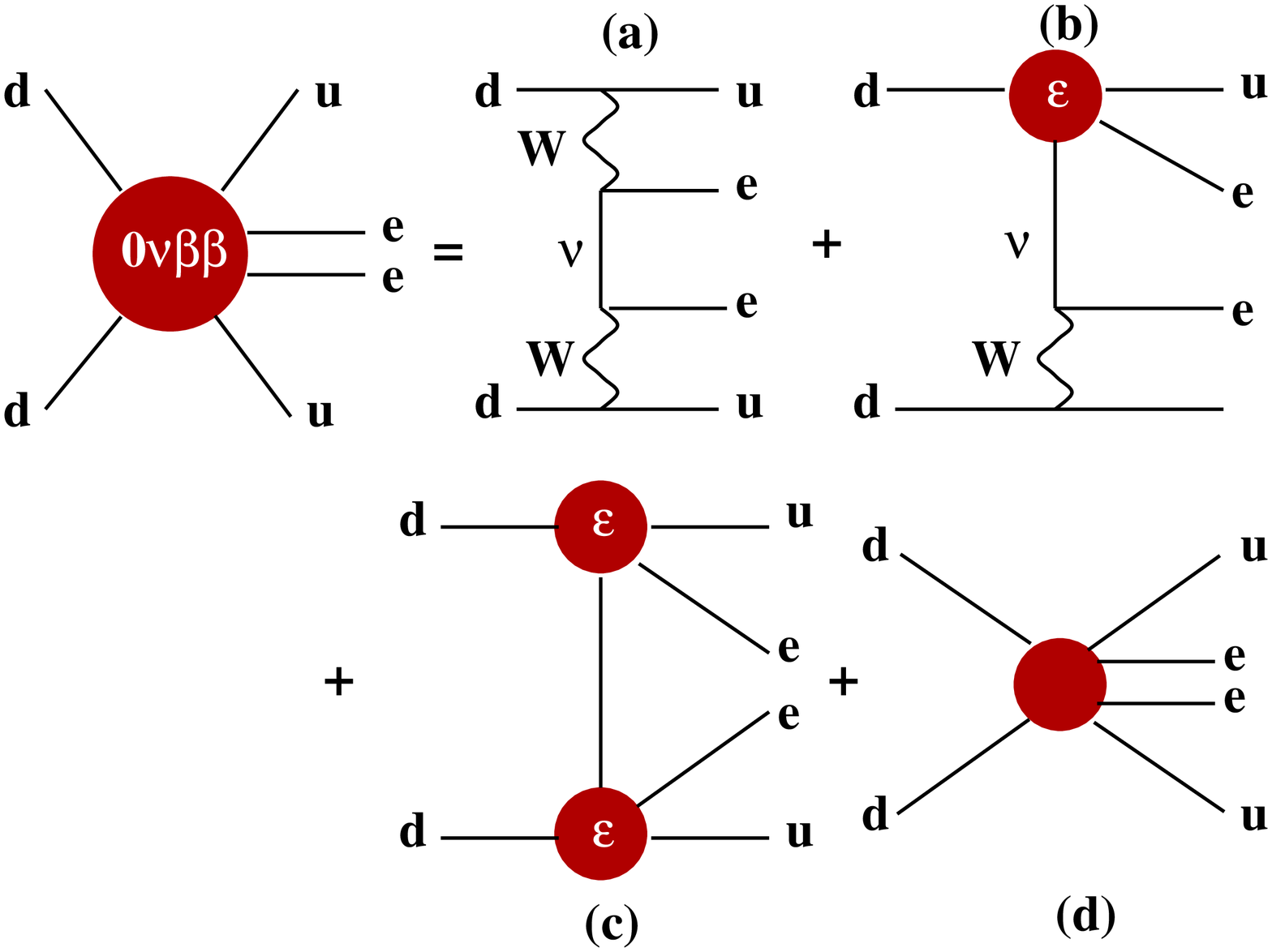}
\end{center}
\caption{Feynman graphs of the general double beta decay with long range 
	(a-c) and short range (d) parts (see 
\cite{KK-Paes99}).
}
\label{Diagr}
\end{figure}

\vspace{-0.3cm}
	While the two-neutrino mode (1)
	is allowed by the Standard Model of particle physics, 
	the neutrinoless mode 
	(\onbb) (2) requires violation of lepton number 
	($\Delta$L=2). 
	This mode is possible only, if the neutrino 
	is a Majorana particle, i.e. the neutrino is its own antiparticle 
	(E. Majorana
\cite{Majorana37},  
	G. Racah 
\cite{Rac37},    
	 for subsequent works we refer to 
\cite{Mcl57,Case57,Ahl96}, 
	 for some reviews see   	
\cite{Doi85,Mut88,Gro89/90,Kla95/98,KK60Y,Vog2001}). 
	First calculations of \onbb decay based 
	on the Majorana theory have been done by W.H. Furry 
\cite{Fur39}.
	The most general Lorentz-invariant parametrization 
	of neutrinoless double beta decay 
	is shown in Fig.
\ref{Diagr}. 

	The usually used assumption is that the first term (i.e. the 
	Majorana mass mechanism) dominates 
	the decay process. 
	However, as can be seen from Fig. 
\ref{Diagr},
	and as discussed elsewhere (see, e.g. 
\cite{KK60Y,KK-LeptBar98,KK-SprTracts00,Faes01}) 
	neutrinoless double beta decay can not only probe a
	Majorana neutrino mass, but various new physics scenarios beyond the
	Standard Model, such as R-parity violating supersymmetric models, 
	R-parity conserving SUSY models,  
	leptoquarks, 
	violation of Lorentz-invariance,  
	and compositeness 
	(for a review see 
	\cite{KK60Y,KK-LeptBar98,KK-SprTracts00}). 
	Any theory containing lepton number violating
	interactions can in principle lead to this process allowing to obtain
	information on the specific underlying theory. 
	It has been pointed out already in 1982, however, 
	that {\it independently} of the mechanism of neutrinoless 
	double decay, the occurence of the latter implies 
	a non-zero neutrino mass and vice versa 
\cite{Sch82a}. 
	This theorem has been extended to supersymmetry. It has been shown 
\cite{H_KK_Kov97-98}
	that if the neutrino has a finite Majorana mass, 
	then the sneutrino necessarily has a (B-L) violating 
	'Majorana' mass, too, independent of the mechanism of mass generation.
	The experimental signature of the neutrinoless mode is a peak at the
	Q-value of the decay.

	Restricting to the Majorana mass mechanism, 
	a measured half-life allows to deduce information on 
	the effective Majorana neutrino mass
$\langle m \rangle $ , 
	which is a superposition of neutrino mass eigenstates 
\cite{Doi85,Mut88}:
	\bea 
[T^{0\nu}_{1/2}(0^+_i \rightarrow 0^+_f)]^{-1}= C_{mm} 
\frac{\langle m \rangle^2}{m_{e}^2}
+C_{\eta\eta} \langle \eta \rangle^2 + C_{\lambda\lambda} 
\langle \lambda \rangle^2 +C_{m\eta} \langle \eta \rangle \frac{\langle m \rangle}{m_e} 
	\nn
	\eea
	\beq
+ C_{m\lambda}
\langle \lambda \rangle \frac{\langle m_{\nu} \rangle}{m_e} 
+C_{\eta\lambda} 
\langle \eta \rangle \langle \lambda \rangle,
	\eeq

\beq{}
\langle m \rangle = 
|m^{(1)}_{ee}| + e^{i\phi_{2}} |m_{ee}^{(2)}|
+  e^{i\phi_{3}} |m_{ee}^{(3)}|~,
\label{mee}
\eeq
	where $m_{ee}^{(i)}\equiv |m_{ee}^{(i)}| \exp{(i \phi_i)}$ 
	($i = 1, 2, 3$)  are  the contributions to $\langle m \rangle$
	from individual mass eigenstates, 
	with  $\phi_i$ denoting relative Majorana phases connected 
	with CP violation, and
	$C_{mm},C_{\eta\eta}, ...$ denote nuclear matrix elements squared, 
	which can be calculated, (see, e.g. 
\cite{Sta90},  
	for a review and some recent discussions see e.g. 
\cite{Tom91,KK60Y,Mut88,Gro89/90,FaesSimc,St-KK01-1,St-KK01-2,Faes01}). 
	Ignoring contributions from right-handed weak currents on the 
	right-hand side of eq.(1), only the first term remains.

	The effective mass is closely related to the parameters of neutrino
	oscillation experiments, as can be seen from the following expressions
\bea{  }
|m^{(1)}_{ee}|&=&|U_{e1}|^2 m_1, \\
|m^{(2)}_{ee}|&=&|U_{e2}|^2 \sqrt{\Delta m^2_{21} + m_1^2},\\
|m^{(3)}_{ee}|&=&|U_{e3}|^2\sqrt{\Delta m^2_{32}+ \Delta m^2_{21} + m_1^2},
\label{gg}
\eea
	Here, $U_{ei}$ are entries of the neutrino mixing matrix, and 
	$\Delta m^2_{ij}~=~ |m^2_i - m^2_j|$,
	with $m_i$ denoting neutrino mass eigenstates.
	$U_{ei}$ and $\Delta m^2$ can be determined from 
	neutrino oscillation experiments.

	The importance of $\langle m \rangle$ for solving the problem 
	of the structure 
	of the neutrino mixing matrix and in particular to fix 
	the absolute scale of the neutrino
	mass spectrum which cannot be fixed by $\nu$ - oscillation 
	experiments alone,  
	has been discussed in detail in e.g. 
\cite{KK-Sar_Evid01,KKPS1-2,Viss01}.

	Double beta experiments to date gave only {\it upper limits} for 
	the effective mass.
	The most sensitive limits 
\cite{HM97a,HM99,HDM01} 
	were {\it already} of striking importance for neutrino physics, 
	excluding for example, in hot dark matter models, 
	the small mixing angle (SMA) MSW solution 
	of the solar neutrino problem 
\cite{Gla00,Ellis99,Adh99,Minak-Nuno,Minak01,Minak97,KKPS1-2,KK60Y}
	in degenerate neutrino mass scenarios.

	The HEIDELBERG-MOSCOW double beta decay experiment in the Gran Sasso
	Underground Laboratory 
\cite{KK-Prop87,KK-Neutr98,Kla96b,KK-LeptBar98,KK-StProc00,KK60Y}
	searches for double beta decay of 
	$^{76}{Ge} \longrightarrow ^{76}{Se}$ + 2 $e^-$ + (2$\bar{\nu}$) 
	since 1990. 
	It is the most sensitive double beta experiment since 
	almost eight years now.	
 	The experiment operates five enriched (to 86$\%$) high-purity 
	$^{76}{Ge}$ detectors, with a total mass of 11.5 kg,  
	the active mass of 10.96 kg being equivalent to a source 
	strength of 125.5 mol $^{76}{Ge}$ nuclei.

	In this paper, we present a new, refined analysis of the data  
	obtained in the HEIDELBERG-MOSCOW experiment during the 
	period August 1990 - May 2000 which have recently been published 
\cite{HDM01}.
	The analysis concentrates on the 
	neutrinoless decay mode which is the one relevant for
	particle physics (see, e.g.
\cite{KK60Y}).
	First evidence for the neutrinoless decay mode will be presented
	(a short communication has been given already in 
\cite{KK-Evid01} (see also 
\cite{KK-China01}), 
	and first reactions have been published already 
\cite{KK-Sar_Evid01,KK3-Sar_Evid01,Ma02-Ev,Barg02-Ev,Ueh02-Ev,Min-Evid02,Chat-Evid02,Xing-Evid02,Ham-Evid02,Ahl-Kirch-Evid02,Nature02-Evid,Ma-Evid2-02,Hab-Suz02,Mohap02-Evid,Kays02-Evid,Fodor02-Evid,Pakv02-Evid,Niels02,He02-Evid,Josh02-Evid,Bran02-Evid,Cheng02-Evid,Smirn02-Evid,Mats02-Evid,Wiss-Evid02}). 
	For the results concerning 2$\nu\beta\beta$ decay and 
	Majoron-accompanied decay we refer to 
\cite{HDM01}.
	The results will be put into the context of other 
	present $\beta\beta$ activities.

\section{Experimental Set-up and Results}
\label{EXP}

	A detailed description of the HEIDELBERG-MOSCOW 
	experiment has been given recently in 
\cite{HDM97}, and in 
\cite{HDM01}.  
	Therefore only some important features will be given here. 
	We start with some general notes.

	1. Since the sensitivity for the \onbb half-life is
\beq{}
	T^{0\nu}_{1/2} \sim a  \times \epsilon \sqrt{\frac{Mt}{\Delta EB}} 
\label{T0nu}
\eeq 
	 (and 
	$\frac{1}{\sqrt{T^{0\nu}}} \sim \langle m_\nu \rangle$), 
	with a denoting the degree of enrichment, $\epsilon$ 
	the efficiency of the detector for detection of a double beta event, 
	M the detector (source) mass, $\Delta E$ the energy resolution, 
	B the background and t the measuring time, 
	the sensitivity of our 11\,kg {\it of enriched} $^{76}{Ge}$ 
	experiment corresponds to that of an at last 1.2\,ton 
	{\it natural} Ge experiment. After enrichment, energy resolution, 
	background and source strength are the most important 
	features of a $\beta\beta$ experiment.
	
	2. The high energy resolution of the Ge detectors of 0.2$\%$ 
	or better, assures that there 
	is no background for a \onbb line from the two-neutrino 
	double beta decay in this experiment, in contrast to most 
	other present experimental approaches, 
	where limited energy resolution is a severe drawback.

	3. The efficiency of Ge detectors for detection 
	of \onbb decay events is close to 100\,$\%$ (95\%, see 
\cite{Diss-Dipl}).

	4. The source strength in this experiment of 11\,kg 
	is the largest source strength ever operated in 
	a double beta decay experiment.

	5. The background reached in this experiment is, 
	with 0.17\,events/kg\,y\,keV in the \onbb decay region 
	(around 2000-2080\,keV), the lowest limit ever obtained 
	in such type of experiment.

	6. The statistics collected in this experiment during 
	10 years of stable running is the largest ever collected 
	in a double beta decay experiment.
	
	7. The Q value for neutrinoless double beta decay has been 
	determined recently with very high precision 
	to be Q$_{\beta\beta}$=2039.006(50)\,keV 
\cite{New-Q-2001,Old-Q-val}.


\begin{table}[b]
\begin{center}
\newcommand{\m}{\hphantom{$-$}}
\renewcommand{\arraystretch}{1.}
\setlength\tabcolsep{7.7pt}
\begin{tabular}{c|c|c|c|c|c}
\hline
			&	Total		&	
Active			&	Enrichment	&	
FWHM	1996	&	FWHM	2000		\\
\hline
	Detector	&	Mass		&
	Mass		&	in $^{76}{Ge}$	&
	at 1332 keV	&	at 1332 keV		\\
	Number		&	[kg]		&
	[kg]		&	[$\%$]		&
	[keV]		&	[keV]		\\
\hline
	No. 1		&	0.980		&
	0.920		&	85.9 $\pm$ 1.3		&
	2.22 $\pm$ 0.02	&	2.42 $\pm$ 0.22\\
	No. 2		&	2.906		&
	2.758		&	86.6 $\pm$ 2.5		&
	2.43 $\pm$ 0.03	&	3.10 $\pm$ 0.14\\
	No. 3		&	2.446		&
	2.324		&	88.3 $\pm$ 2.6		&
	2.71 $\pm$ 0.03	&	2.51 $\pm$ 0.16\\
	No. 4		&	2.400		&
	2.295		&	86.3 $\pm$ 1.3		&
	2.14 $\pm$ 0.04	&	3.49 $\pm$ 0.24\\
	No. 5		&	2.781		&
	2.666		&	85.6 $\pm$ 1.3		&
	2.55 $\pm$ 0.05	&	2.45 $\pm$ 0.11\\
\hline
\end{tabular}
\end{center}
\caption{\label{Techn_Param}Technical parameters of the five enriched 
	detectors. (FWHM - full width at half maximum.)}
\end{table}

\begin{figure}[t]
\begin{center}
\includegraphics[height=4.3cm,width=6.cm]{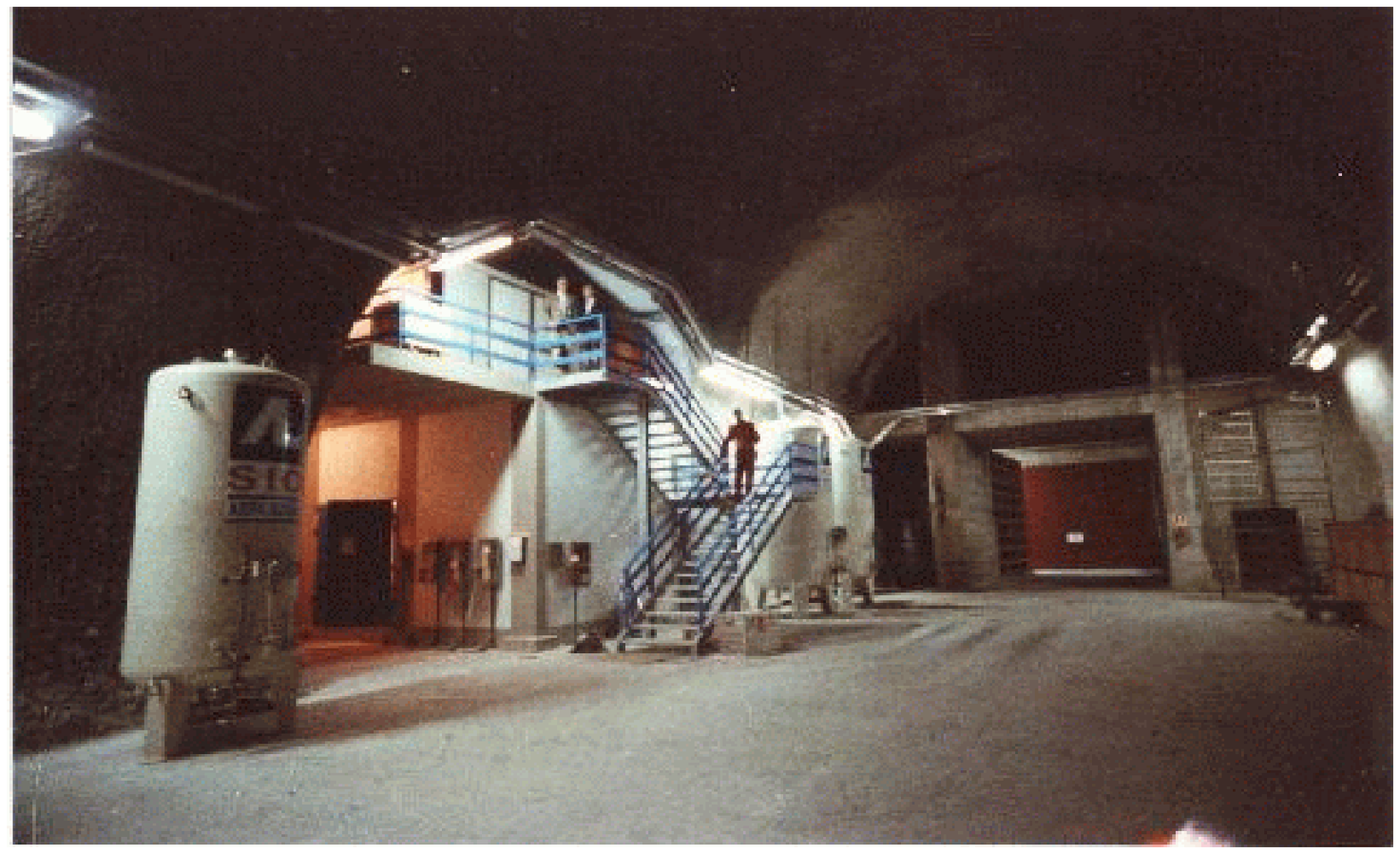}
\includegraphics[height=4.3cm,width=6.cm]{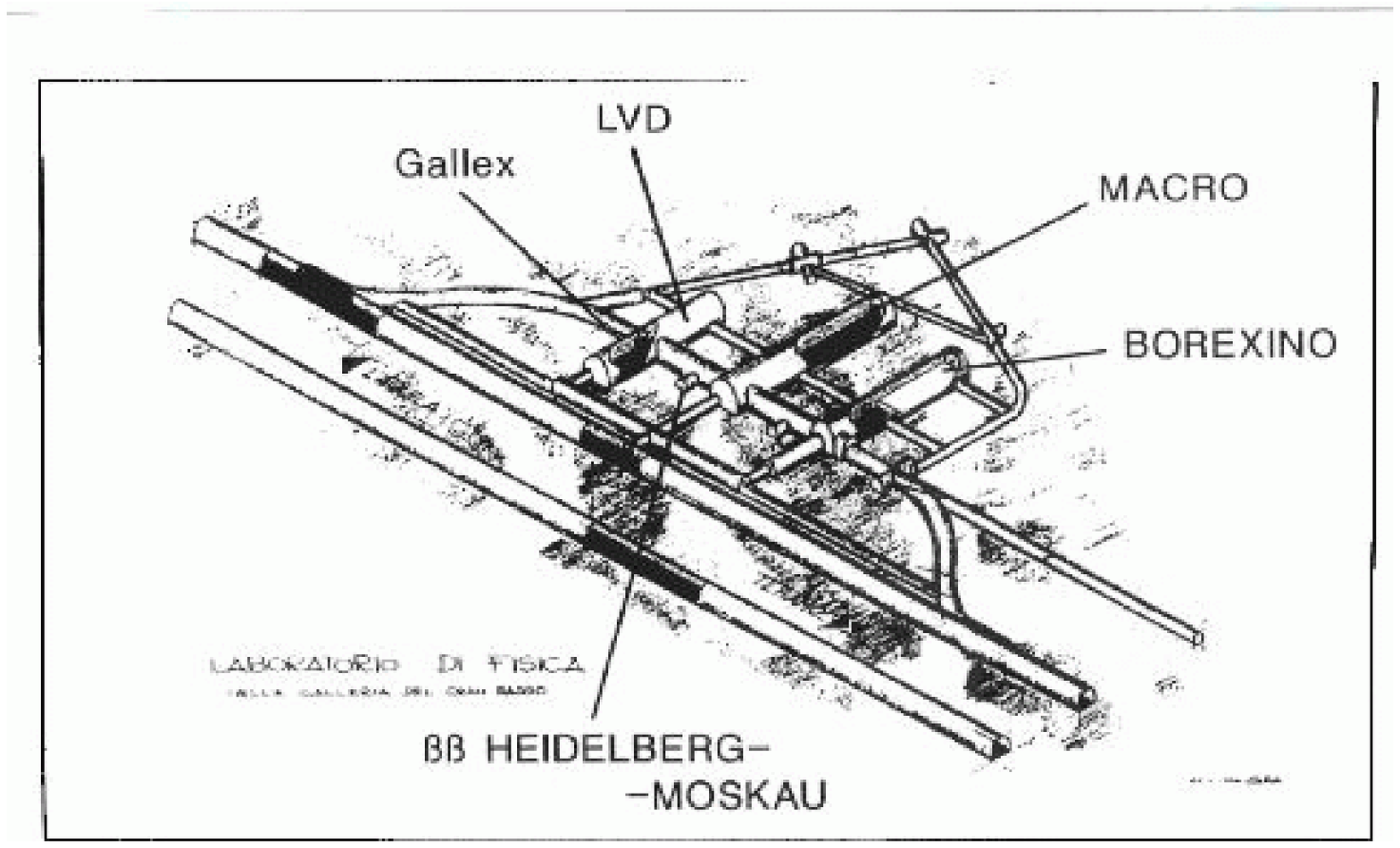}
\includegraphics[height=4.3cm,width=5cm]{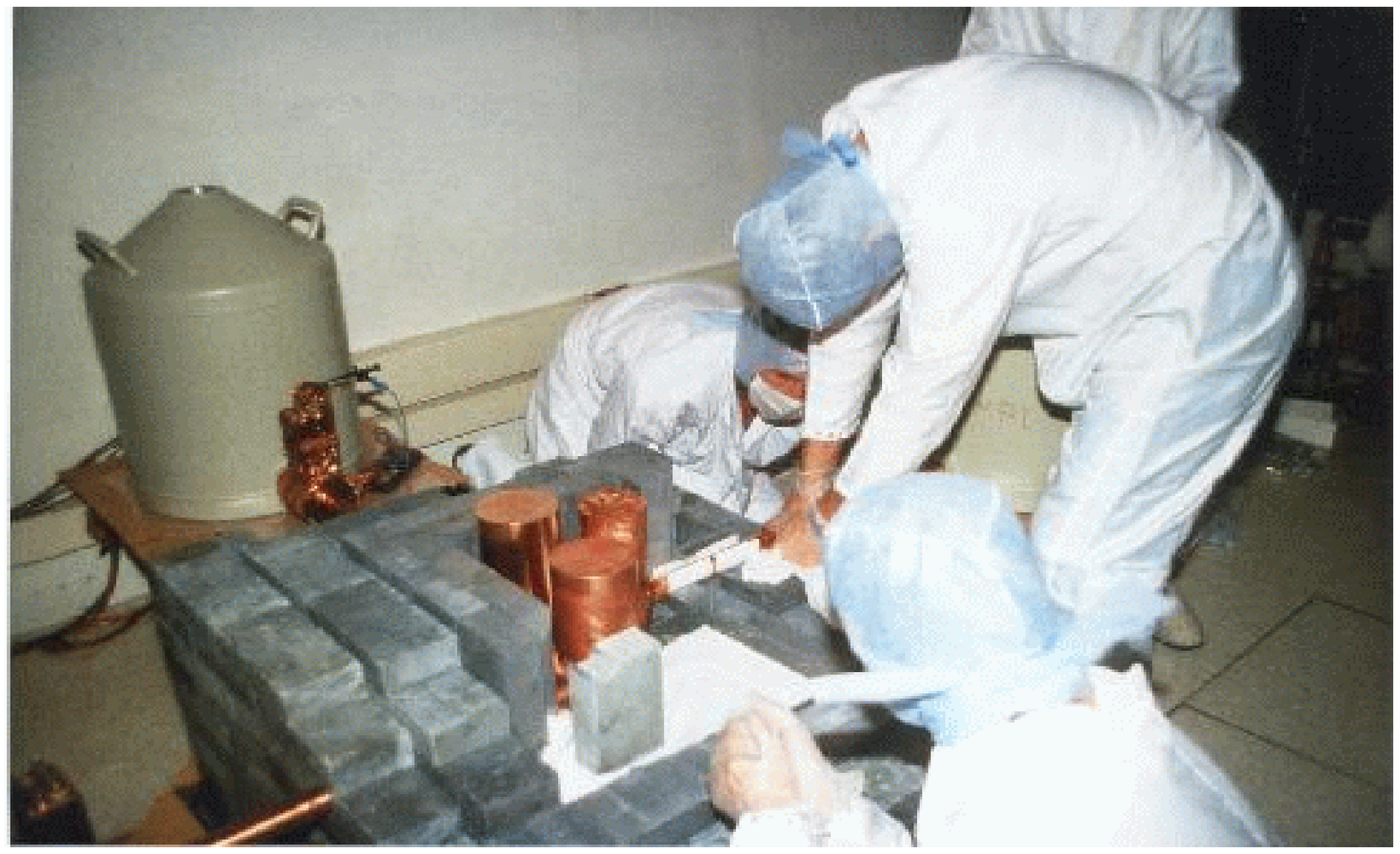}
\includegraphics[height=4.3cm,width=7cm]{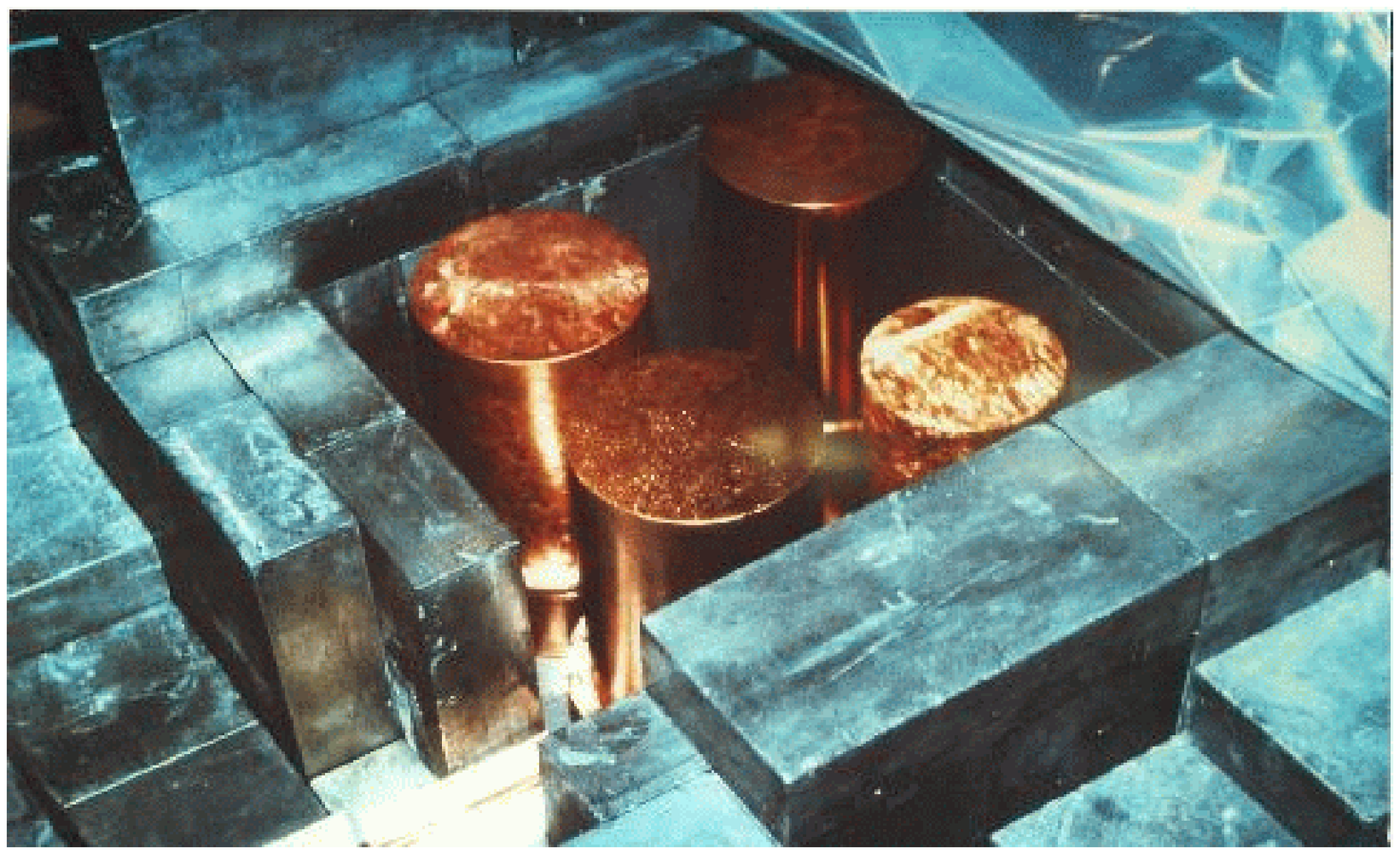}
\end{center}
\caption{The $\beta\beta$-laboratory of the HEIDELBERG-MOSCOW 
	experiment in the Gran Sasso (upper left) 
	and its location between halls A and B (upper right), 
	and four of the enriched detectors during installation (lower part).
}
\label{Set-up}
\end{figure}

	We give now some experimental details. 
	All detectors (whose technical parameters 
	are given in Table 1 (see    
\cite{HDM97})), 
	except detector No. 4, are operated 
	in a common Pb shielding of 30 cm, which consists of 
	an inner shielding of 10 cm radiopure LC2-grade Pb followed 
	by 20 cm of Boliden Pb. The whole setup is placed 
	in an air-tight steel box 
	and flushed 
	with radiopure nitrogen in order to suppress the $^{222}{Rn}$ 
	contamination of the air. The steel box is centered 
	inside a 10-cm boron-loaded polyethylene shielding 
	to decrease the neutron flux from outside. 
	An active anticoincidence shielding is placed on the top 
	of the setup to reduce the effect of muons.
	Detector No. 4 is installed in a separate setup, which has 
	an inner shielding of 27.5 cm electrolytical Cu, 20 cm lead, 
	and boron-loaded polyethylene shielding below the steel box,  
	but no muon shielding. 
	Fig.
\ref{Set-up}
	gives a view of the experimental setup.

	To check the stability of the experiment, a calibration 
	with a $^{228}{Th}$ and a 
	$^{152}{Eu}$+$^{228}{Th}$, and a $^{60}{Co}$ source is done weekly.
	High voltage of the detectors, temperature in the detector 
	cave and the computer room, the nitrogen flow  
	in the detector boxes, the muon anticoincidence signal,  
	leakage current of the detectors, overall and individual 
	trigger rates are monitored daily.  
	The energy spectrum is taken in 8192 channels in the  
	range from threshold up to about 3 MeV, and in 
	a parallel spectrum up to about 8 MeV.  

	Because of the big peak-to-Compton ratio of 
	the large detectors, external $\gamma$ activities are 
	relatively easily 
	identified, since their 
	Compton continuum is to a large extent shifted into the peaks. 
	The background identified 
	by the measured $\gamma$ lines in the background 
	spectrum consists of:  
\vspace{-0.1cm}
\begin{enumerate}
\item 
	primordial activities of the natural 
	decay chains from $^{238}{U}$, $^{232}{Th}$, and $^{40}{K}$  
\item 	
	anthropogenic radio nuclides, like $^{137}{Cs}$, 
	$^{134}{Cs}$, $^{125}{Sb}$, $^{207}{Bi}$ 
\item 
	cosmogenic isotopes, produced by activation 
	due to cosmic rays. 

	The activity of these sources 
	in the setup is measured directly and can be located due 
	to the measured and simulated relative peak intensities 
	of these nuclei. 

	Hidden in the continuous background 
	are the contributions of 
\item 	
	the bremsstrahlungs
	spectrum of $^{210}{Bi}$ (daughter of $^{210}{Pb}$), 
\item 	
	elastic and inelastic neutron scattering, and 
\item 
	direct muon-induced events.
\end{enumerate}		
	External $\alpha$ and $\beta$ 
	activities are shielded by the 0.7-mm inactive zone 
	of the p-type Ge detectors on the outer layer of the crystal. 
	The enormous radiopurity of HP-germanium is proven by the fact 
	that the detectors No. 1, No. 2 and No. 3 show 
	no indication of any $\alpha$ peaks in the measured data. 
	Therefore no 
	contribution of the natural decay 
	chains can be located inside the crystals. 
	Detectors No. 4 and 
	No. 5 seem to be slightly contaminated with $^{210}{Pb}$ 
	on the level of few $\mu$Bq/kg,  
	most likely surface contaminations at the inner contact. 
	This contamination was identified 
	by a measured $\alpha$ peak in the background spectrum 
	at 5.305\,MeV of the daughter $^{210}{Po}$ and the constant 
	time development of the 
	peak counting rate. There is no contribution to the background 
	in  the interesting evaluation areas of the experiment 
	due to this activity. 
	For further details about the experiment and background we refer to 
\cite{HDM97,Diss-Dipl} (see also Table 2).

	In the vicinity of the Q-value of the double beta decay of
	Q$_{\beta\beta}$~=~2039.006(50) \,keV,
	very weak lines at 2034.744 and 2042\,keV  
	from the cosmogenic nuclide $^{56}{Co}$, 
	and from $^{214}{Bi}$ 
	($^{238}U$-decay chain) at 2010.7, 2016.7, 2021.8 and 2052.9\,keV,  
	may be expected.

	On the other hand, there are no background $\gamma$-lines 
	at the position of an expected \onbb line, 
	according to our Monte Carlo analysis 
	of radioactive impurities in the experimental setup 
\cite{Diss-Dipl}
	and according to the compilations in 
\cite{Tabl-Isot96}.

	In total 55 possible $\gamma$-lines from various isotopes in the
	region between 2037 and 2041\,keV are known 
\cite{Tabl-Isot96}. 
	Only 5 of the isotopes responsible for them (\ch{102}{Rh},
	\ch{146}{Eu}, \ch{124}{I}, \ch{124}{Sb} and \ch{170}{Lu}) 
	have half-lifes larger than 1 day. 
	However, some of the isotopes yielding lines in this energy range 
	can in principle be produced by inelastic hadron
	reactions (induced by muons or neutrons).

	Therefore each of these 55 isotopes was checked for the existence of a
	$\gamma$-line from the isotope which has a high emission 
	probability $I_h$. 
	A search was made for this $\gamma$-line 
	in the {\em measured} spectrum 
	to obtain its intensity ($S_h$) or an upper limit for it.
	Then the adopted intensity $S_0$ for a $\gamma$-line 
	from the same isotope 
	in the area around $\sim$2039\,keV can be calculated by
	using the emission probability $I_0$ for the line at $\sim$2039\,keV.
	Different absorption for gammas of different energies 
	are taken into account in a schematic way.


\begin{table}[h]
\begin{center}
\renewcommand{\arraystretch}{.8}
\setlength\tabcolsep{1.pt}
\begin{tabular}{c|c|c|p{1.5cm}p{1.cm}p{1.cm}|p{2.cm}|p{1.5cm}}
\hline
	Detec-		&	Life				&
	Date		&	\multicolumn{3}{c}{Shielding}	&
Background${^*)}$ 	&	~~~~PSA	\\

			&					&
			&					&
			&					&
{~~[counts/}		&		\\

	tor		&	Time				&
			&					&
			&					&
{~{keV y kg]}}		&		\\
\cline{4-8}
\hphantom{A}
			&					&
			&					&	
			&	boron-				&
~2000.-2100.		&		\\

		Number	&	[ days ]			&
	Start End	&	~Cu			&	
		Pb	&	poly.			&
	~~~~~keV	&		\\
\hline
\hline
	No.1		&		387.6			&
	8/90-8/91	&	~yes				&
			&					&
~~\hphantom{A}0.56	&	~~\hphantom{A}no	\\
\hline
			&					&
	1/92-8/92	&					&
			&					&
			&	~~\hphantom{A}no	\\
\hline
	No.2		&	225.4				&
	9/91 - 8/92	&					&
	yes		&					&
	~~\hphantom{A}0.29	&	~~\hphantom{A}no	\\
\hline
\multicolumn{8}{c}{\bf Common shielding for three detectors}	\\
\hline
	No.1		&	382.8	&
	9/92 - 1/94	&&	yes	&
			&	
	~~\hphantom{A}0.22		&	~~\hphantom{A}no	\\
\hline
	No.2		&	383.8	&
	9/92 - 1/94	&&	yes	&
			&	
	~~\hphantom{A}0.22		&	~~\hphantom{A}no	\\
\hline
	No.3		&	382.8	&
	9/92 - 1/94	&&	yes	&
			&		
	~~\hphantom{A}0.21		&	~~\hphantom{A}no	\\
\hline
	No.1		&	263.0	&
	2/94 - 11/94	&&	yes	&
	yes		&	
	~~\hphantom{A}0.20		&	~~\hphantom{A}no	\\
\hline
	No.2		&	257.2	&
	2/94 - 11/94	&&	yes	&
	yes		& 	
	~~\hphantom{A}0.14		&	~~\hphantom{A}no	\\
\hline
	No.3		&	263.0	&
	2/94 - 11/94	&&	yes	&
	yes		&		
	~~\hphantom{A}0.18		&	~~\hphantom{A}no	\\
\hline
\multicolumn{8}{c}{\bf Full Setup}\\
\multicolumn{8}{c}{\bf Four detectors in common shielding, 
one detector separate}\\
\hline
	No.1		&	203.6	&
	12/94 - 8/95	&&
	yes		&	yes		&	
	~~\hphantom{A}0.14		&	~~\hphantom{A}no	\\
\hline
	No.2		&	203.6	&
	12/94 - 8/95	&&
	yes		&	yes		&	
	~~\hphantom{A}0.17		&	~~\hphantom{A}no	\\
\hline
	No.3		&	188.9	&
	12/94 - 8/95	&&
	yes		&	yes		&	
	~~\hphantom{A}0.20		&	~~\hphantom{A}no	\\
\hline
	No.5		&	48.0	&
	12/94 - 8/95	&&
	yes		&	yes		&	
	~~\hphantom{A}0.23		&	~since 2/95	\\
\hline
	No.4		&	147.6	&
	1/95 - 8/95	&	~yes		&
&&	
	~~\hphantom{A}0.43		&	~~\hphantom{A}no	\\
\hline
\hline
	No.1		&	203.6	&
	11/95 - 05/00	&&
	yes		&	yes		
&	~~\hphantom{A}0.170		&	~~\hphantom{A}no	\\
\hline
	No.2		&	203.6	&
	11/95 - 05/00	&&
	yes		&	yes		
&	~~\hphantom{A}0.122		&	~~\hphantom{A}yes	\\
\hline
	No.3		&	188.9	&
	11/95 - 05/00	&&
	yes		&	yes		
&	~~\hphantom{A}0.152		&	~~\hphantom{A}yes\\
\hline
	No.5		&	48.0	&
	11/95 - 05/00	&&
	yes		&	yes		
&	~~\hphantom{A}0.159		&	~~\hphantom{A}yes	\\
\hline
	No.4		&	147.6	&
	11/95 - 05/00	&	~yes		
&&	~
&	~~\hphantom{A}0.188		&	~~\hphantom{A}yes	\\
\hline
\end{tabular}
\end{center}
\caption{\label{activities}Development of the experimental set-up and of the 
	background numbers 
	in the different data acquisition periods for the enriched detectors
	of the HEIDELBERG-MOSCOW experiment. 
	${^*)}$ Without PSA method.}
\end{table}


	If the calculated intensity $S_l<1$ for the $\gamma$-line in the
	interesting area, this isotope can be safely excluded to contribute a
	significant part to the background.

	For example, the isotope \ch{139}{Xe} possesses a $\gamma$-line at
	2039.1\,keV with an emission probability of 0.078\%.
	This isotope also possesses a $\gamma$-line at 225.4\,keV 
	with an emission probability of 3.2\%. 
	The intensity of the line at 225.4\,keV in our spectrum
	was measured to be $<$6.5 counts. This means that the $\gamma$-line at
	2039.1\,keV has $< 2\frac{0.078}{3.2}6.5=0.32$ counts, and
	therefore can be excluded.

	Only eight isotopes could contribute a few counts according to the
	calculated limits, in the
	interesting area for the \onbb-decay area:
	\ch{52m}{Fe}, \ch{93m}{Ru}, \ch{120}{In}, \ch{131}{Ce}, 
	\ch{170}{Lu}, \ch{170m}{Ho}, \ch{174}{Ta} and \ch{198}{Tl}.
	Most of them have a half-life of a few seconds, only
	\ch{170}{Lu} has a half-life of 2.01 days, and no one of them
	has a longer living mother isotope.
	To contribute to the background they must be
	produced with a constant rate, e.g. by inelastic neutron and/or muon
	reactions.
	Only 5 isotopes can be produced in a reasonable way, 
	by the reactions listed in Tabl. 
\ref{isotops}.


\begin{table}
\renewcommand{\arraystretch}{1.}
\setlength\tabcolsep{1.pt}
\begin{center}
\begin{tabular}{l|l|l|l|l|l}
\hline
isotope & 	\ch{52m}{Fe} &	\ch{93m}{Ru} &	\ch{120}{In} 
& 	\ch{170m}{Ho} &	\ch{198}{Tl} \\
\hline
&  &  &&  &  \\
production & \ch{50}{Cr}($\alpha,2n\gamma$)	
& \ch{92}{Mo}($\alpha,3n$)	& \ch{120}{Sn}($n,p$),	
& \ch{170}{Er}($n,p$) 	
& \ch{197}{Au}($\alpha,3n\gamma$) \\
 reaction &  &  &  \ch{123}{Sb}($n,\alpha$)  &   &   \\
\hline 
\end{tabular}
\end{center}
\caption{\label{isotops}
	Possible reactions (second line) which could produce the
  	isotopes listed in the first line (see 
\cite{toi}).} 
\end{table}


	Except \ch{120}{Sb} each of the target nuclides is stable. 
	All reactions induced with $\alpha$-particles can be excluded 
	due to the very short interaction length of $\alpha$-particles.
	Two possibilities remaining to explain possible events in the
	\onbb-decay area would be:

\begin{itemize}
\item \ch{120}{Sn}($n,p$)\ch{120}{In}:\\
	The cross section for this reaction is 2.5$\pm$1 mb for $E_n=14.5$\,MeV
\cite{ref1}. 
	Assuming a neutron flux of (0.4$\pm$0.4)$\times
	10^{-9}\frac{1}{cm^2\;s}$ for neutrons with an energy 
	between 10-15\,MeV as measured in the Gran Sasso 
\cite{Arn99} 
	the rate of \ch{120}{In} atoms 
	produced per year is about $2\times10^{-5}$ when there are 50\,g of
	\ch{120}{Sn} in the detector setup.
	Even when the cross-section is larger for lower energies, this can not
	contribute a significant number of counts to the background.

\item \ch{170}{Er}($n,p$)\ch{170m}{Ho}:\\
	The cross section for this reaction is about 
	1.13$\pm$1\,mb for $E_n=14.8$\,MeV
\cite{ref2}. 
	Assuming again a neutron flux of (0.4$\pm$0.4)$\times
	10^{-9}\frac{1}{cm^2\;s}$ for neutrons 
	with an energy between 10-15\,MeV the rate 
	of \ch{170}{Er} atoms produced per year is even less
	when assuming 50\,g of \ch{120}{Er} in the detector-setup.
\end{itemize}

	In both cases it would be not understandable, 
	how such large amounts of 
	\ch{120}{Sn} or \ch{170}{Er} could have come 
	into the experimental setup.
	Concluding we do not find indications for any nuclides, 
	that might produce
	$\gamma$-lines with an energy around 2039\,keV 
	in the experimental setup.

\begin{figure}[t]
\begin{center}
\includegraphics[height=5.7cm,width=11cm]{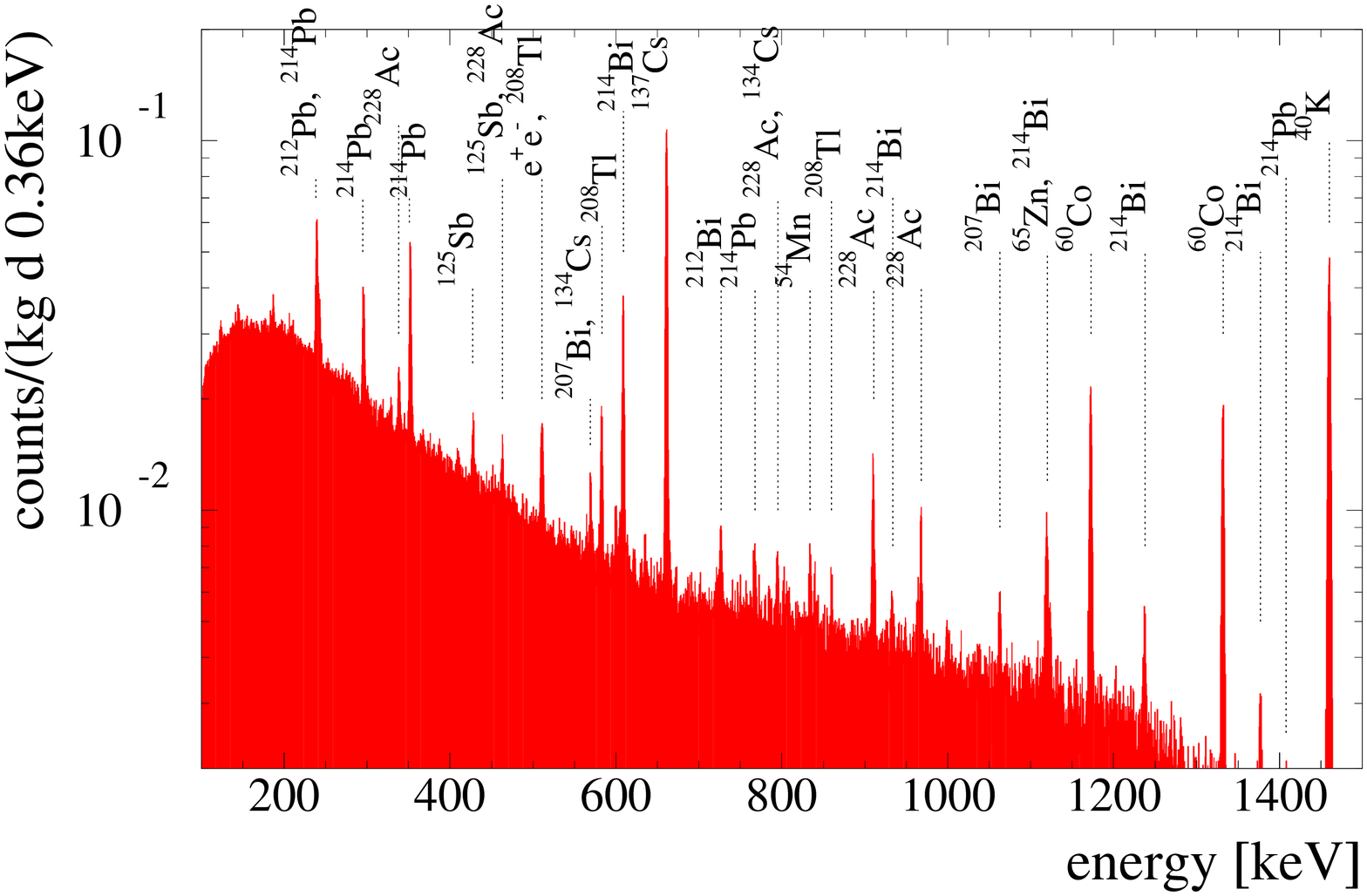}
\includegraphics[height=5.7cm,width=11cm]{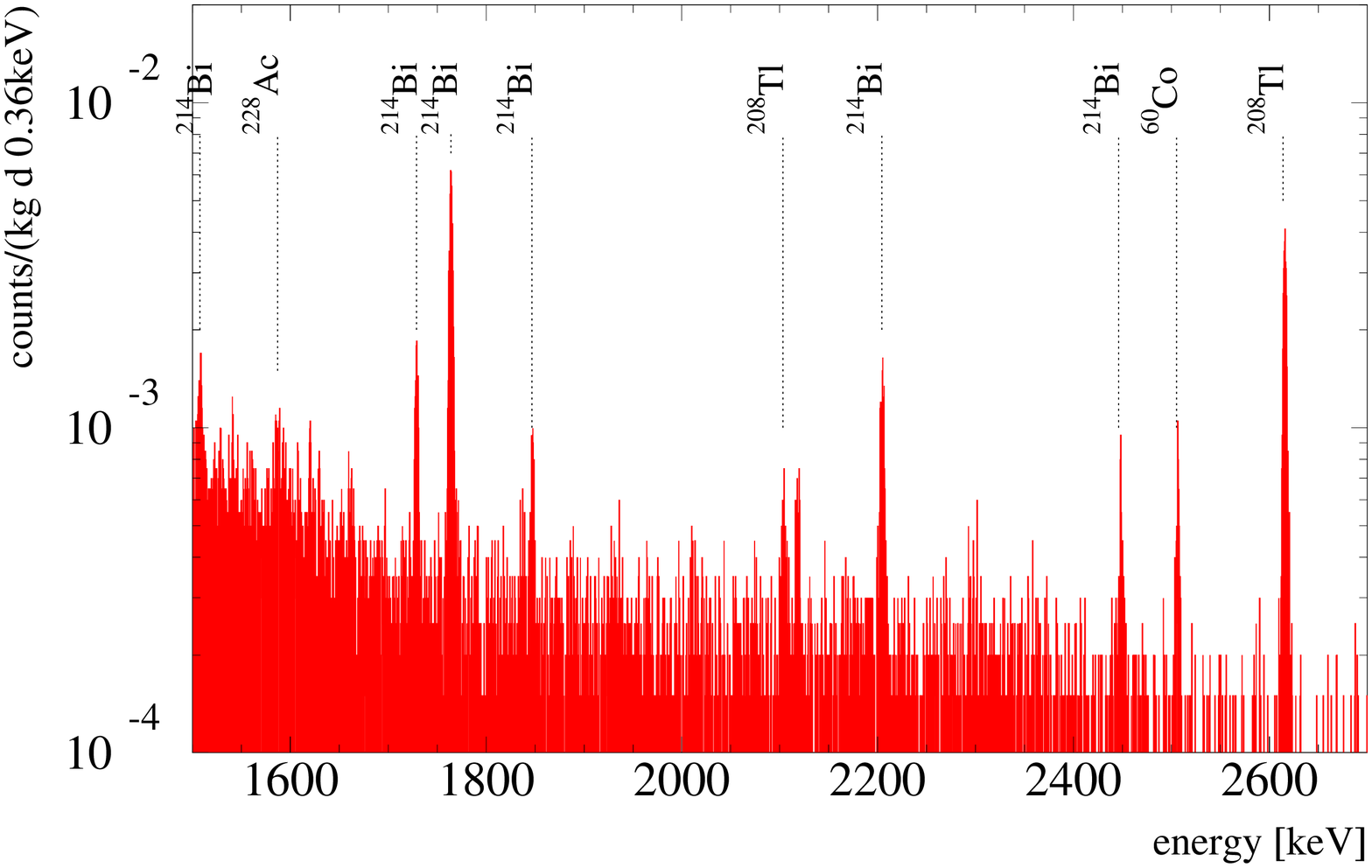}
\end{center}
\caption{Sum spectrum of  
	enriched detectors Nr. 1,2,3,4,5 over the period 
	August 1990 - May 2000 (54.9813\,kg\,y) measured 
	in the HEIDELBERG-MOSCOW experiment (binning 0.36\,keV). 
	The sources of the main identified 
	background lines are noted.
}
\label{Full}
\end{figure}

\begin{figure}[t]
\begin{center}
\includegraphics[scale=0.30]
{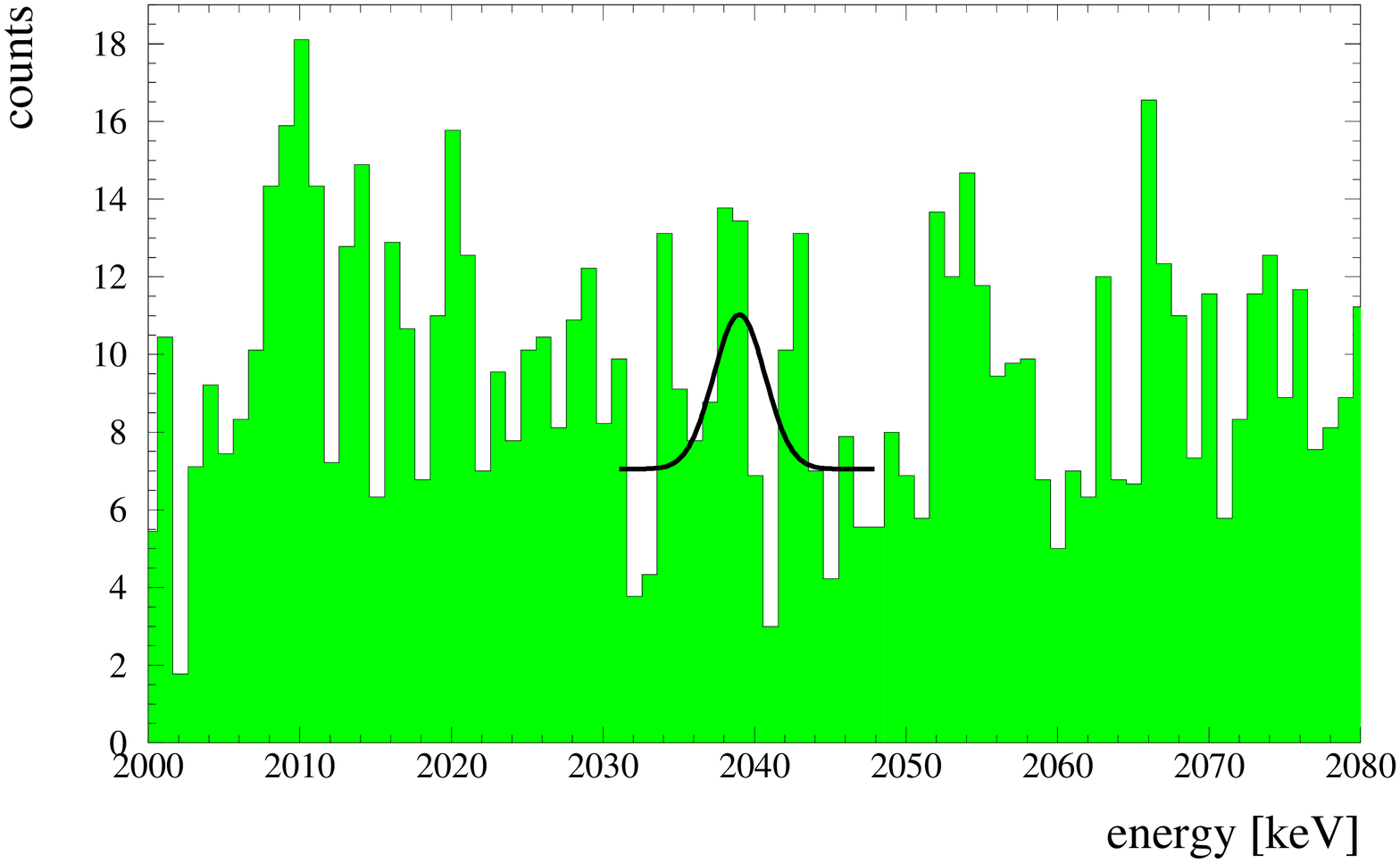} 
\end{center}

\vspace{-0.8cm}
\caption{Sum spectrum of the 
	$^{76}{Ge}$ detectors Nr. 1,2,3,4,5 over the period 
	August 1990 to May 2000, 
	(54.9813\,kg\,y) in the energy interval 2000 - 2080 keV, around the 
	Q$_{\beta\beta}$ value of double beta decay 
	(Q$_{\beta\beta}$~=~2039.006(50) keV) 
	summed to 1\,keV bins.
	The curve results from Bayesian inference in the 
	way explained in sec.3. It corresponds to a half-life 
	T$_{1/2}^{0\nu}$=(0.80 - 35.07)$\times~10^{25}$~y (95\% c.l.) 
}
\label{Sum_spectr_Alldet_1990-2000}
%
%
%
%
%
\begin{center}
\includegraphics[scale=0.33]
{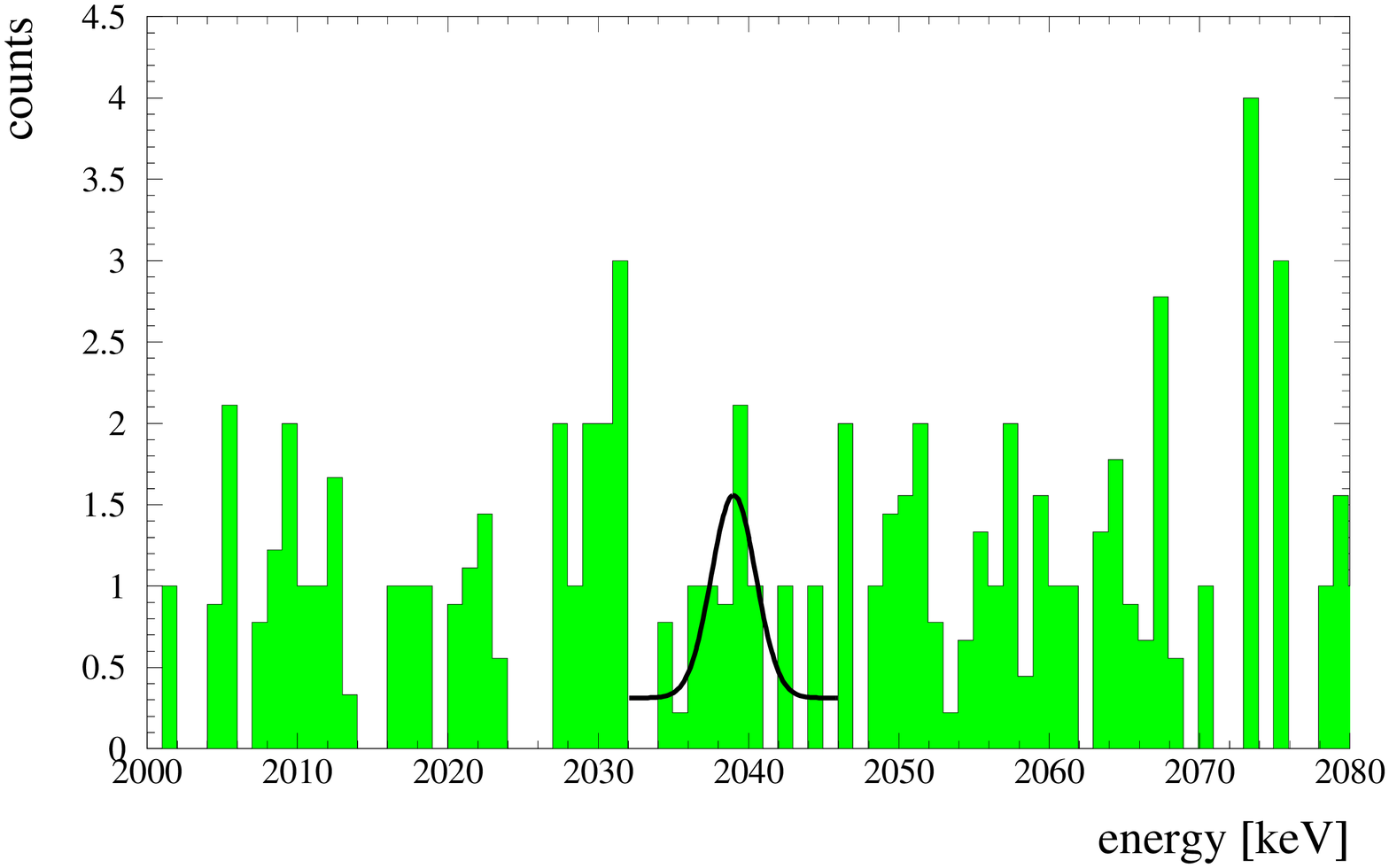} 
\end{center}

\vspace{-0.8cm}
\caption{Sum spectrum of single site events, 
	measured with the detectors 
	Nr. 2,3,5 operated with pulse shape analysis in the period 
	November 1995 to May 2000 (28.053\,kg\,y),  
	summed to 1\,keV bins. 
	Only events identified as single site events (SSE) 
	by all three pulse shape analysis methods 
\cite{HelKK00,Patent-KKHel,KKMaj99}
	have been accepted.
	The curve results from Bayesian inference in the 
	way explained in sec.3. 
	When corrected for the efficiency 
	of SSE identification (see text), 
	this leads to the following value for the half-life:
	T$_{1/2}^{0\nu}$=(0.88 - 22.38)$\times~10^{25}$~y (90\% c.l.).
}
\label{Sum_spectr_5det}
\end{figure}

	Fig. 
\ref{Full} 
	shows the combined spectrum of 
	the five enriched detectors obtained over the period 
	August 1990 - May 2000, with a statistical significance 
	of 54.981 kg\,y (723.44\,molyears). 
	(Note that in Fig. 1 of 
\cite{HDM01})
	only the spectrum of the first detector is shown, 
	but normalized to 47.4\,kg\,y 
\cite{Errat-ZPh01}).
	The identified background lines give an indication 
	of the stability of the electronics over a decade of measurements. 
	The average rate (sum of all detectors) 
	observed over the measuring time, 
	has proven to be constant 
	within statistical variations.
	Fig. 
\ref{Sum_spectr_Alldet_1990-2000} 
	shows the part of 
	the spectrum shown in Fig. 
\ref{Full},
	in more detail around the Q-value of double beta decay.   
	Fig. 
\ref{Sum_spectr_5det} 
	shows the spectrum of single site events (SSE) 
	obtained for detectors 2,3,5 
	in the period November 1995 - May 2000, 
	under the restriction that the signal simultaneously fulfills  
	the criteria of {\it all three} methods 
	of pulse shape analysis we have recently developed 
\cite{HelKK00,KKMaj99} 
	and with which we operate all detectors except detector 1
	(significance 28.053\,kg\,y) since 1995.


	Double beta events are single site events 
	confined to a few mm region in the detector 
	corresponding to the track length of the emitted electrons.
	In all three methods, mentioned above, 
	the output of the charge-sensitive preamplifiers 
	was differentiated with 10-20\,ns sampled with 250\,MHz and 
	analysed off-line. The methods differ in the analysis 
	of the measured pulse shapes. 
	The first one relies on the broadness 
	of the charge pulse maximum, the second 
	and third one are based on neural 
	networks. All three methods are 'calibrated' with known 
	double escape (mainly SSE) and total absorption 
	(mainly MSE) $\gamma$-lines
\cite{HelKK00,Patent-KKHel,KKMaj99,Diss-Dipl}. 
	They allow to achieve about 80$\%$ detection 
	efficiency for both interaction types.

	The expectation for a \onbb signal would be a line of single 
	site events on some background of multiple site events 
	but also single site events, the latter coming to a large 
	extent from the continuum of the 2614\,keV $\gamma$-line from 
	$^{208}{Tl}$ (see, e.g., the simulation in 
\cite{HelKK00}).
	From simulation we expect that about 5$\%$  
	of the double beta single site events should be seen as MSE. 
	This is caused by bremsstrahlung of the emitted electrons
\cite{Diss-Dipl}.

	Installation of PSA has been performed in 1995 for the  
	four large detectors. Detector 
	Nr.5 runs since February 1995, detectors 2,3,4 since 
	November 1995 with PSA. 
	The measuring time with PSA  
	from November 1995 until May 2000 is 36.532\,kg years, 
	for detectors 2,3,5 it is 28.053\,kg\,y.

\begin{figure}[h]
\begin{center}
\includegraphics[scale=0.21]{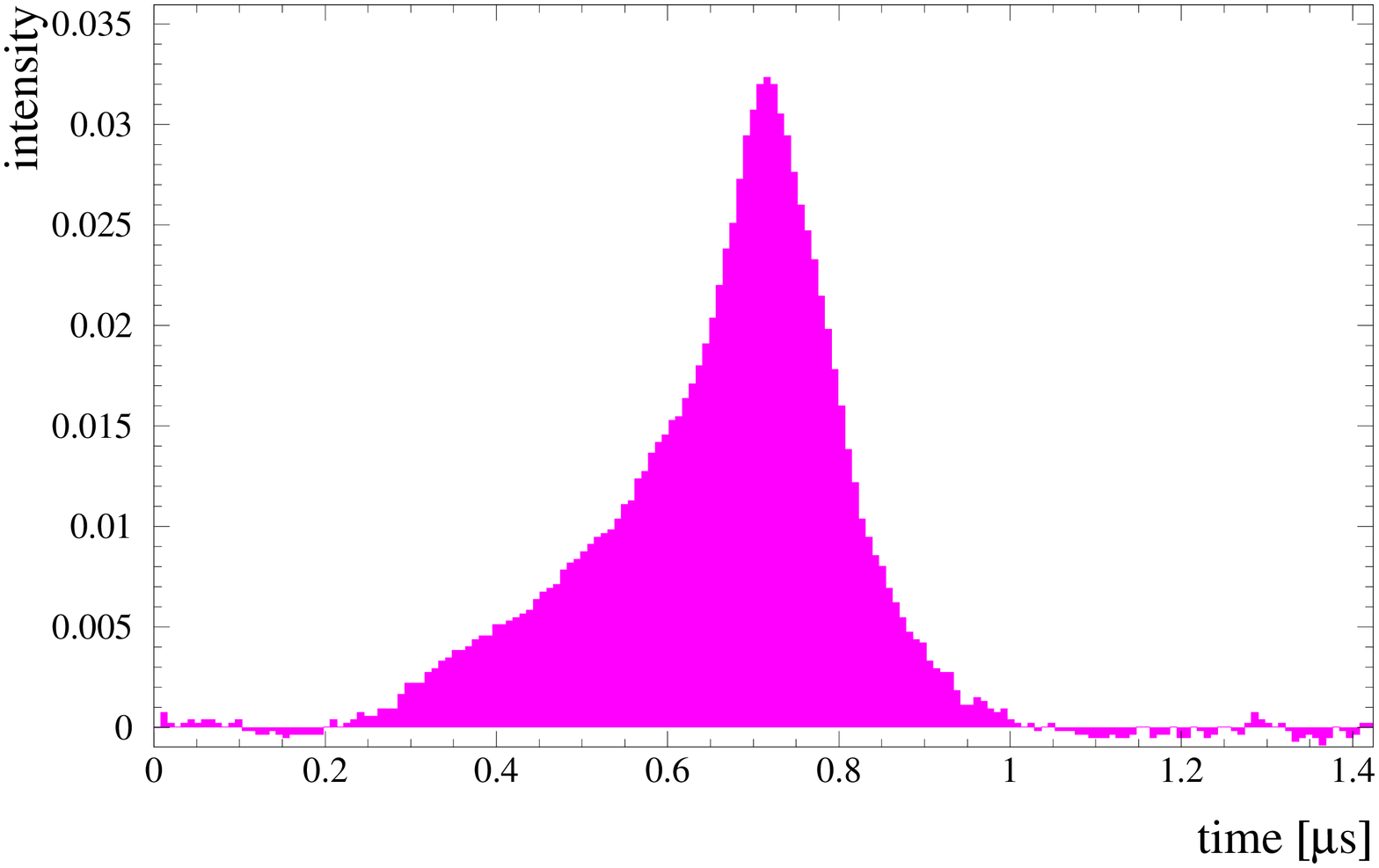} 
\includegraphics[scale=0.21]{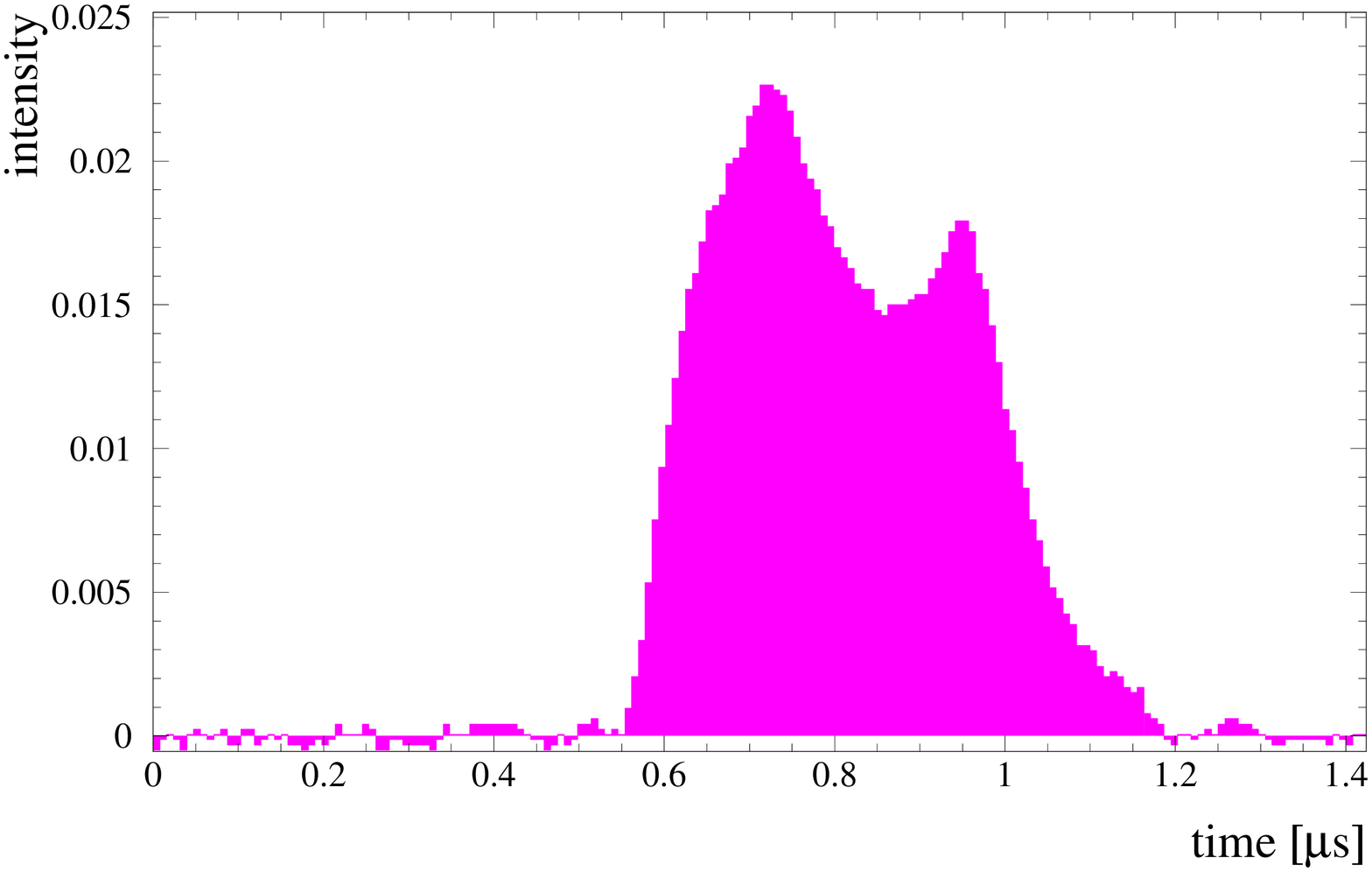} 
\end{center}
\caption{Left: Shape of one candidate for \onbb decay (energy 
	2038.61\,keV) classified as SSE by all three methods 
	of pulse shape discrimination.  
	Right: Shape of one candidate (energy 
	2038.97\,keV) 
	classified as MSE by all three methods.}
\label{Shape}
\end{figure}

	Fig. 
\ref{Shape} 
	shows typical SSE and MSE events from our spectrum.


	All the spectra are obtained 
	after rejecting coincidence events between different 
	Ge detectors and events coincident with activation 
	of the muon shield. 
	The spectra, which are taken in bins of 0.36\,keV, are shown
	in Figs. 
\ref{Sum_spectr_Alldet_1990-2000},\ref{Sum_spectr_5det}, 
	Fig.2 of 
\cite{KK-Evid01} 
	in 1\,keV bins, which explains 
	the broken number in the ordinate. 
	We do the analysis of the measured spectra 
	with (Fig. 
\ref{Sum_spectr_Alldet_1990-2000}) 
	and without the data of detector 4 (see Fig.2 in 
\cite{KK-Evid01}, 46.502\,kg\,y) 
	since the latter 
	does not have a muon shield and has the weakest 
	energy resolution. 
	The 0.36\,keV bin spectra are used in all analyses 
	described in this work.
	We ignore for each detector the first 200\,days of operation, 
	corresponding to about three half-lives of $^{56}{Co}$  
	($T_{1/2}$ = 77.27\,days), 
	to allow for some decay of short-lived radioactive impurities.


\begin{figure}[h]
\vspace{-0.3cm}
\begin{center}
\includegraphics[scale=0.45]{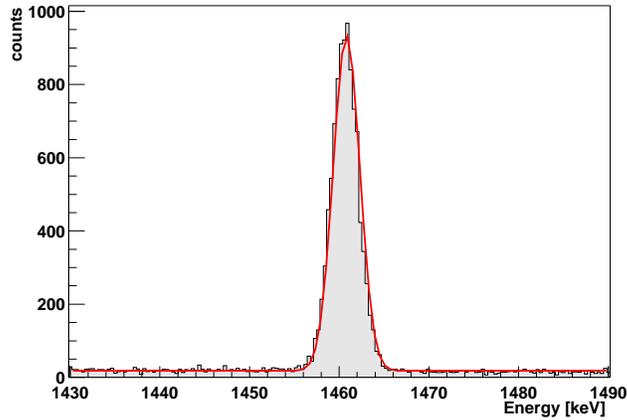}
\end{center}
\caption{Fit of one the lines (here the 1460.81\,keV line from $^{40}{K}$) 
	in the sum spectrum of all five detectors, used for 
	the calibration and the determination of the energy 
	resolution (see Table 4).}
\label{Calibr-1460}
\end{figure}


	The background rate in the energy range 2000 - 2080\,keV 
	is found to be (0.17 $\pm$ 0.01)\,events/\,kg\,y\,keV 
	({\it without} pulse shape analysis)  
	considering {\it all} data as background. 
	This is the lowest value ever obtained in such type of experiments.
	The energy resolution at the Q$_{\beta\beta}$ value 
	in the sum spectra 
	is extrapolated from the strong lines in the spectrum 
	to be (4.00 $\pm$ 0.39)\,keV in the spectra with detector 4, 
	and (3.74 $\pm$ 0.42)\,keV (FWHM) in the spectra without detector 4 
	(see Fig. 
\ref{Calibr-1460} and Table 
\ref{Calibr}).	
	The energy calibration of the experiment has an uncertainty
	 of 0.20\,keV (see Table  
\ref{Calibr}).


\begin{table}
\renewcommand{\arraystretch}{1.}
\setlength\tabcolsep{2.pt}
\begin{center}
\begin{tabular}{c|c|c|c}
\hline
energy [keV] & energy [keV] & width [keV] & width [keV] \\ 
fit          & from \protect\cite{Tabl-Isot96}  & fit         & from calc.  \\ \hline
1460.81 $\pm$ 0.02   & 1460.81 & 1.49 $\pm$ 0.01 & 1.49 $\pm$ 0.13 \\
1764.56 $\pm$ 0.05 & 1764.49 & 1.70 $\pm$ 0.05 & 1.59 $\pm$ 0.15\\
2103.31 $\pm$ 0.45 & 2103.53 & 1.86 $\pm$ 0.35 & 1.71 $\pm$ 0.16\\
2204.12 $\pm$ 0.14 & 2204.19 & 1.89 $\pm$ 0.13 & 1.74 $\pm$ 0.17\\
2447.73 $\pm$ 0.26 & 2447.86 & 1.82 $\pm$ 0.33 & 1.82 $\pm$ 0.18\\
2614.48 $\pm$ 0.07 & 2614.53 & 1.80 $\pm$ 0.06 & 1.88 $\pm$ 0.18\\
\hline
\end{tabular}
\end{center}
\caption{\label{Calibr}
	Energies and widths of some prominent lines in the sum spectrum 
	of all five detectors determined by our energy calibration 
	and peak fit methods, and comparison with the energy given 
	in the literature 
\cite{Tabl-Isot96},
	and the fitted dependence of the width as function of energy.} 
\end{table}




\section{Data Analysis}

	We analyse the measured spectra 
	with the following methods:

	1. {\sf Bayesian inference}, which is used widely
	at present in nuclear and astrophysics (see, e.g. 
\cite{Bayes_Method-General,RPD00,Hagan94,Dietz-Diss03}). 
	This method is particularly suited for low
	counting rates, where the data follow 
	a Poisson distribution, that cannot be approximated by a Gaussian. 

	2. {\sf Feldman-Cousins Method} 
\cite{Feldm98,RPD00}.

	3. {\sf Maximum Likelihood Method (see 
\cite{RPD00,Kchi-kvadr}).}

	The Bayesian method 
	is described in the next subsection, 3.1. 
	In subsection 3.2 we give some numerical examples 
	of the sensitivities of this method and 
	of the Maximum Likelihood Method in the 
	search for events of low statistics 
\cite{Dietz-Diss03}.

\subsection{The Bayesian Method}

	We first describe the procedure summarily 
	and then give some mathematical details (see 
\cite{Dietz-Diss03}).

	One knows that the lines in the spectrum 
	are Gaussians with the standard deviation 
	$\sigma$=1.70\,keV in Fig. 
\ref{Sum_spectr_Alldet_1990-2000} 
	and $\sigma$=1.59\,keV in Fig. 
\ref{Sum_spectr_5det}. 
	This corresponds to 4.0(3.7)\,keV FWHM. 
	Given the position of a line, 
	we used Bayes theorem to infer the contents 
	of the line and the level of a constant background.

	Bayesian inference yields 
	the joint probability distribution for both parameters. 
	Since we are interested in the contents of the line, 
	the other parameter was integrated out.
	This yields 
	the distribution of the line contents. 
	If the distribution has its maximum at zero contents, 
	only an upper limit for the contents can be given 
	and the procedure does not suggest the existence of a line. 
	If the distribution has its maximum at non-zero contents, 
	the existence of a line is suggested and one can define 
	the probability K$_E$ that there is a line with non-zero contents.

	We define the Bayesian procedure in some more detail. 
	It starts from the distribution 
	$p(x_1...x_M|\rho,\eta)$
	of the count rates $x_1 ...x_M$ in the bins 1..M of the spectrum - 
	given two parameters $\rho,\eta$.
	The distribution $p$ is the product 
\beq{}
	p(x_1...x_M|\rho,\eta)~
	=~\prod_{k=1}^{M} \frac{{\lambda_k}^{x_k}}{x_k!}~e^{-\lambda_k}
\label{probab}
\eeq	
	of Poissonians for the individual bins. 
	The expectation value $\lambda_k$ is the superposition 
\beq{}
	\lambda_k~=~\rho~[\eta~ f_1(k) + (1 - \eta)~ f_2(k)]
\label{lambda-h}
\eeq
	of the form factors f$_1$ of the line 
	and f$_2$ of the background; 
	i.e. f$_1$(k) is the Gaussian centered 
	at E with the above-mentioned standard deviation
	value and f$_2$(k)$\equiv$f$_2$ is a constant.
	Note that the model allows for a spectrum 
	of background {\it only}, i.e. $\eta$=0, 
	and in this sense also tests the hypothesis 'only background'.

	Since
\beq{}
	\sum_{k=1}^{M} f_\nu(k)~=~1   \hspace{1.5cm}for~~ \nu=1,2,
\label{sum-f}
\eeq
	one has
\beq{}
	\sum_{k=1}^{M} \lambda_k~=~\rho. 
\label{sum-l}
\eeq
 
	Hence, $\rho$ parametrizes the total intensity in the spectrum, 
	and $\eta$ is the relative
	intensity in the Gaussian line.

	The total intensity $\rho$ shall be integrated out. 
	For this purpose, one needs the prior distribution 
	$\mu(\rho|\eta)$ of $\rho$ for fixed $\eta$. We obtain it from 
	Jeffreys' rule 
	($\S$5.35 of 
\cite{Hagan94}). 
\beq{}
	\mu\,(\rho\,|\,\eta)~=~{\left|\,\overline{ \frac{\partial^2}{\partial{\rho}^2}\,\ln p} \,\right|}^{1/2}.
\label{mu}
\eeq

\noindent
	The overline denotes the expectation value 
	with respect to $x_1...x_M$.

\noindent
	The integration
\beq{}	
	p_1\,(x_1...x_M\,|\,\eta) 
~\sim~\int d\rho\, p\,(x_1...x_M\,|\,\rho,\eta)~\mu\,(\,\rho\,|\,\eta) 
\label{integr}
\eeq
	then yields the model $p_1$ conditioned by $\eta$ alone. 
	It is normalized to unity and the prior distribution
\beq{}
	\mu_1\,(\eta)~=~{\left|\, \overline{\frac{\partial^2}{\partial{\eta}^2}\,\ln{p_1}} \,\right|}^{1/2}
\label{mu-1}
\eeq
	of $\eta$ is obtained by application of Jeffrey's rule 
	to $p_1$.\footnote{We have done the analysis 
	(sections 3.3, 3.4) also using other prior distributions 
	and found only little effect on the result (see 
\cite{KK02}).}
	Bayes' theorem yields the posterior distribution
\beq{}
	P_1(\eta|x_1...x_M)~=~
	\frac{p_1\,(x_1...x_M\,|\,\eta)\,\mu_1(\eta)}{\int d\eta\,p_1\,\mu_1}
\label{grosseP}
\eeq
	of $\eta$.

	From the posterior the 'error interval' for $\eta$ is obtained. 
	It is the shortest interval in which $\eta$ lies 
	with probability K. 
	The length of an interval is defined by help 
	of the measure $\mu_1(\eta)$.
	We call this the Bayesian interval for the probability K in order 
	to distinguish it from a confidence interval 
	of classical statistics. 
	There is a limit, where Bayesian intervals agree 
	with confidence intervals. See below.

\begin{figure}[h]
\begin{center}
\includegraphics[scale=0.3]
{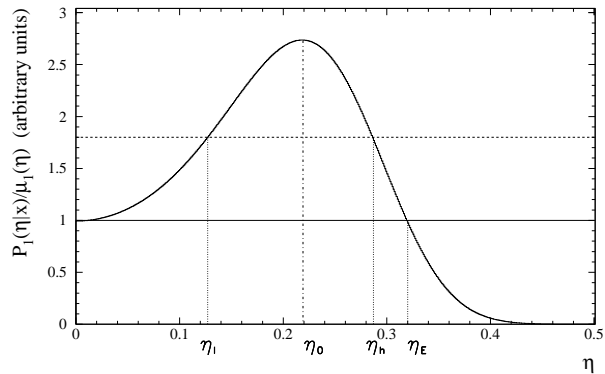} 
\end{center}
%
\caption{The figure shows the relation between P$_1$, K and K$_E$. 
	K is the integral over P$_1$ in the limits [$\eta_l,\eta_h$]. 
	The integral over P$_1$ in the interval [0,$\eta_E$] is K$_E$.
}
\label{Statistics}
\end{figure}

	The borders of a Bayesian interval are given by the 
	intersections of the likelihood function 
	$P_1\,(\eta\,|\,x)/\mu_1(\eta)\sim p_1(x|\eta)$ with
	a horizontal line at $\eta_l,\,\eta_h$ (see Fig. 
\ref{Statistics}).
	The probability K is obtained by integrating $P_1$ 
	from $\eta_l$ to $\eta_h$.


	When the likelihood function has its maximum at $\eta=0$, 
	then the Bayesian interval will - for every K - include $\eta=0$.
	Then this value cannot be excluded and only 
	an upper limit for the contents of the line can be given.
	
	When the maximum of the likelihood function is 
	at a point different from $\eta=0$ - as it is in Fig. 
\ref{Statistics}- 
	then there is a range 
	of K-values such that the associated interval 
	excludes the point $\eta=0$.
	Under this condition let us construct the interval 
	that has its lower border at $\eta=0$. 
	It extends up to $\eta_E$. 
	The associated probability is called K$_E$. 
	The point $\eta=0$ now limits the possible $\eta$-values 
	in a non-trivial way because for every K\,$<$\,K$_E$, 
	the associated error interval {\it excludes} zero. 
	We call K$_E$ the probability that there exists a line.

	The above considerations lead to a peak finding procedure 
\cite{Dietz-Diss03}. 
	One can prescribe a line at an arbitrary energy E 
	of the spectrum - say of Fig. 
\ref{Sum_spectr_Alldet_1990-2000} - 
	and determine the probability K$_E$ that there is one. 
	Such searches lead to the results given in the next section.
	
	Let us note that classical and Bayesian statistics 
	become equal to each other when the likelihood function 
	is well approximated by a Gaussian. 
	In this case, the probability K is the same as classical confidence.

	Note that the method of minimum ${\chi}^2$ 
	is based on an even more stringent limit. 
	It requires Gaussian distributions of the {\it data}. 
	Since the Gaussian is defined everywhere on the real axis, 
	the method can yield negative values of the parameters, 
	especially negative $\eta$ in the present case. 
	The Bayesian method respects the natural limitations 
	of the parameters because it accepts non-Gaussian distributions.

	The method of maximum likelihood is, roughly spoken, 
	the Bayesian method with the prior distribution set constant. 
	This is a useful approximation when the posterior 
	is sufficiently narrow.
	Then the posterior becomes approximately Gaussian. 
	In this sense, the method of maximum likelihood is based 
	on a hidden Gaussian approximation.

\subsection{Numerical Simulations}

	To check the methods of analysing the measured data 
	(Bayes and Maximum Likelihood Method), 
	and in particular to check the programs we wrote, we have
	generated spectra and lines with a random number generator and
	performed then a Bayes and Maximum Likelihood analysis (see also 
\cite{KK-Found02}).
	The length of each generated spectrum is 8200 channels, with a line
	located at bin 5666, the width of the line (sigma) being 4 channels 
	(These special values have been choosen so that every spectrum is
	analogue to the measured data).
	The creation of a simulated spectrum is executed in two steps, first
	the background and second the line was created, using random number
	generators available at CERN (see 
\cite{clhep}).
	In the first step, a Poisson random number generator was used to
	calculate a random number of counts for each channel, using a mean
	value of $\mu$=4 or $\mu=0.5\;$, respectively, 
	in the Poisson distribution
\beq
P(n)=\frac{\mu^n}{n!}\;e^{-\mu}
\eeq
	These mean values correspond
	roughly to our mean background measured in the spectra with or without 
	pulse shape analysis.

	In the second step, a Gaussian random number generator was used to
	calculate a random channel number for a Gaussian distribution with a
	mean value of 5666 (channel) and with a sigma of 4 (channels). 
	The contents of this channel then is increased by one count.
	This Gaussian distribution filling procedure was repeated for n times, 
	n being the number of counts in this line.

	For each choice of $\mu$ and n, 100 different spectra were created, and
	analysed subsequently with two different methods: 
	the maximum likelihood
	method (using the program set of 
\cite{Kchi-kvadr}) 
	and the Bayes-method. Each method, 
	when  analysing a spectrum, gives a lower and an upper limit
	for the number of counts in the line for a given
	confidence level $c$ (e.g. $c=95$\%) (let us call it {\em confidence
  	area A}).
	A confidence level of 95\% means, that in 95\% of all cases the true
	value should be included in the calculated confidence area. 
	This should be exactly correct when analysing an infinite number of
	created spectra. 
	When using 100 spectra, as done here, 
	it should be expected that this
	number is about the same.
	Now these 100 spectra with a special $n$ and $\mu$ are taken to
	calculate a number $d$, which is the number of that cases, 
	where the true
	value $n$ is included in the resulting confidence area A.

	This number $d$ is given in Table 
\ref{Simul-Sp} for
	the results of the two different analysis methods and for various
	values for $\mu$ and $n$. 
	It can be seen, that the Bayes method reproduces even 
	the smallest lines properly, while the Maximum Likehood 
	method has some limitations there.


\begin{table}[h]
\vspace{-0.3cm}
\renewcommand{\arraystretch}{.9}
\setlength\tabcolsep{7.1pt}
\begin{center}
\begin{tabular}{|l|ll|ll||ll|ll|}
\hline
	& \multicolumn{4}{c}{4 counts} & \multicolumn{4}{c}{0.5 counts}\\
\hline
counts & \multicolumn{2}{c}{Bayes} & \multicolumn{2}{c}{Max. Lik.}
& \multicolumn{2}{c}{Bayes} & \multicolumn{2}{c}{Max. Lik.}\\
in line &  68\% & 95\% & 68\% & 95\% &  68\% & 95\% & 68\% & 95\%\\ \hline
0   & 81 & 98 & 60 & 85 & 81 & 99 & 62 & 84 \\ \hline
5   & 88 & 98 & 68 & 80 & 75 & 100 & 82 & 98 \\ \hline
10  & 74 & 97 & 74 & 90 & 
86 & 100 & 84 & 100 \\ \hline
20  & 73 & 96 & 77 & 94  & 90 & 100 & 92 & 100 \\ \hline
100 & 90 & 98 & 87 & 99 & 95 & 100 & 99 & 100 \\ \hline
200 & 83 & 99 & 78 & 99 & 92 & 100 & 100 & 100 \\ \hline
\end{tabular}
\end{center}
\caption{\label{Simul-Sp}Results from the analysis of the simulated
	spectra using a mean background of 4 and 0.5 counts, 
	and different line intensities. The number $d$ of cases, 
	where the true number of counts in the line 
	(given in the left column), is found in the calculated 
	confidence area is given in the second or third column 
	for the Bayes method, and in the fourth and fifth column 
	for the maximum likelihood method. 
	For details see text.}
\end{table}


	Another test has been performed. 
	We generated 1000 simulated spectra containing {\it no} line. 
	Then the probability has been calculated with the Bayesian method 
	that the spectrum {\it does} contain a line at a given energy. 
	Table  
\ref{Number} 
	presents the results: 
	the first column contains the corresponding confidence 
	limit $c$ (precisely the parameter K$_E$ defined earlier), 
	the second column contains the {\it expected} number 
	of spectra indicating existence of a line 
	with a confidence limit above the value $c$ 
	and the third column contain the number of spectra 
	with a confidence limit above the value $c$, 
	found in the simulations.
	The result underlines that K$_E$ here is equivalent 
	to the usual confidence level of classical statistics.


\begin{table}[h]
\vspace{-0.3cm}
\renewcommand{\arraystretch}{1.1}
\setlength\tabcolsep{11.7pt}
\begin{center}
\begin{tabular}{|l|l|l|}
\hline
	C.L.		&  Expected	&	Found	 \\
\hline
 	90.0$\%$	&  	100 $\pm$ 31	& 	96	\\
 	95.0$\%$	 &  	50 $\pm$ 7	& 	42	\\
 	99.0$\%$	 &  	10 $\pm$ 3	& 	12	\\
  	99.9$\%$	&  	1 $\pm$ 1	& 	0	\\
\hline
\end{tabular}
\end{center}
\caption{\label{Number}
	Number of spectra with a calculated confidence limit 
	above a given value.
	For details see text.}
\end{table}


\begin{figure}[h]

\begin{center}
\includegraphics[height=6.cm,width=10.5cm]{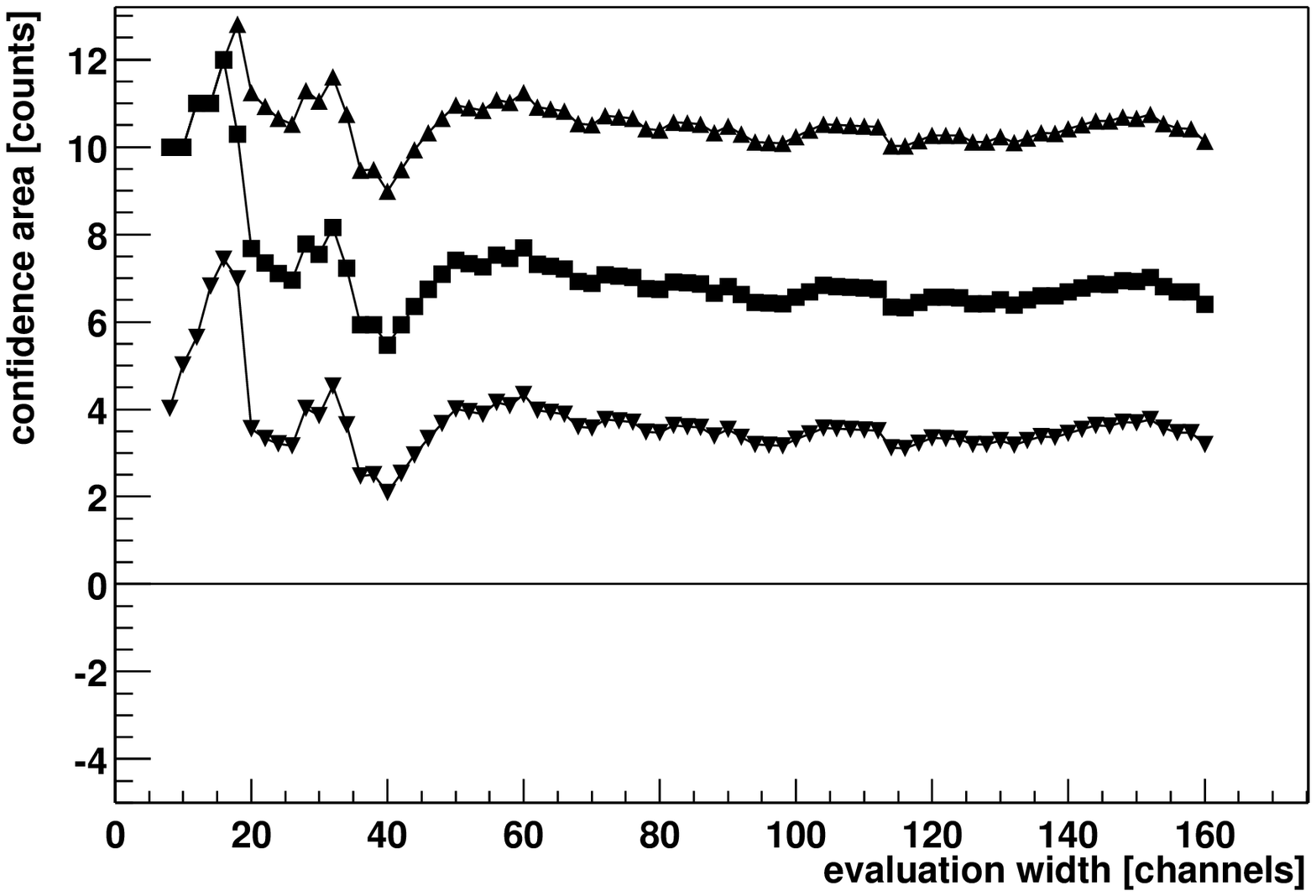}
\includegraphics[height=6.cm,width=10.5cm]{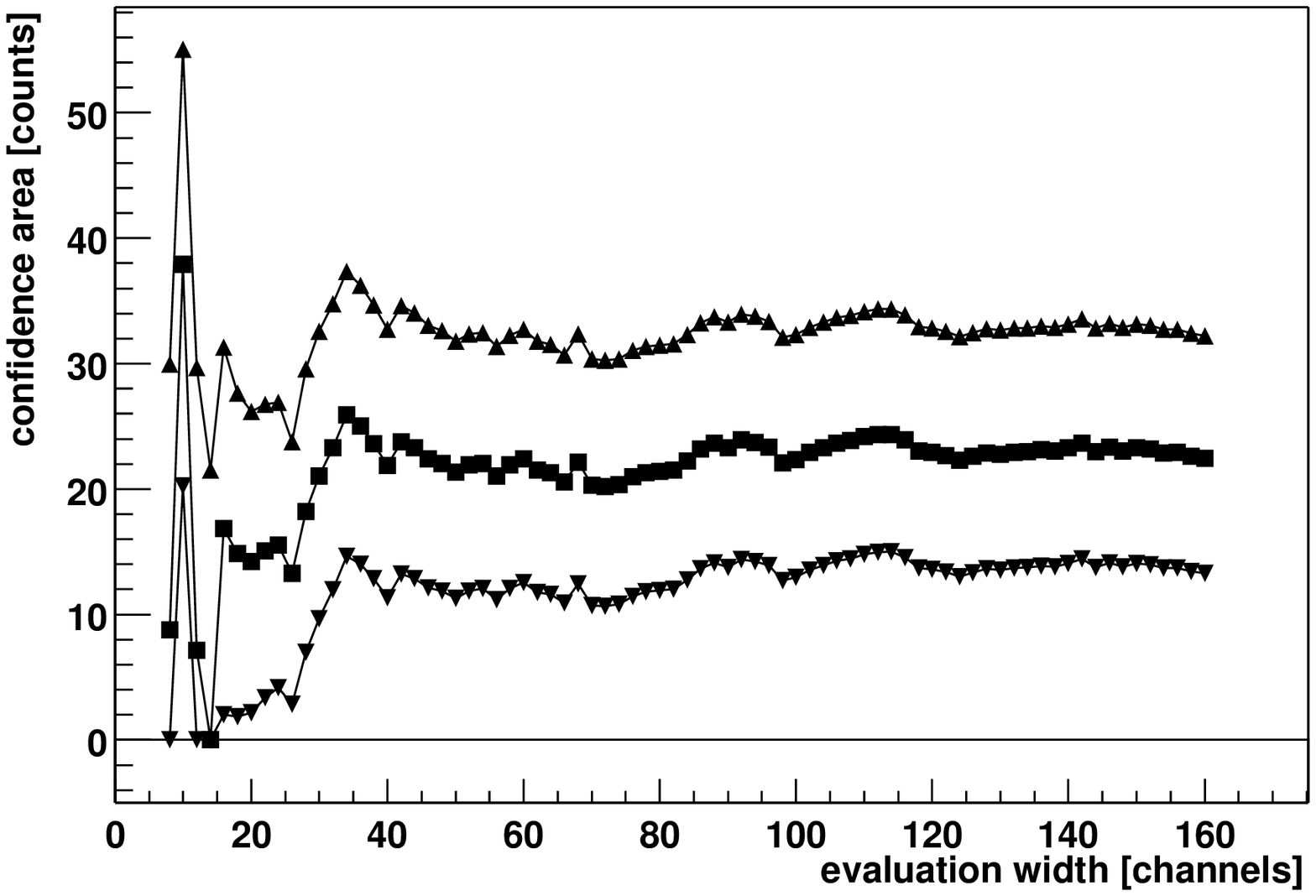}
\end{center}
\caption{\underline{Upper part:} Analysis of simulated spectrum with 
	Gaussian peak of 5\,events and FWHM of 9.4\,channels 
	on a Poisson-distributed background spectrum of 0.5\,events/channel, 
	as function of interval of analysis. 
	The middle line is the best value. 
	Upper and lower lines correspond to the 68.3$\%$ confidence limits. 
	\underline{Lower part:}
	 The same as above, but the peak contains 20\,events, 
	the background is 4\,events/channel. 
	One channel corresponds to 0.36\,keV in our measured spectra.
}
\label{Spect1}
\end{figure}


	We further investigated with the computer-generated spectra 
	the dependence of the peak analysis 
	on the width of the energy range of evaluation. 
	Two examples are shown in Fig. 
\ref{Spect1}.
	Here the contents of the simulated peak found with the Bayes method 
	is shown as function of the analysis interval given 
	in channels 
	(one channel corresponds to 0.36\,keV in our measured spectra). 
	The line in the middle is the best-fit value of the method, 
	the upper and lower lines correspond to the upper 
	and lower 68.3$\%$ confidence limits.
	In the upper figure the true number of counts in the simulated 
	line was 5 events, on a Poisson-distributed background 
	of 0.5\,events/channel, in the lower figure it was 20\,events 
	on a background of 4\,events/channel. 
	It can be seen, that the analysis gives safely the correct 
	number of counts, when choosing an analysis interval of not less 
	than 40\,channels.


\subsection{Analysis of the Full Data}

	We first concentrate on the full 
	spectra (see Fig. 
\ref{Sum_spectr_Alldet_1990-2000},  
	and Fig.2 in 
\cite{KK-Evid01}), 
	 without {\it any} data manipulation 
	(no background subtraction, no pulse shape analysis).
	For the evaluation, we consider the {\it raw data} of 
	the detectors. 

	The Bayesian peak finding procedure described in the 
	last section leads to the 
	result shown on the left hand sides of Figs. 
\ref{Bay-Alles-90-00-gr-kl-Bereich},\ref{Bay-Chi-all-90-00-gr}.
	For every energy E of the spectral range 2000 -2080\,keV, 
	we have determined the probability K$_E$ that there is a line at E. 
	All the remainder of the spectrum was considered 
	to be background in this search.

\begin{figure}[t]
%
\begin{center}
%
\includegraphics[height=10.5cm,width=7.5cm]{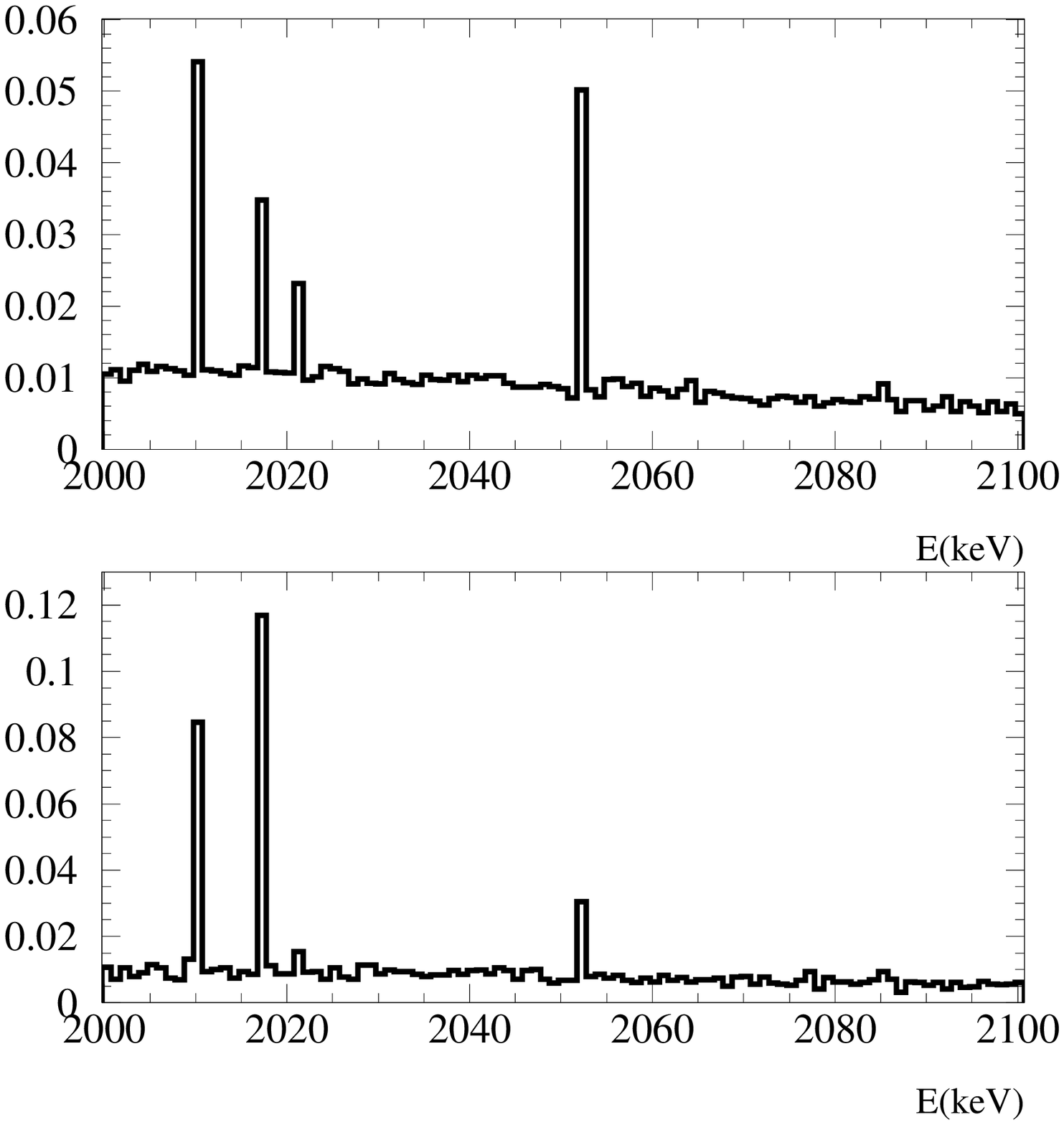}
\includegraphics[height=9.5cm,width=3.5cm]{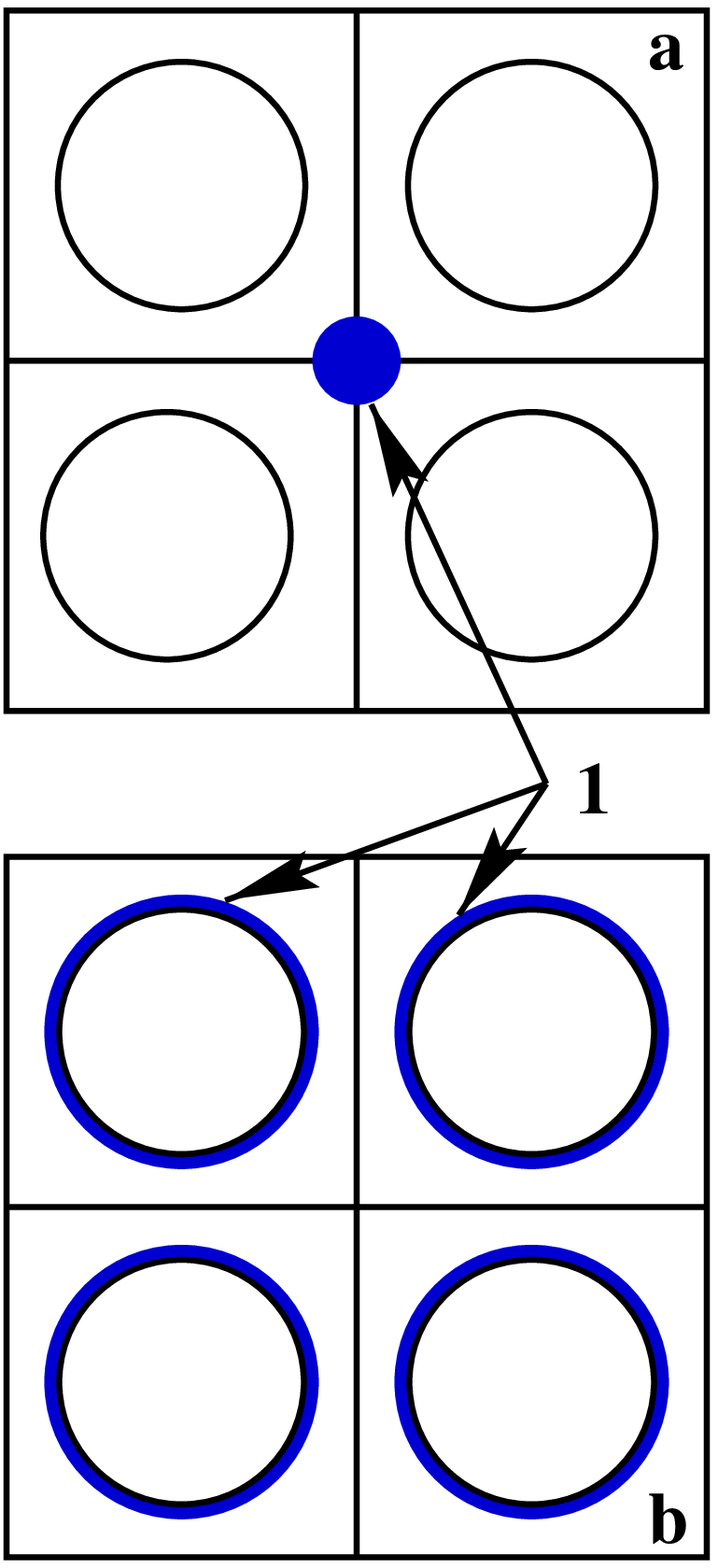}
\end{center}
%
\caption[]{Demonstration of the dependence of the relative intensities 
	of the weak $^{214}{Bi}$ lines on the location of the impurity 
	in the experimental setup. 
	The right side shows two different locations 
	of the impurity (1) relative to the detectors. 
	The left side shows the corresponding different relative 
	intensities to be expected. 
	\underline{Upper part:} Source in some distance from 
	the detectors (as in a) right part). 
	\underline{Lower part:} Source in copper cap very near around 
	the detectors (as in b) right part). 
	The ordinate shows relative intensities (see 
\cite{KK02}).} 
\label{Bi-simul}
\end{figure}

\begin{figure}[h]
\begin{center}

\vspace{-0.3cm}
\includegraphics[scale=0.21]
{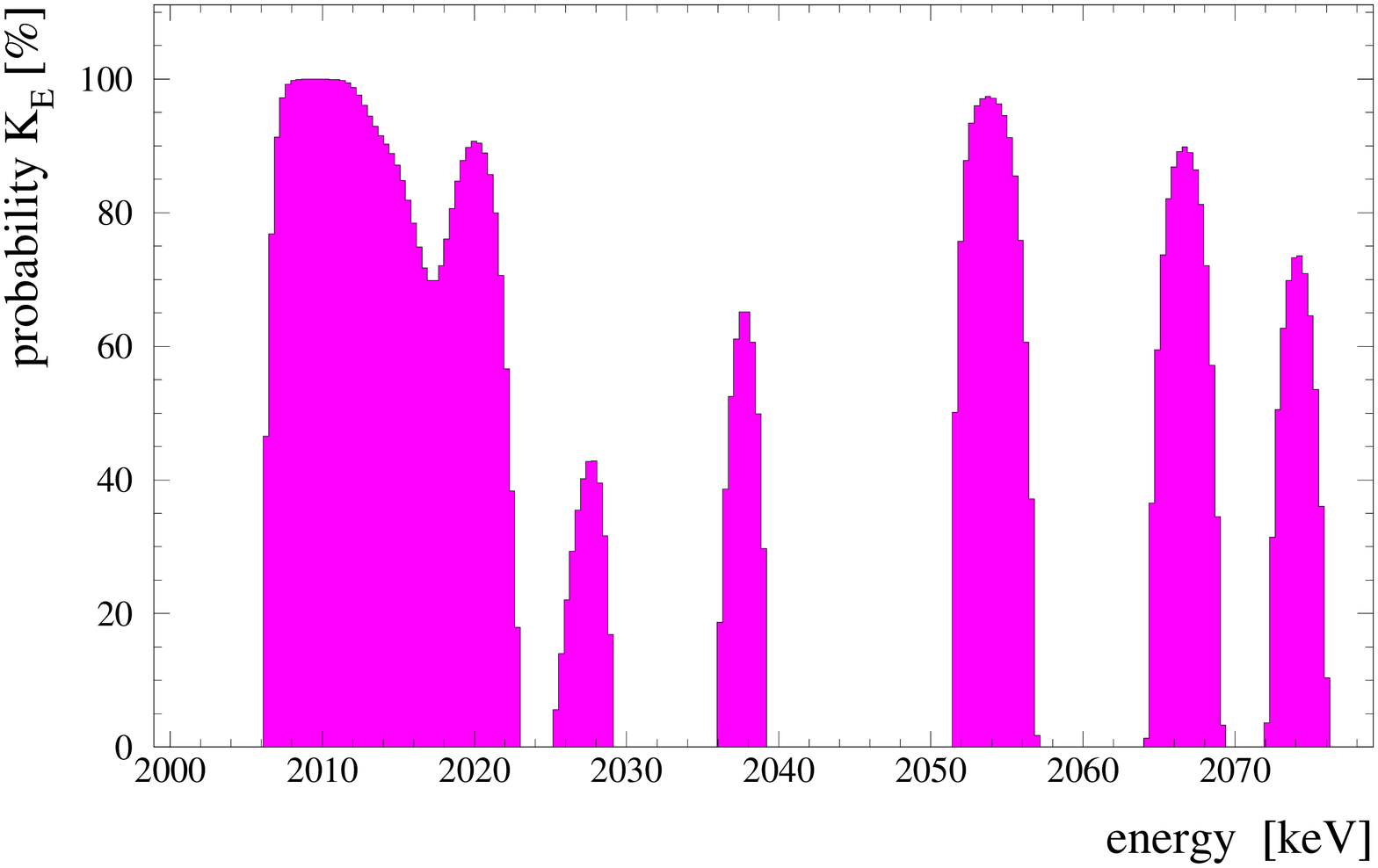} 
\includegraphics[scale=0.21]
{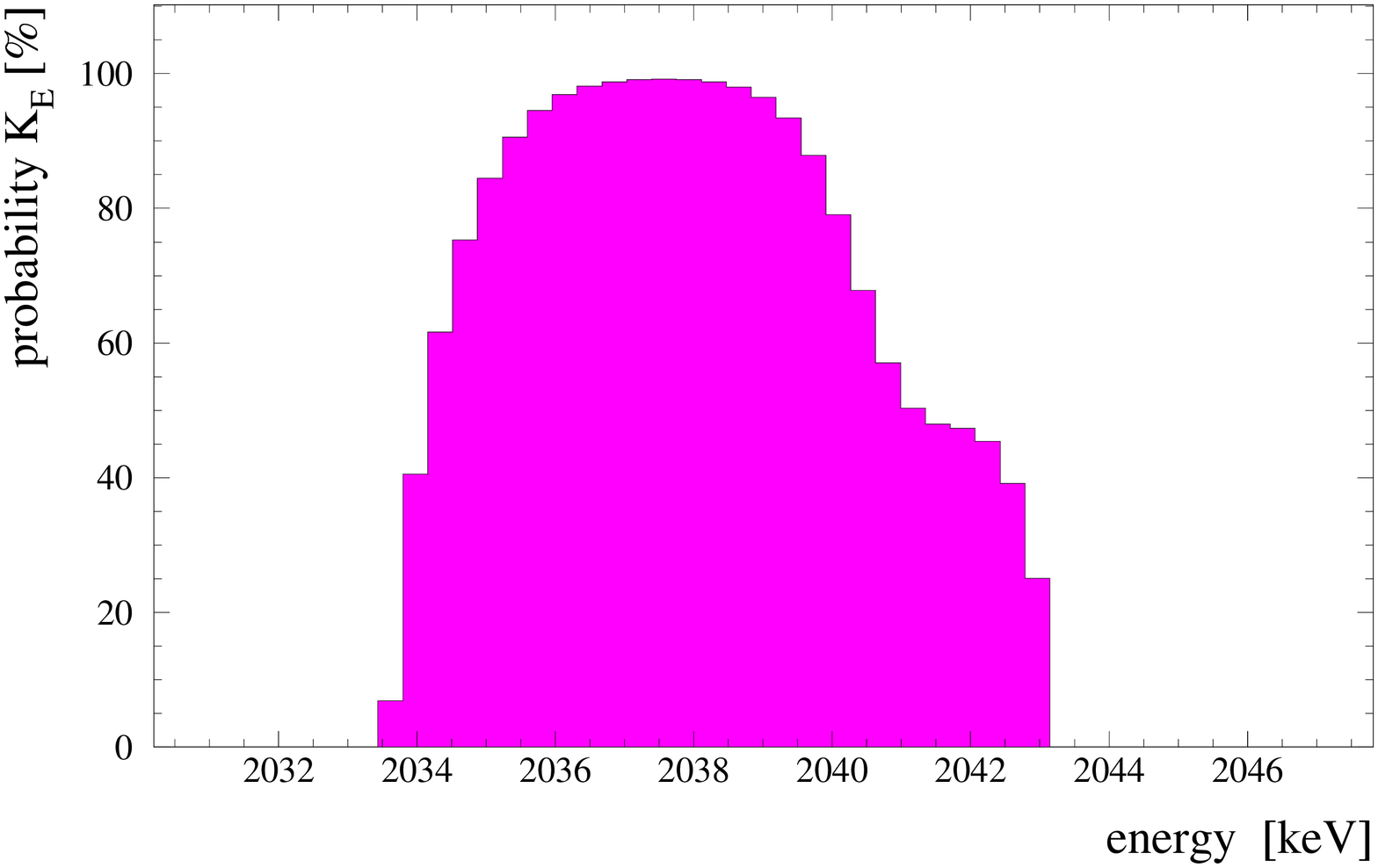}
\end{center}
\vspace{-0.3cm}
\caption{Scan for lines in the full spectrum 
	taken from 1990-2000 with detectors Nr. 1,2,3,4,5, 
	(Fig.
\ref{Sum_spectr_Alldet_1990-2000}),  
	with the Bayesian method of sec. 3.1. 
	The ordinate is the probability K$_E$ that a line 
	exists as defined in the text.  
	\underline{Left:} Energy range 2000 - 2080\,keV. 
	\hspace{2.cm} \underline{Right:} Energy range of analysis 
	$\pm\,5\sigma$ around Q$_{\beta\beta}$.}
\label{Bay-Alles-90-00-gr-kl-Bereich}
%
%
%
\vspace{-0.2cm}
\begin{center}

\vspace{-0.3cm}
\includegraphics[scale=0.21]
{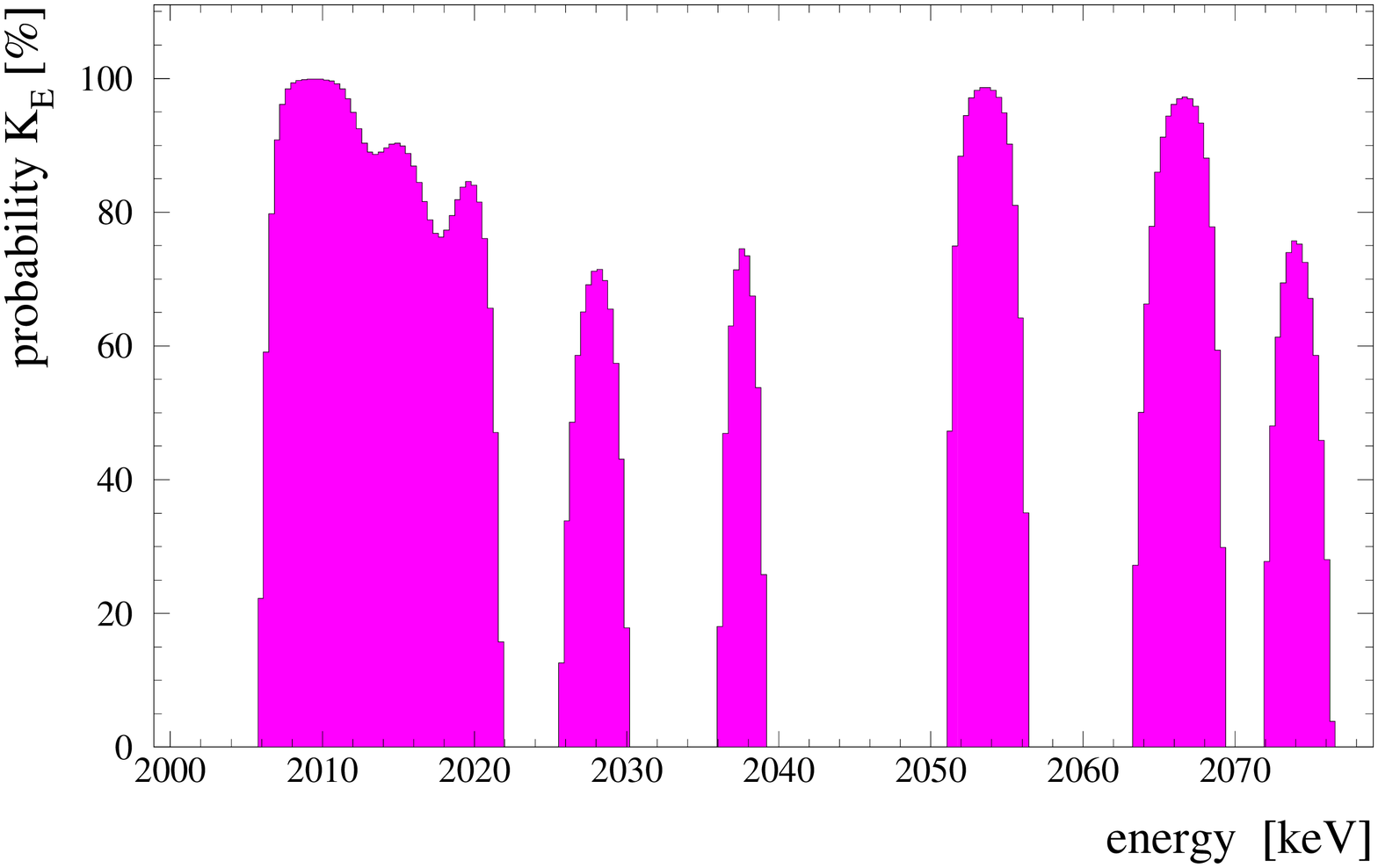} 
\includegraphics[scale=0.21]
{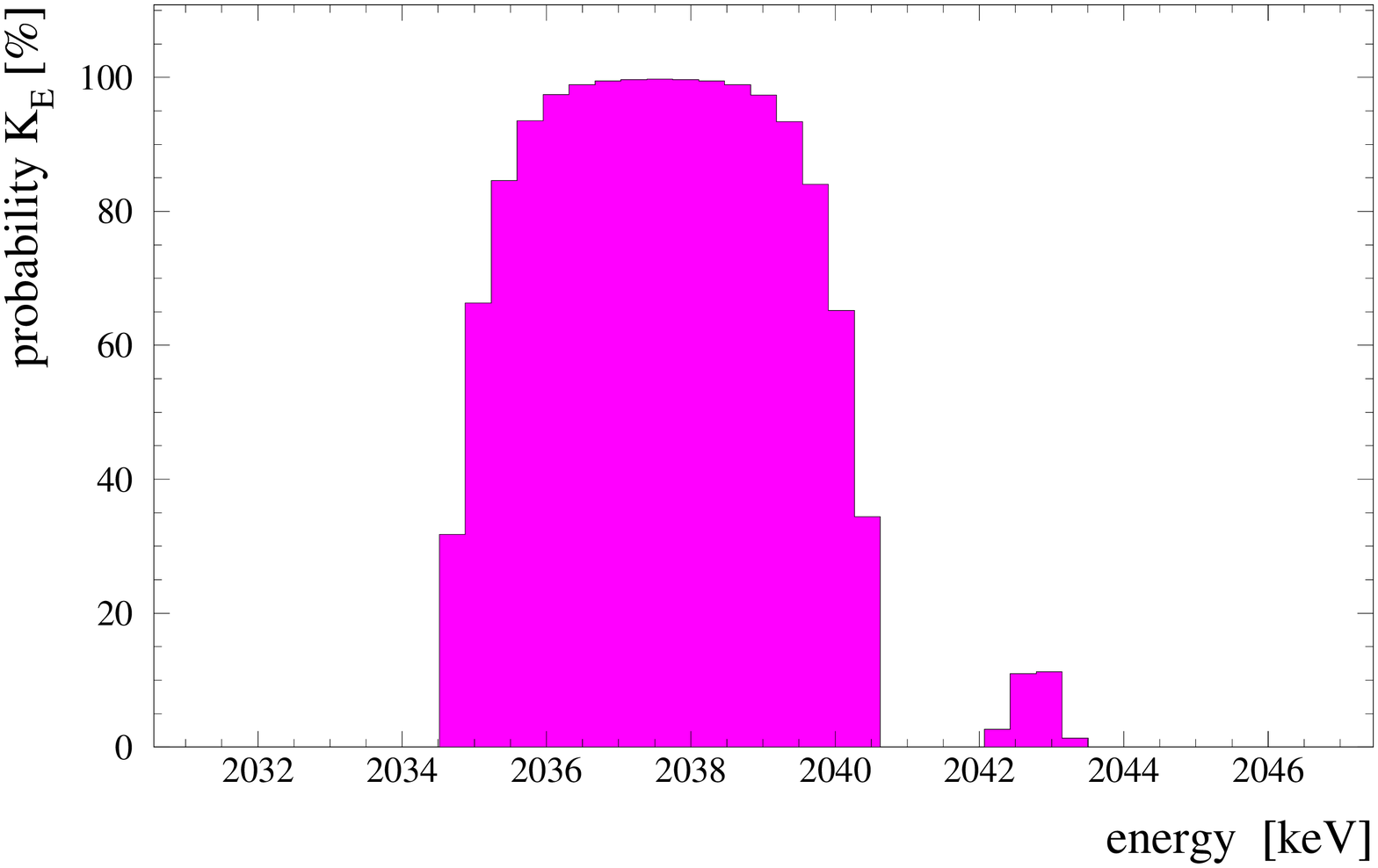} 
\end{center}

\vspace{-0.3cm}
	\caption{\underline{Left:} 
	Probability K$_E$ that a line exists at a given energy in the 
	range of 2000-2080\,keV derived via Bayesian inference 
	from the spectrum 
	taken with detectors Nr. 1,2,3,5 over the period August 1990 
	to May 2000, 46.502\,kg\,y (see Fig.2 from 
\cite{KK-Evid01}.) 
	\underline{Right:} 
	Result of a Bayesian scan for lines as in 
	the left part of this figure,  
	but in the energy range of analysis 
	$\pm\,5\sigma$ around Q$_{\beta\beta}$.} 
\label{Bay-Chi-all-90-00-gr}
%
\vspace{-0.2cm}
\begin{center}

\vspace{-0.3cm}
\includegraphics[scale=0.21]
{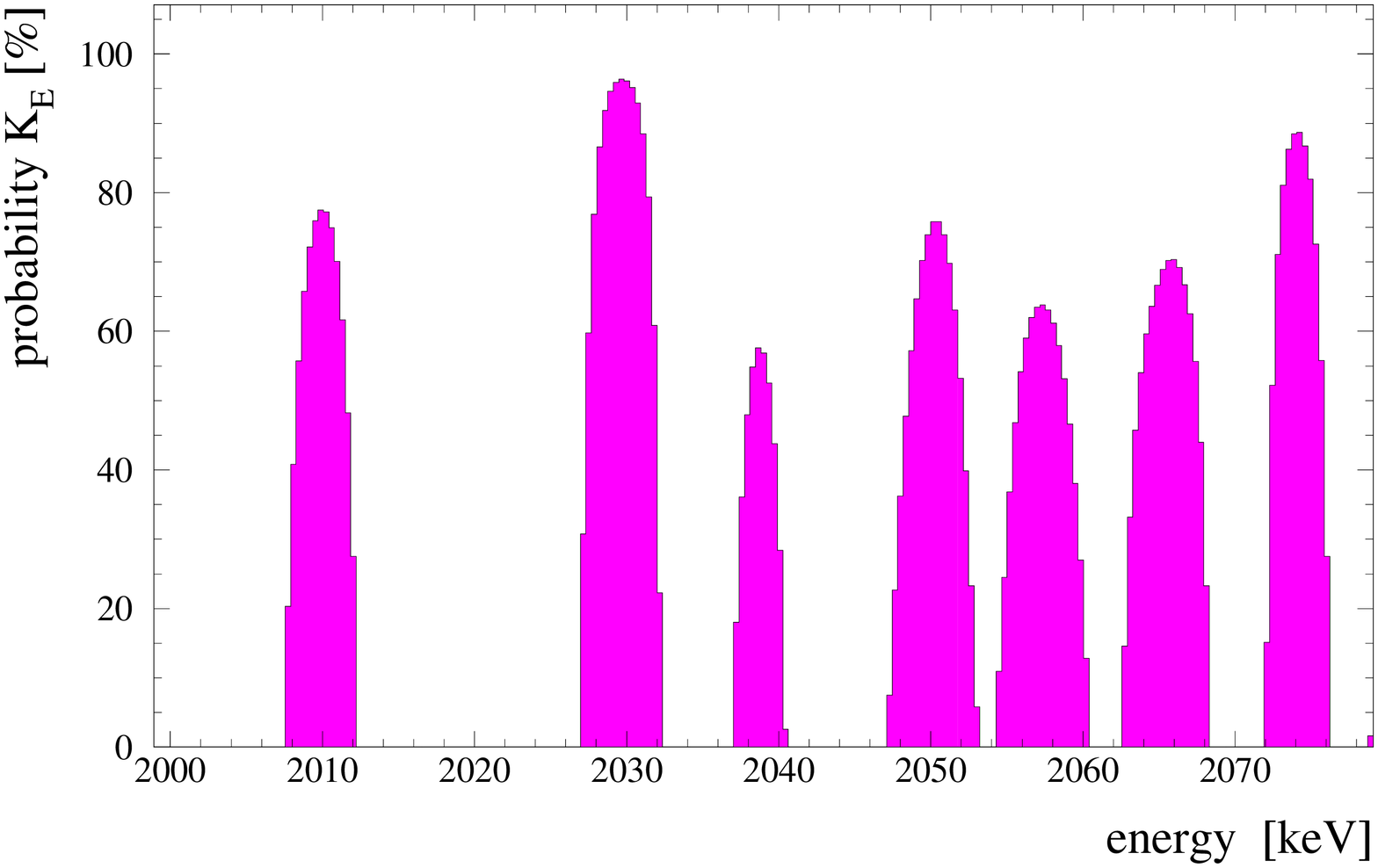} 
\includegraphics[scale=0.21]
{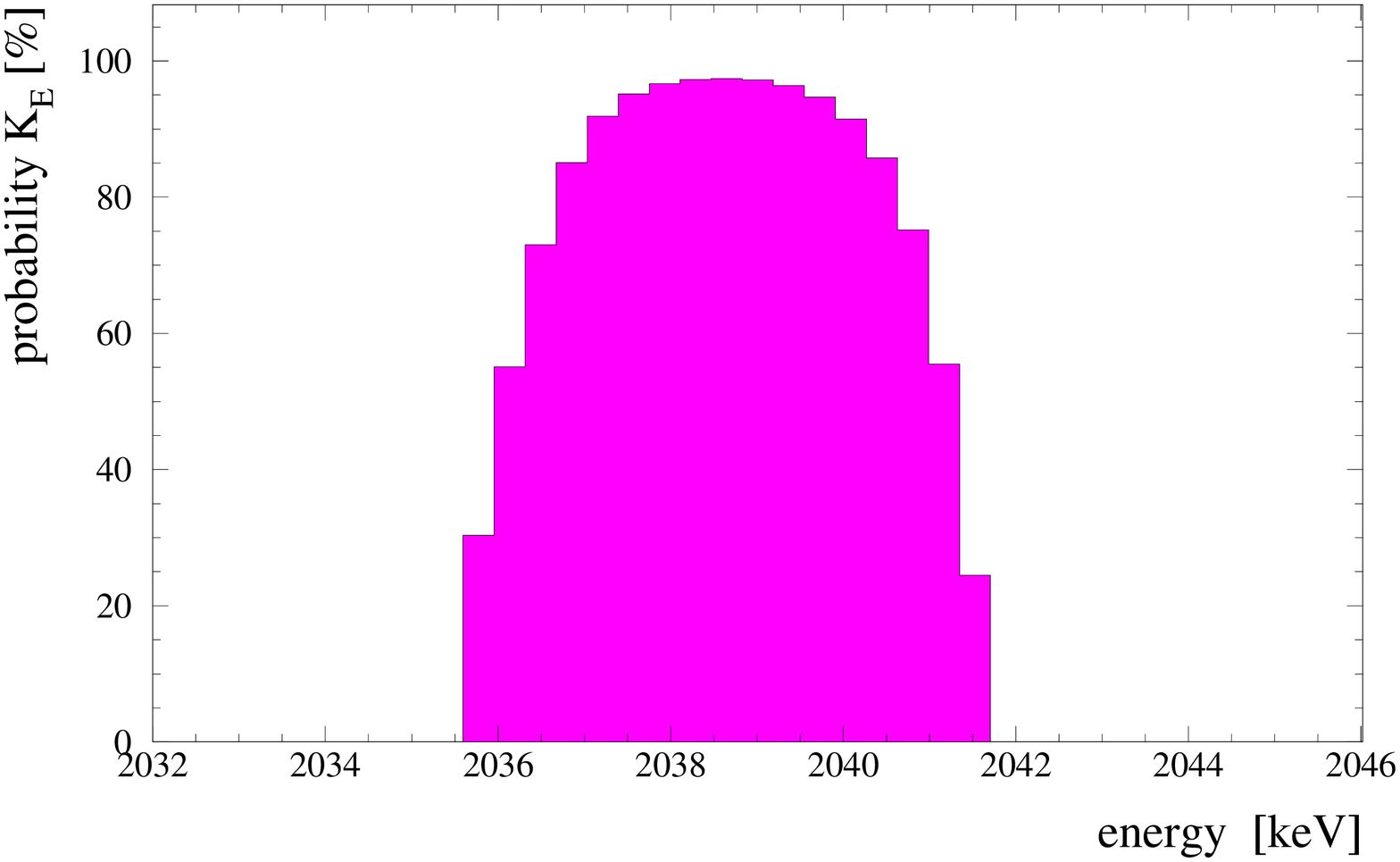} 
\end{center}

\vspace{-0.3cm}
\caption[]{Scan for lines in the single site event spectrum 
	taken from 1995-2000 with detectors Nr. 2,3,5, 
	(Fig. 
\ref{Sum_spectr_5det}), 
	with the Bayesian method (as in Figs.
\ref{Bay-Alles-90-00-gr-kl-Bereich},\ref{Bay-Chi-all-90-00-gr}).
	\underline{Left:} Energy range 2000 -2080\,keV. 
	\underline{Right:} Energy range of analysis 
	$\pm\,4.4\sigma$ around Q$_{\beta\beta}$.} 
\label{Bay-Hell-95-00-gr-kl-Bereich}
\end{figure}

	The peak detection procedure yields lines 
	at the positions of known 
\cite{Tabl-Isot96}
	weak $\gamma$-lines 
	from the decay of $^{214}{Bi}$ at 2010.7, 2016.7, 2021.8 
	and 2052.9\,keV.  
	The lines at 2010.7 and 2052.9\,keV are observed 
	at a confidence level of 3.7 and 2.6$\sigma$, respectively.
	The observed intensities are consistent with the expectations 
	from the strong Bi lines in our spectrum, 
	and the branching ratios given in 
\cite{Tabl-Isot96},
	within about the 2$\sigma$ experimental error (see Table 
\ref{Bi-lines} 
	and  
\cite{KK-new02}).
	The expectations here are calculated including summing effects, 
	by Monte-Carlo simulation of our set-up. 
	{\it Only} in this way the strong dependence 
	of the relative intensities on the location 
	of the impurities in the set-up can be properly taken into account
	(see Fig. 
\ref{Bi-simul}).	
	(A separate measurement with a $^{226}_{88}{Ra}$ source being 
	in progress, will allow to study the intensities of the weak 
	$^{214}{Bi}$ lines in the setup with high statistics).


\begin{table}[h]
\begin{center}
\newcommand{\m}{\hphantom{$-$}}
\renewcommand{\arraystretch}{1.35}
\setlength\tabcolsep{.6pt}
\begin{tabular}{c|c|c|c|c|c|c|c}
\hline
\hline
        	          &   Intensity        &
                        &             &
        	        &      Expect.   &  Expect.   &  Aalseth\\
        Energy          &       of    &
                        &       Branching       &
        Simul. of        &      rate		&	rate	
&	et al.	\\
                (keV)   &       Heidelberg-             &
        $\sigma$        &       Ratios\protect\cite{Tabl-Isot96}        &
        Experim.       &       accord.     &  accord.     
& (see \protect\cite{Reply-Evid01})	\\
                $*)$   &       Mos.Exper.     &
                        &       [$\%$]          &
        Setup +)       &       to sim.**) &
to\protect\cite{Tabl-Isot96}++)    &    ***)   \\
\hline
        609.312(7)      & 4399$\pm$92 &  & 44.8(5) & 
5715270$\pm$2400	&  &  &\\
        1764.494(14)    & 1301$\pm$40 &  & 15.36(20) & 
	1558717$\pm$1250	&  & &\\
        2204.21(4)      & 319$\pm$22 &  & 4.86(9) & 
	429673$\pm$656	&  & &\\
        2010.71(15)     &       37.8$\pm$10.2         &
        3.71            &       0.05(6)                 &
	15664$\pm$160	&  
	12.2$\pm$0.6	&    4.1$\pm$0.7           &       0.64    \\
        2016.7(3)       &       13.0$\pm$8.5          &
        1.53            &       0.0058(10)              &
	20027$\pm$170	&  15.6$\pm$0.7
&    0.5$\pm$0.1           &       0.08    \\
        2021.8(3)       &       16.7$\pm$8.8          &
        1.90            &       0.020(6)                &
	1606$\pm$101	&  
	1.2$\pm$0.1	&    1.6$\pm$0.5           &       0.25    \\
        2052.94(15)     &       23.2$\pm$9.0          &
        2.57            &       0.078(11)               &
	5981$\pm$115	&  
	4.7$\pm$0.3	&    6.4$\pm$1             &       0.99    \\
        2039.006        &       12.1$\pm$8.3          &
        1.46            &                               &
                        &                               &       \\
\hline
\end{tabular}
\end{center}
\vspace{0.3cm}
\caption{\label{Bi-lines}$^{214}\rm{Bi}$ is product
        of the $^{238}\rm{U}$ natural decay chain through ${\beta}^-$
        decay of $^{214}\rm{Pb}$ and $\alpha$ decay
        of $^{218}\rm{At}$.
        It decays to $^{214}\rm{Po}$ by ${\beta}^-$ decay.
        Shown in this Table are the measured intensities 
	of $^{214}\rm{Bi}$ lines
        in the spectrum shown in Fig.1 of Ref.
\protect\cite{KK-Evid01}
        in the energy window 2000 - 2060\,keV, our calculation
        of the intensities expected on the basis of the branching
        ratios given in Table of Isotopes
\protect\cite{Tabl-Isot96},
        with and without simulation of the experimental setup,
        and the intensities expected by Aalseth et al. hep-ex/0202018,
	who do not simulate the setup and thus ignore summing of the
        $\gamma$ energies (see 
\protect\cite{Reply-Evid01}).
\protect\newline
        $*)$ We have considered for comparison the 3 strongest $^{214}\rm{Bi}$
        lines, leaving out the line at 1120.287\,keV (in the measured
        spectrum this line is partially overimposed on the 1115.55 keV 
        line of $^{65}\rm{Zn}$). The number of counts in each line have been
        calculated by a maximum-likelihood fit of the line with a 
        gaussian curve plus a constant background.
\protect\newline
        $+)$ The simulation is performed assuming that the impurity
        is located in the copper part of the detector chamber (best
        agreement with the intensities of the strongest lines in the
        spectrum). The error of a possible misplacement is not
        included in the calculation. The number of simulated events is 
        $10^8$ for each of our five detectors. 
\protect\newline
        $**)$ This result is obtained normalizing the simulated
        spectrum to the experimental one using the 3 strong lines
        listed in column one. Comparison to the neighboring column 
	on the right shows that the expected rates for the weak 
        lines can change strongly if we take into account the simulation.
        The reason is that the 
	line at 2010.7\,keV can be produced by summing of the 
        1401.50\,keV (1.55\%)
        and 609.31\,keV (44.8\%) lines, the one at 
        2016.7 keV by summing of the 1407.98 (2.8\%) and 609.31 (44.8\%) lines;
        the other lines at 2021.8\,keV and 2052.94\,keV 
	do suffer only very weakly from the summing effect 
	because of the different decay schemes.
\protect\newline
        $++)$ This result is obtained 
	using the number of counts for the three strong lines observed 
        in the experimental spectrum and the branching ratios from
        \protect\cite{Tabl-Isot96}, but {\it without} simulation. 
	For each
        of the strong lines the expected number of
        counts for the weak lines is calculated and then an average of 
        the 3 expectations is taken.
\protect\newline
        ***)
        Without simulation of the experimental setup.
        The numbers given here are close to those in the neighboring
        left column, when taking into account that Aalseth et al.
        refer to a spectrum which has only $\sim$11\% of the
        statistics of the spectrum shown in Fig.1 of Ref.
\protect\cite{KK-Evid01}.
\protect\newline
}
\end{table}

\begin{figure}[h]
\begin{center}
\includegraphics[scale=0.45, angle=-90]
{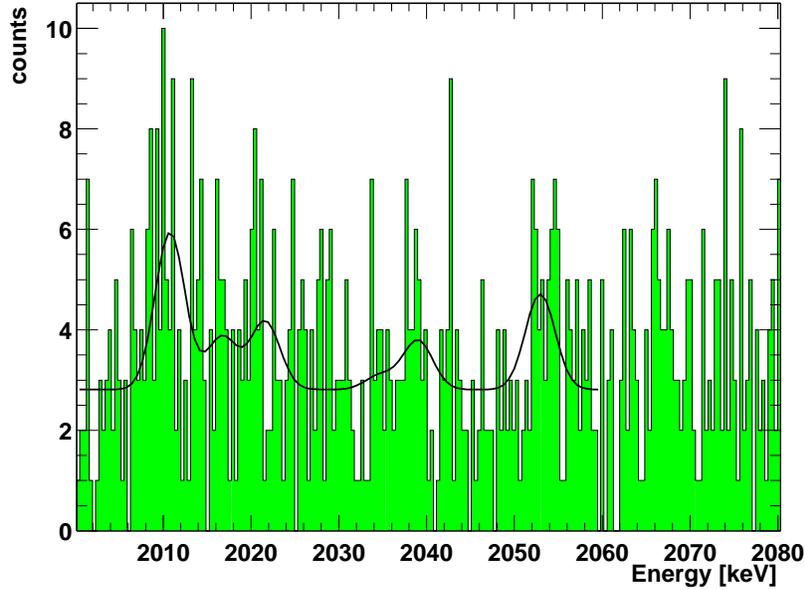} 
\end{center}
\caption[]{Simultaneous analysis of the spectrum measured 
	with the $^{76}{Ge}$ detectors Nr. 1,2,3,4,5 over the period 
	August 1990 - May 2000 (54.9813\,kg\,y) 
	(same as in Fig. 
\ref{Sum_spectr_Alldet_1990-2000}, 
	but here shown in the 0.36\,keV original binning) in the energy 
	range 2000 - 2060\,keV, with the Maximum Likelihood Method. 
	The probability for a line at 2039.0\,keV found 
	is this way is 86\%.} 
\label{Bi-Lines-0_36keV}
\end{figure}

	In addition, a line centered at 2039\,keV shows up. 
	This is compatible with the Q-value 
\cite{New-Q-2001,Old-Q-val}
	of the double beta decay process. We emphasize, 
	that at this energy no $\gamma$-line is expected 
	according to the compilations in 
\cite{Tabl-Isot96}
	(see discussion in section 2). 
Figs. 
\ref{Bay-Alles-90-00-gr-kl-Bereich},\ref{Bay-Chi-all-90-00-gr} 
	do not show 
	indications for the lines from $^{56}{Co}$ at 2034.7\,keV 
	and 2042\,keV 
	discussed earlier  
\cite{Diss-Dipl} 
	(but see also Fig. 
\ref{Bi-Lines-0_36keV}). 
	We have at present no convincing identification 
	of the lines around 2070\,keV indicated 
	by the peak identification procedure.

	It may be important to note, that essentially 
	the same lines as found by the peak scan procedure in Figs.
\ref{Bay-Alles-90-00-gr-kl-Bereich},\ref{Bay-Chi-all-90-00-gr},\ref{Bay-Hell-95-00-gr-kl-Bereich},
	are found 
\cite{KK-Found02,KK-new02}
	when doing the same kind of analysis with the best 
	existing {\it natural} Ge experiment of D. Caldwell et al. 
\cite{Caldw91},
	  which has a by a factor of $\sim$4 better statistics 
	{\it of the background}. This experiment does, however, 
	not see the line at 2039\,keV
	(see section 3.5).

	Bayesian peak detection (the same is true 
	for Maximum Likelihood peak detection) of our data suggests 
	a line at Q$_{\beta\beta}$ 
	whether or not one includes detector Nr. 4 
	without muon shield (Figs. 
\ref{Bay-Alles-90-00-gr-kl-Bereich},\ref{Bay-Chi-all-90-00-gr}). 
	The line is also suggested in Fig. 
\ref{Bay-Hell-95-00-gr-kl-Bereich}
	after removal of multiple site events (MSE), see below.

On the left-hand side of Figs. 
\ref{Bay-Alles-90-00-gr-kl-Bereich},\ref{Bay-Chi-all-90-00-gr},\ref{Bay-Hell-95-00-gr-kl-Bereich},  
	the background intensity (1-$\eta$) 
	identified by the Bayesian procedure 
	is too high because the procedure averages the 
	background over all the spectrum (including lines) 
	except for the line it is trying to single out. 
	Inclusion of the known lines into the fit 
	naturally improves the background. As example, we show in Fig. 
\ref{Bi-Lines-0_36keV}
	the spectrum of Fig. 
\ref{Sum_spectr_Alldet_1990-2000}
	(here in the original 0.36\,keV binning) with a simultaneous 
	fit of the range 2000 - 2060\,keV 
	(assuming lines at 2010.78, 2016.70, 2021.60, 2052.94, 2034.76, 
	2039.0, 2041.16\,keV). The probability for 
	a line in this fit at 2039\,keV is 86\%.

\clearpage



\noindent
	Finally, on the right-hand side of Figs. 
\ref{Bay-Alles-90-00-gr-kl-Bereich},\ref{Bay-Chi-all-90-00-gr}
	(and also Fig. 
\ref{Bay-Hell-95-00-gr-kl-Bereich}) 
	the peak detection procedure is carried out within 
	an energy interval that seems to not contain 
	(according to the left-hand side) lines other 
	than the one at Q$_{\beta\beta}$. 
	This interval is broad enough 
	(about $\pm$ 5 standard deviations of 
	the Gaussian line, i.e. as typically used in search 
	for resonances in high-energy physics) 
	for a meaningful analysis (see Fig.
\ref{Spect1}
	in section 3.2). 
	We find, with the Bayesian method, 
	the probability K$_E$ = 96.5$\%$ that 
	there is a line at 2039.0\,keV is the spectrum shown in Fig.
\ref{Sum_spectr_Alldet_1990-2000}.
	This is a confidence level of 2.1 $\sigma$ 
	in the usual language.  
	The number of events is found to be 0.8 to 32.9 
	(7.6 to 25.2) with 95\% (68\%) c.l., 
	with best value  of 16.2 events.
	For the spectrum shown in Fig. 2 in
\cite{KK-Evid01},  
	we find a probability for a line at 2039.0\,keV 
	of 97.4$\%$ (2.2 $\sigma$). 
	In this case
	the number of events is found to be 
	1.2 to 29.4 with 95$\%$ c.l..  
	It is 7.3 to 22.6\,events with 68.3$\%$ c.l.. 
	The most probable number of events (best value) is  
	14.8 events. 
	These values are stable against small variations 
	of the interval of analysis, as expected from Fig. 
\ref{Spect1}
	in section 3.2.
	For example, 
	changing the lower and upper limits of the interval 
	of analysis between 2030 and 2032 and 2046 
	and 2050 yields consistently values of K$_E$ 
	between 95.3 and 98.5$\%$ 
	(average 97.2$\%$) for the spectrum of Fig.2 of 
\cite{KK-Evid01}.

	We also applied the Feldman-Cousins method recommended 
	by the Particle Data Group 
\cite{Feldm98,RPD00}. 
	This method 
	(which does not use the information 
	that the line is Gaussian) finds 
	a line at 2039 keV on a confidence level of 
	3.1 $\sigma$ (99.8$\%$ c.l.).  

\subsection{Analysis of Single Site Events Data}

\begin{figure}[h]
\centering{
\includegraphics*[scale=0.15]{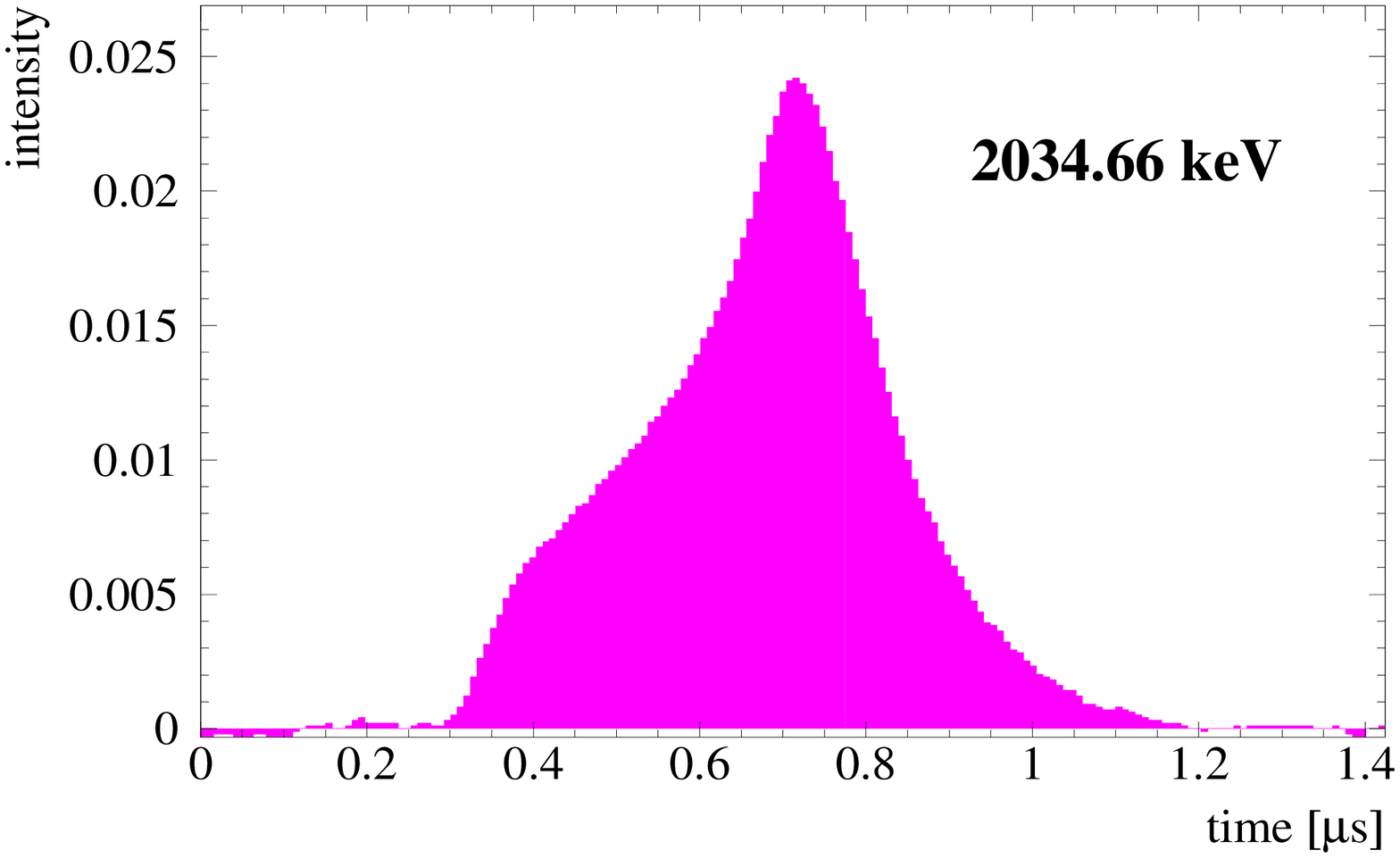}
\includegraphics*[scale=0.15]{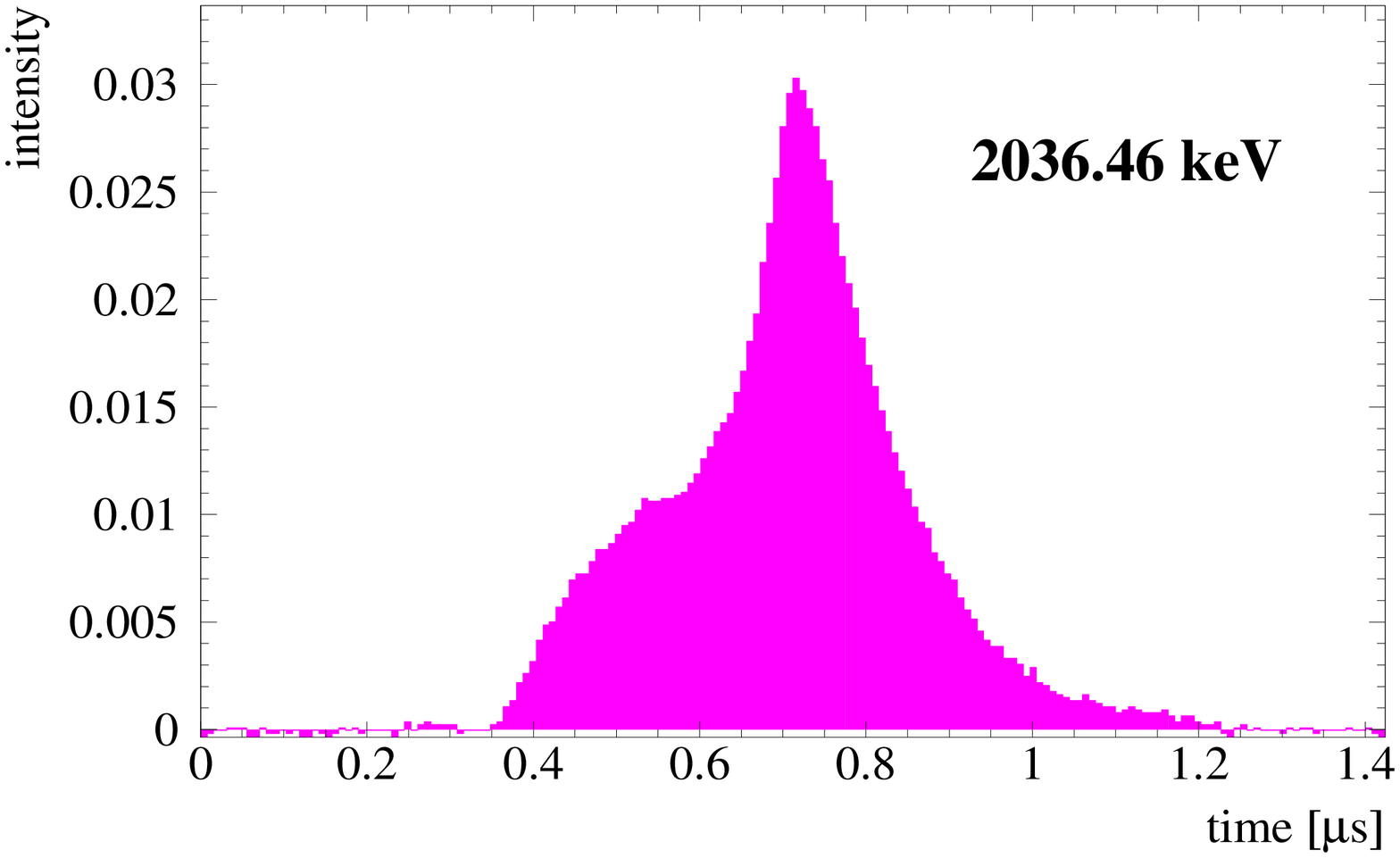}
\includegraphics*[scale=0.15]{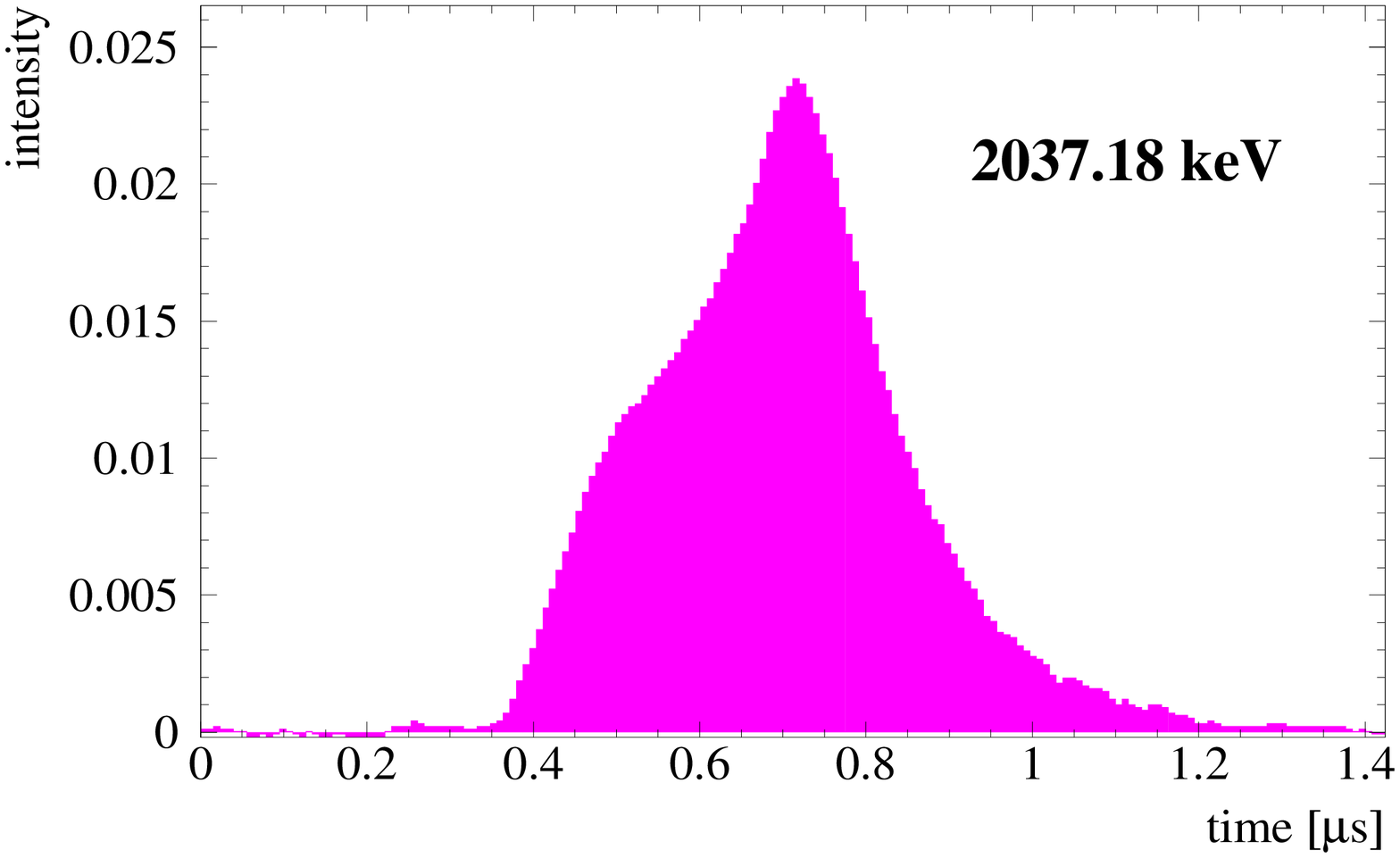}
\includegraphics*[scale=0.15]{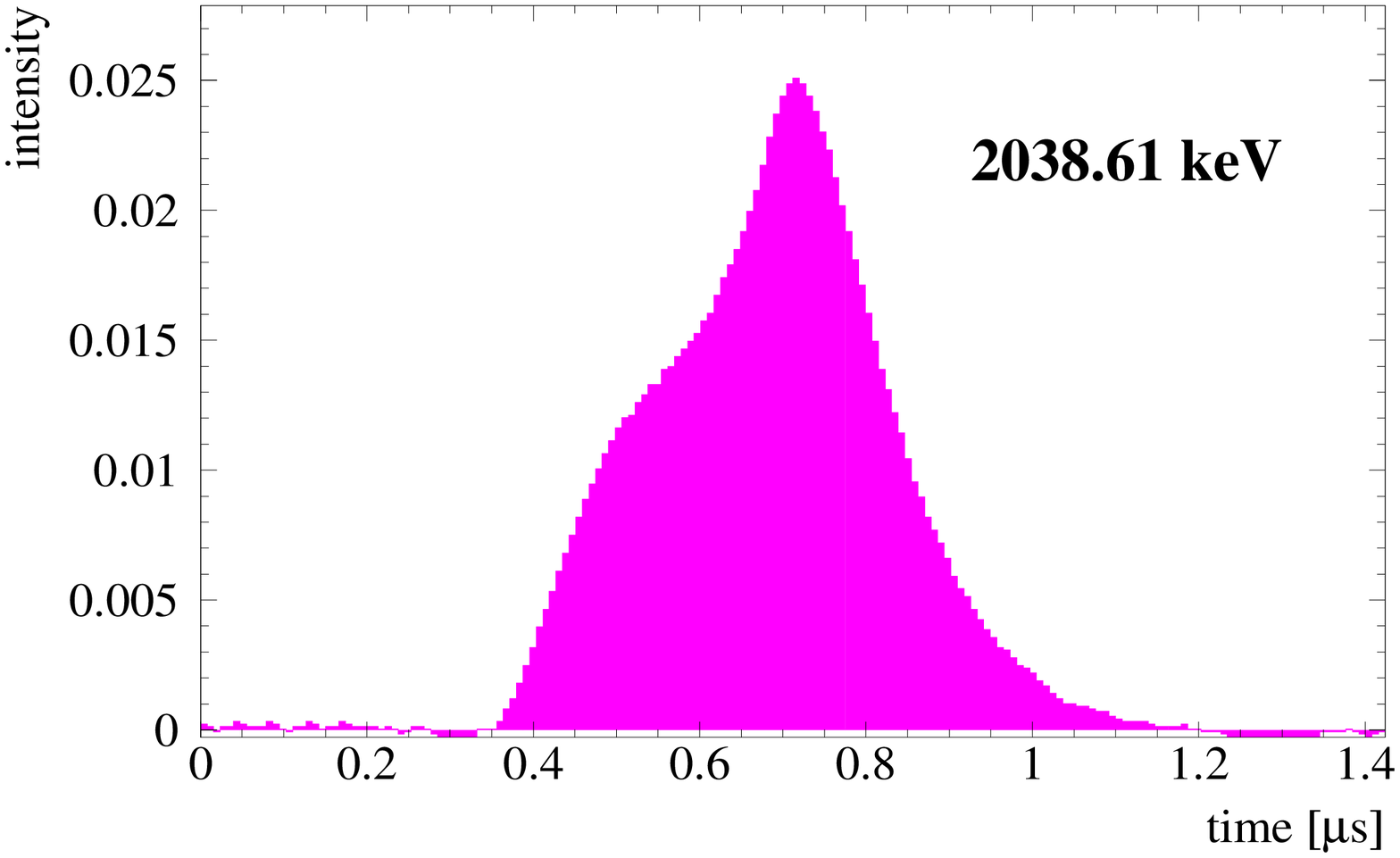}
\includegraphics*[scale=0.15]{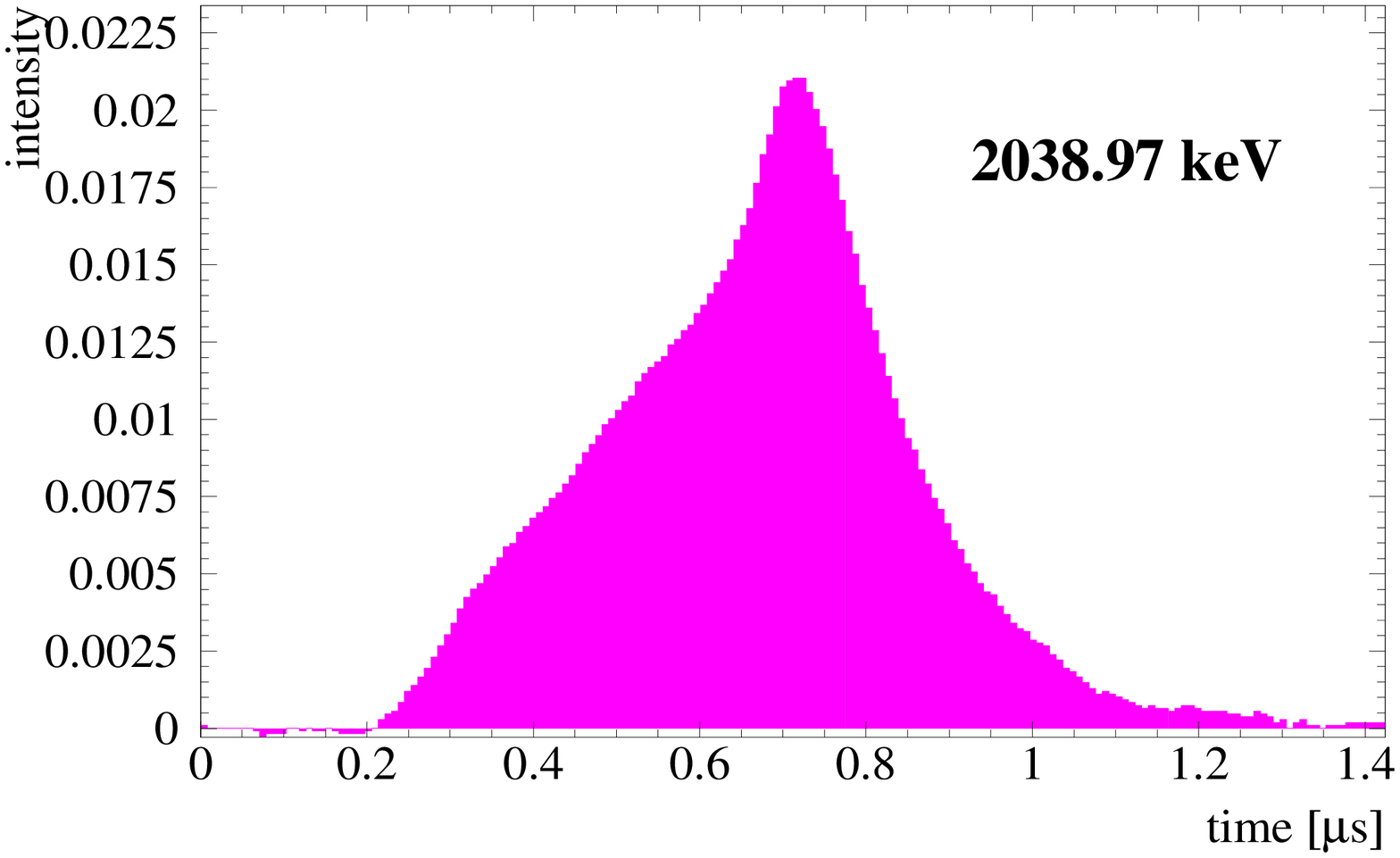}
\includegraphics*[scale=0.15]{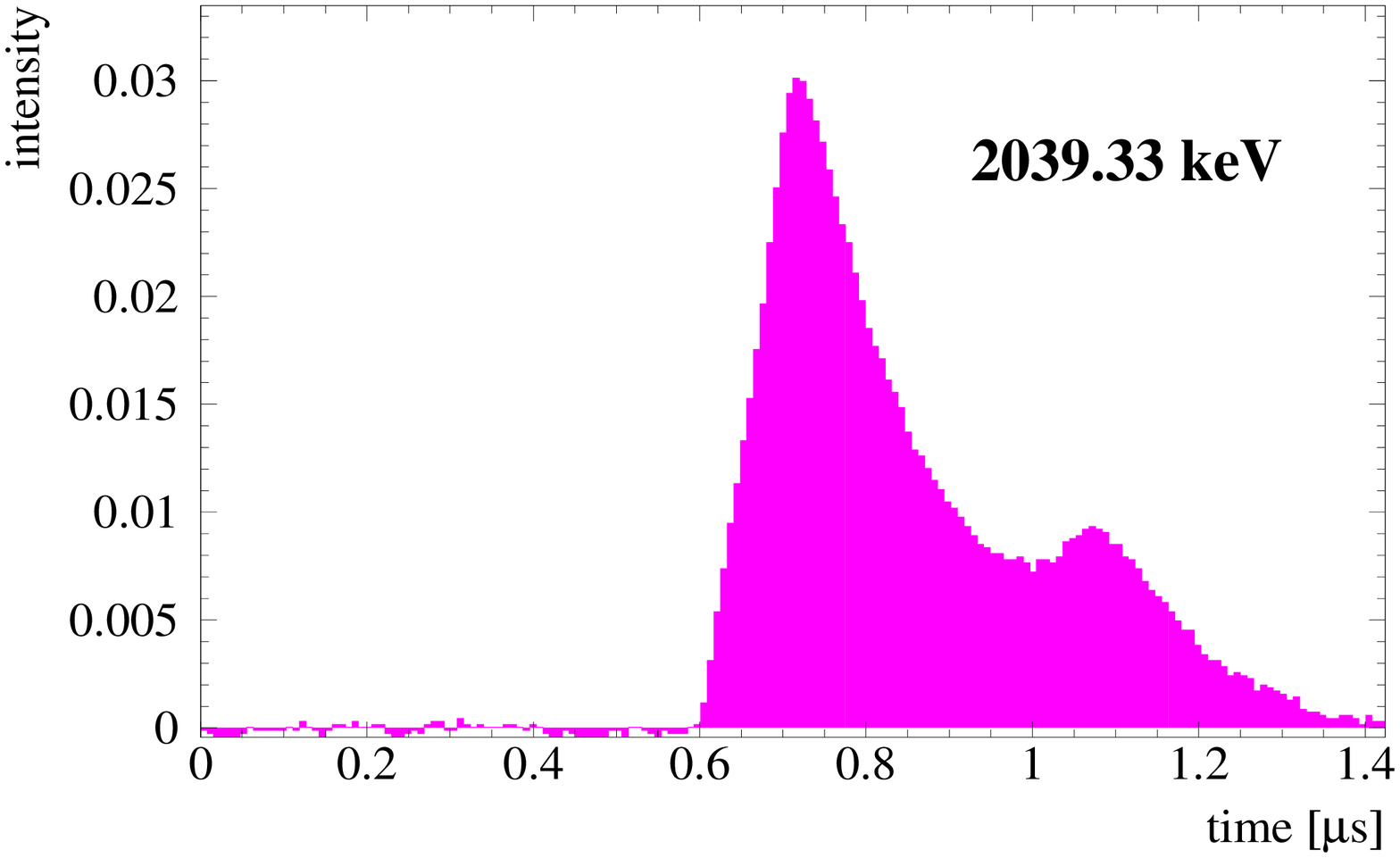}
\includegraphics*[scale=0.15]{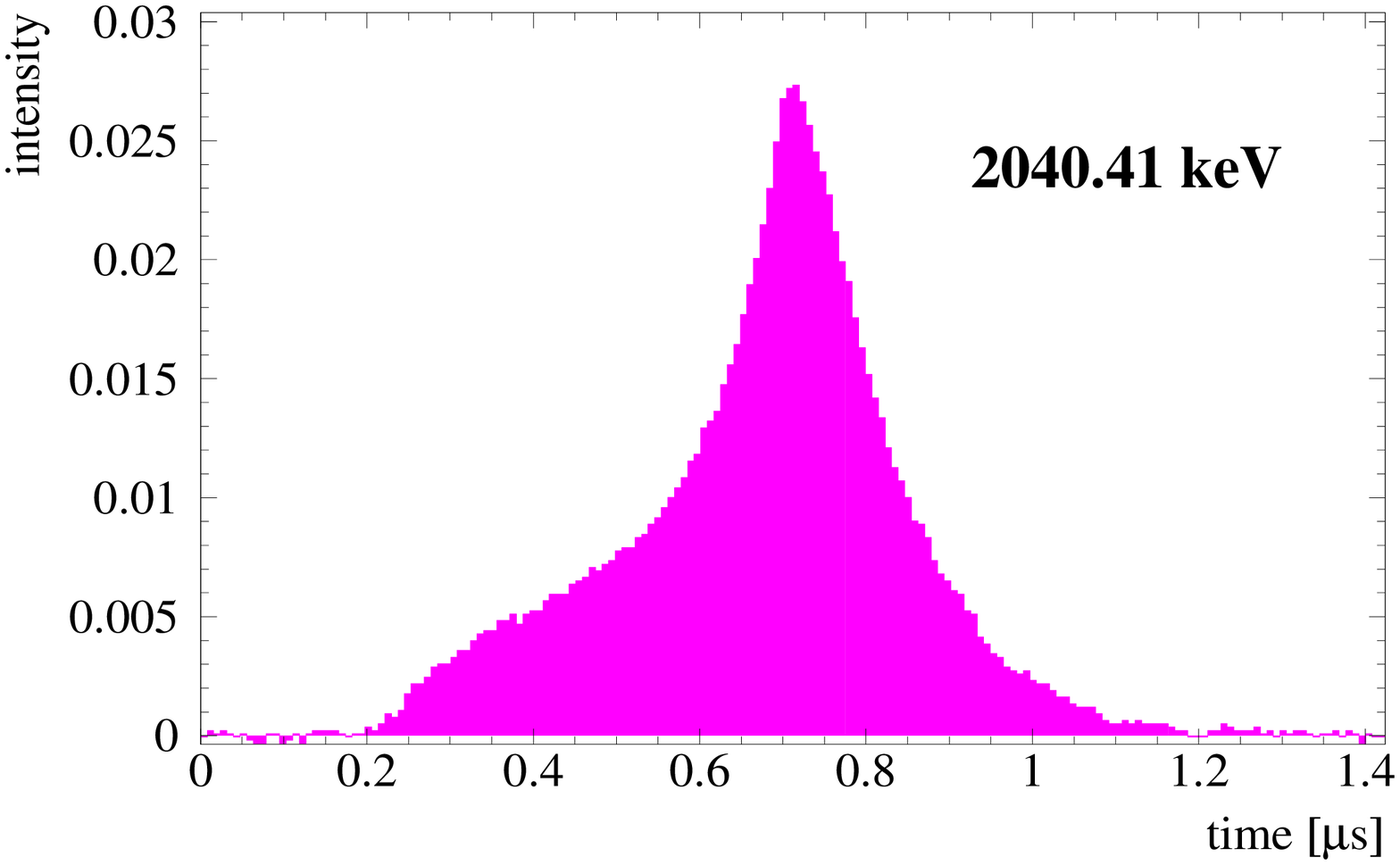}
\includegraphics*[scale=0.15]{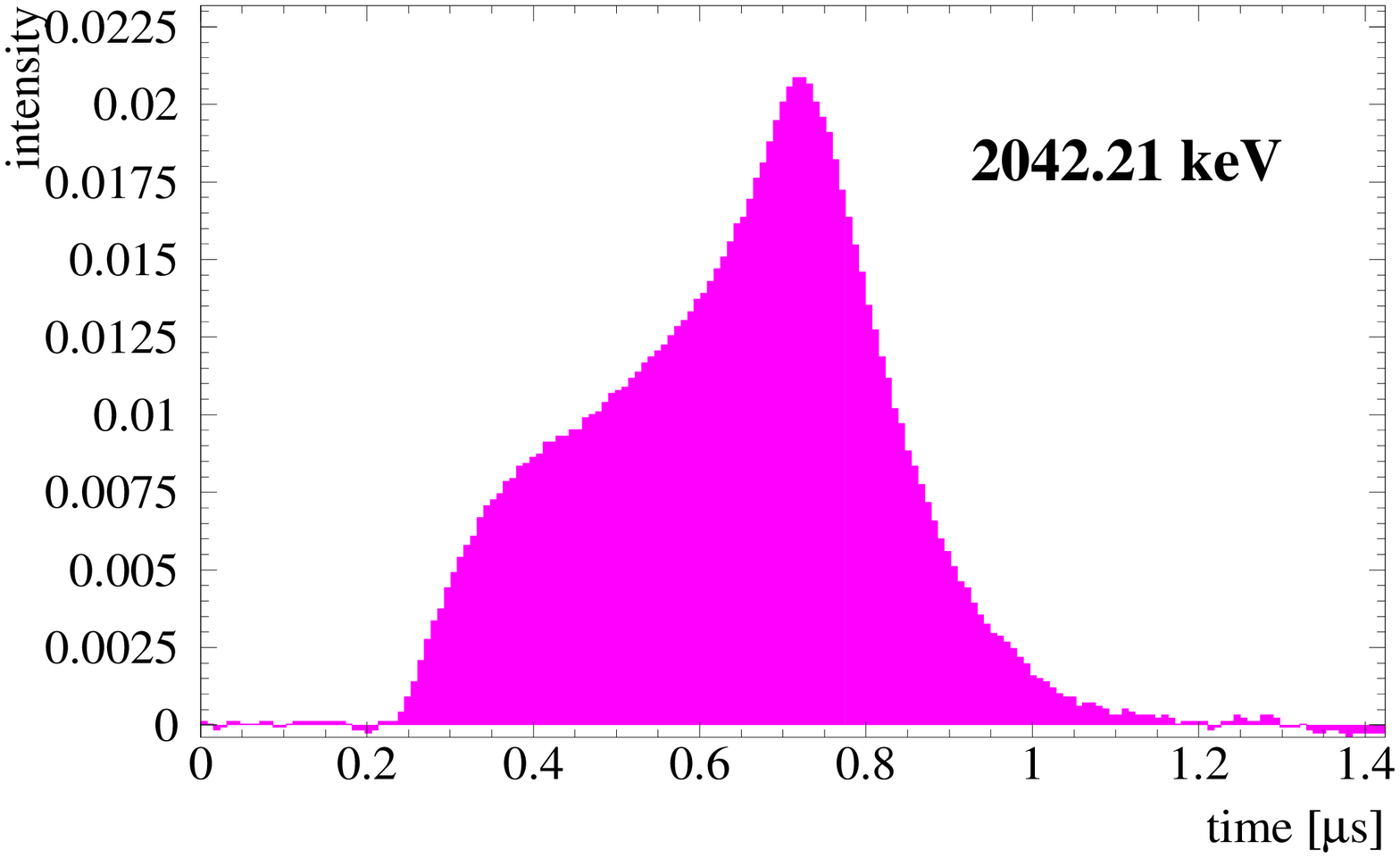}
\includegraphics*[scale=0.15]{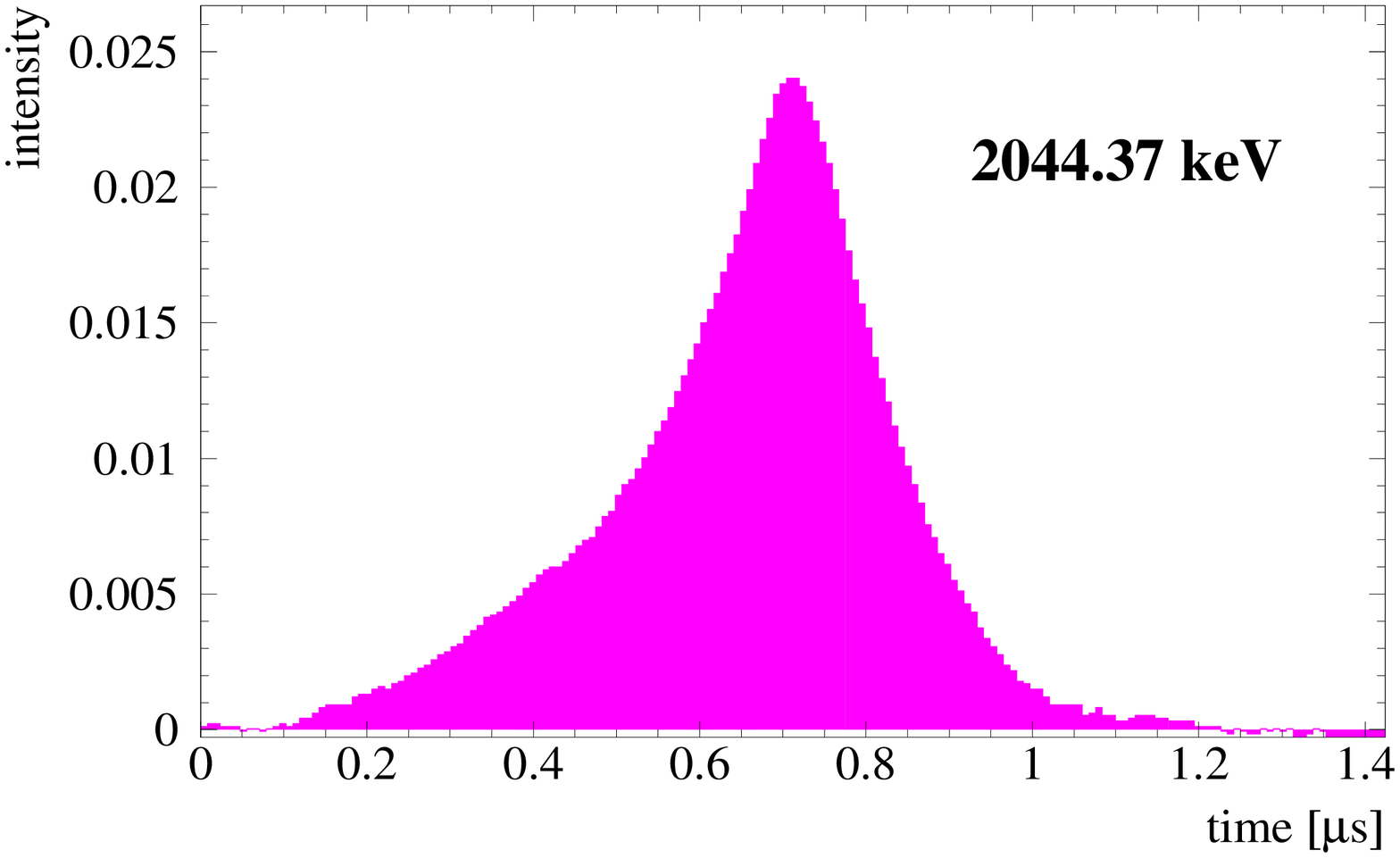}
}
\caption[]{
	Events classified to be single site events (SSE) 
	by all three methods of PSA in the range 2034.1 - 2044.9\,keV, 
	in the measurement period 10.1995 - 05.2000, 28.053\,kg\,y.
}
\label{All-Shapes}
\end{figure}

\begin{figure}[h]
\begin{center}
\includegraphics
[height=7.cm,width=11.cm]{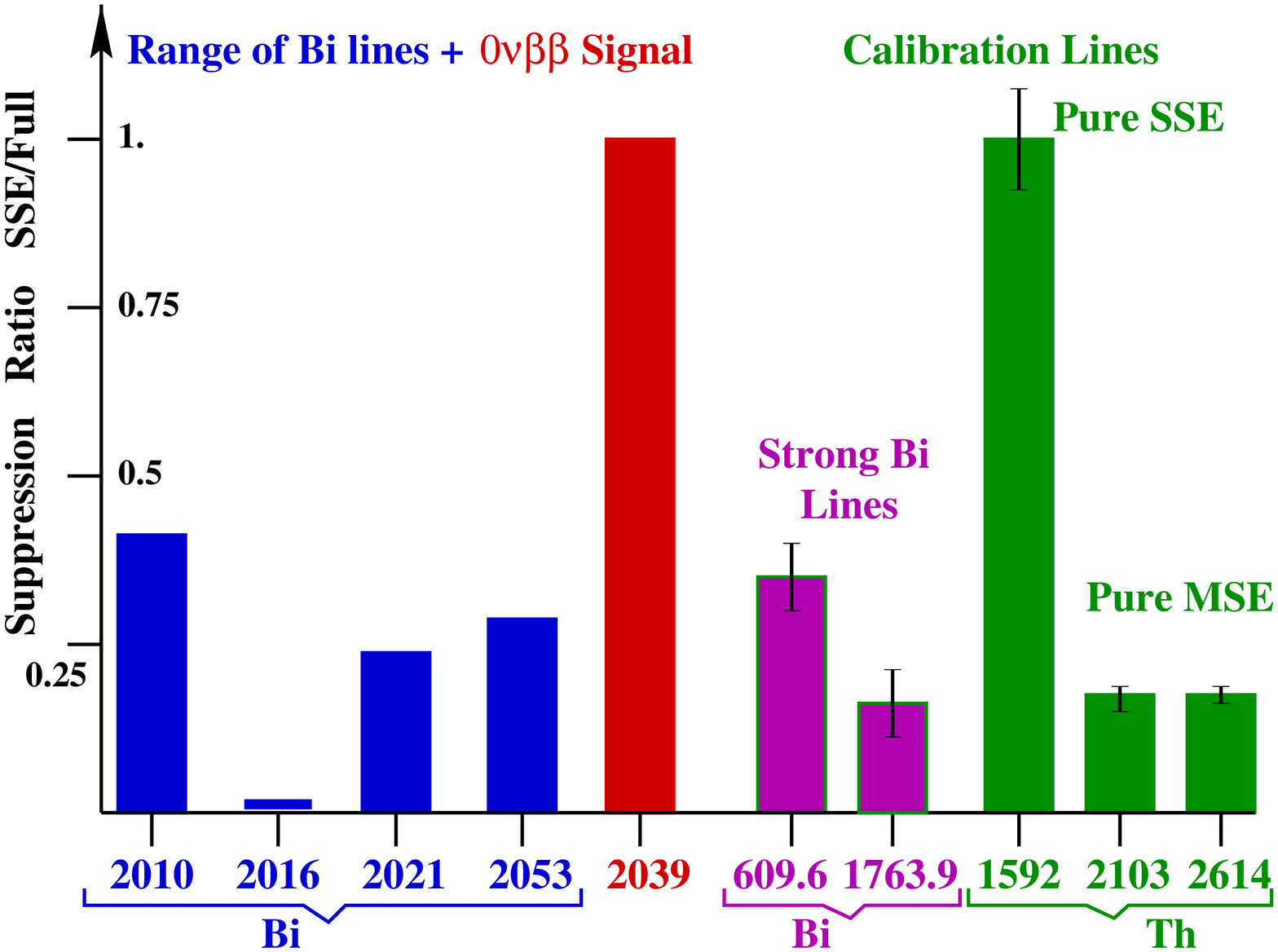}
\end{center}
\caption[]{Relative suppression ratios: 
	Remaining intensity after pulse shape analysis compared 
	to the intensity in the full spectrum. 
\underline{Right:} 
	Result of a calibration measurement with a Th source - 
	ratio of the intensities of the 1592\,keV line 
	(double escape peak, known to be 100\% SSE), set to 1. 
	The intensities of the 2203\,keV line (single escape peak, 
	known to be 100\% MSE) as well as of the 2614\,keV line 
	(full energy peak, known to be dominantly MSE) 
	are strongly reduced (error bars are $\pm 1\sigma$). 
	The same order of reduction is found for the strong Bi lines 
	occurring in our spectrum (see Fig. 3) - 
	shown in this figure are the lines at 609.4 and 1763.9\,keV. 
\underline{Left:} 
	The lines in the range of weak statistics around the line 
	at 2039\,keV (shown are ratios of best fit values). 
	The Bi lines are reduced compared to the line at 2039\,keV (set to 1), 
	as to the 1592\,keV SSE Th line.}  
\label{Suppresion}
\end{figure}

	From the analysis of the single site events (Fig. 
\ref{Sum_spectr_5det}),
	we find after 28.053\,kg\,y of measurement 
	9 SSE events in the region 2034.1 - 2044.9\,keV 
	($\pm$ 3$\sigma$ around Q$_{\beta\beta}$)    
	(Fig. 
\ref{All-Shapes}).
	Analysis 
	of the single site event spectrum (Fig. 
\ref{Sum_spectr_5det}),  
	as described in


\noindent
	 section 3.1, shows again evidence for a line   
	at the energy of Q$_{\beta\beta}$ (Fig. 
\ref{Bay-Hell-95-00-gr-kl-Bereich}). 
	Analyzing the range of 2032 - 2046\,keV, 
	we find 
	the probability of 96.8$\%$ that there is a line at 2039.0\,keV.
	We thus see a signal of single site events,  
	as expected 
	for neutrinoless double beta decay, 
	precisely at the Q$_{\beta\beta}$ value obtained 
	in the precision experiment of 
\cite{New-Q-2001}. 	
	The analysis of the line at 2039.0\,keV 
	before correction for the efficiency yields 
	4.6\,events (best value) or 
	(0.3 - 8.0)\,events within 95$\%$ c.l. 
	((2.1 - 6.8)\,events within 68.3$\%$ c.l.). 
	Corrected for the efficiency to identify 
	an SSE signal by successive application of all 
	three PSA methods, which is 0.55 $\pm$ 0.10, 
	we obtain a \onbb signal with 92.4$\%$ c.l.. 
	The signal is 
	(3.6 - 12.5)\,events with 68.3$\%$ c.l.
	 ~(best value 8.3\,events).
	Thus, with proper normalization concerning the running 
	times (kg\,y) of the full and the SSE spectra, 
	we see that almost the full signal remains after 
	the single site cut (best value), 
	while the $^{214}{Bi}$ lines (best values) are considerably reduced.
	The reduction is comparable to the reduction of the 2103\,keV 
	and 2614\,keV $^{228}{Th}$ lines (known to be multiple site 
	or mainly multiple site), relative to the 1592\,keV 
	$^{228}{Th}$ line (known to be single site), see Fig. 
\ref{Suppresion}.
	The same reduction is also found for the strong $^{214}{Bi}$ 
	lines (e.g. at 609.6 and 1763.9\,keV (Fig. 
\ref{Suppresion})).


\begin{figure}[h]
\begin{center}
\includegraphics
[height=4.6cm,width=13.cm]{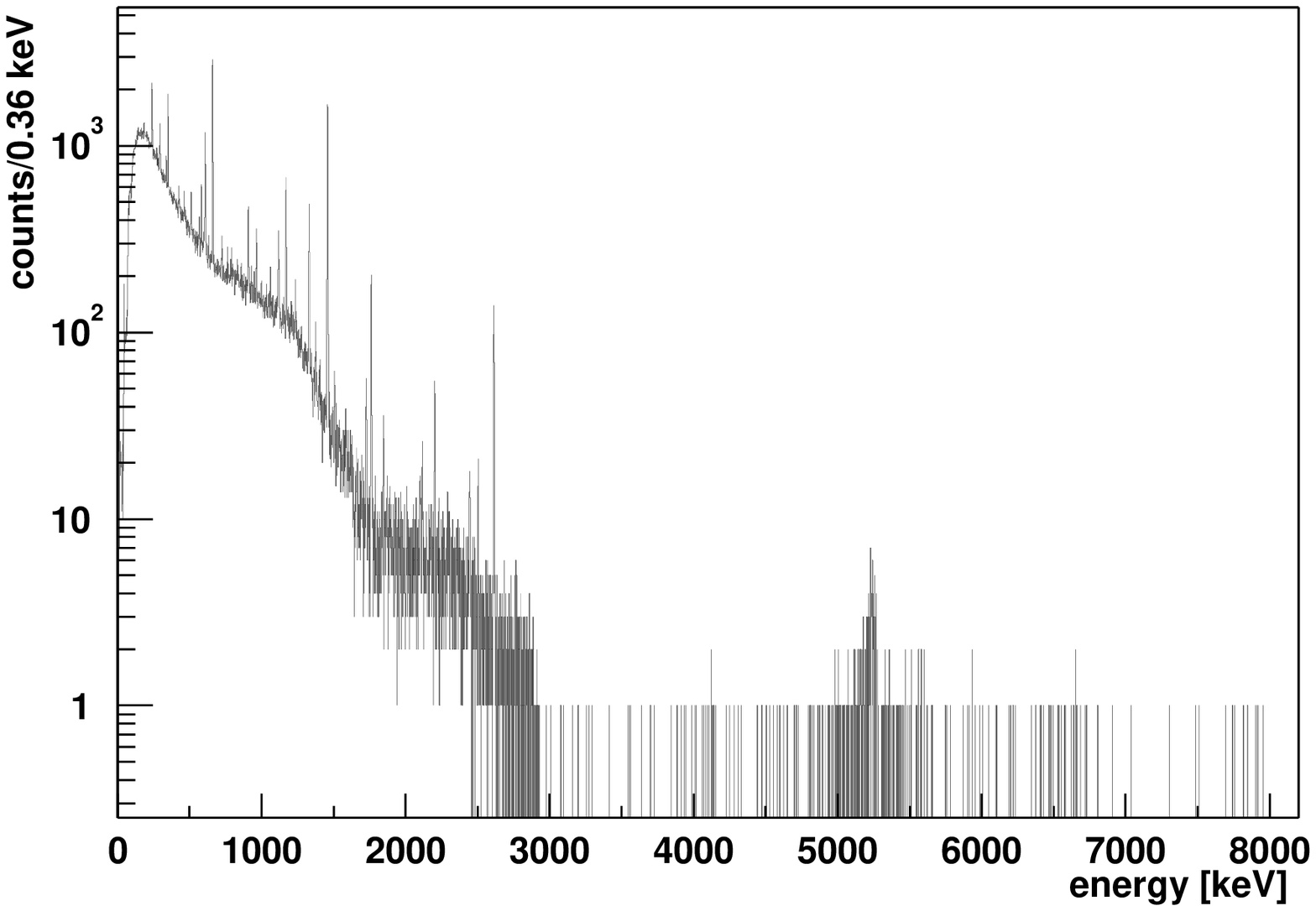}
\includegraphics
[height=4.6cm,width=13.cm]{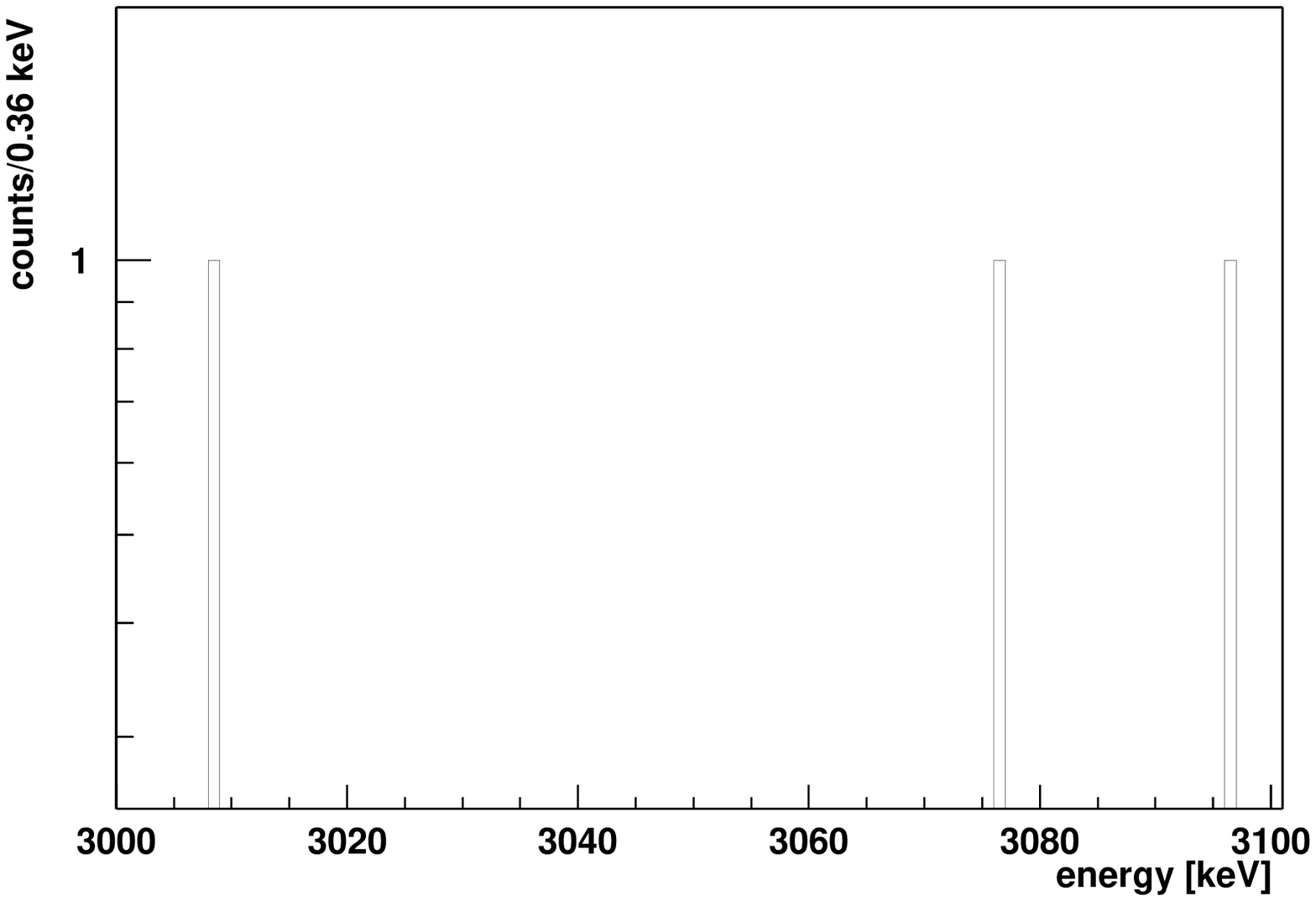}
\end{center}
\caption[]{The measured sum spectrum up to $\sim$ 8\,MeV (upper part) 
	(22.36\,kg\,y). 
	No lines can be identified in the range 3-8\,MeV, 
	except a peak at 5.24$\pm$0.03\,MeV, 
	which we identify as $\alpha$ decay 
	from $^{210}{Po}$ to $^{206}{Pb}$ (see 
\cite{Diss-Dipl}).
	\underline{Lower part:} the range 3000 - 3100\,keV, 
	where the full energy peak should occur at 3061\,keV, 
	in case the 2039\,keV signal would be the double 
	escape peak of a $\gamma$-line.} 
\label{Doubl-Esc-P}
\end{figure}


	The Feldman-Cousins method gives a signal 
	at 2039.0\,keV of 2.8 $\sigma$ (99.4$\%$). 
	The possibility, that the single site signal at 2039.0\,keV 
	is the double 
	escape line corresponding to a (much more intense!) 
	full energy peak of a $\gamma$-line at 2039+1022=3061\,keV 
	is excluded from the high-energy part of our spectrum (see Fig. 
\ref{Doubl-Esc-P}).

\subsection{Comparison with Earlier Results}

	We applied the same methods of peak search as used 
	in sections 3.3, 3.4, to the spectrum measured 
	in the Ge experiment by Caldwell et al. 
\cite{Caldw91}
	more than a decade ago. 
	These authors had the most sensitive experiment using 
	{\it natural} Ge detectors (7.8\% abundance of $^{76}{Ge}$). 
	With their background being a factor of 9 higher than 
	in the present experiment, and their measuring time 
	of 22.6\,kg\,y, they have a statistics 
	for the background larger by a factor of almost 4 
	in their (very similar) experiment. 
	This allows helpful conclusions about the nature of the background.

\begin{figure}[h]
\begin{center}
\includegraphics
[height=4.3cm,width=10.5cm]{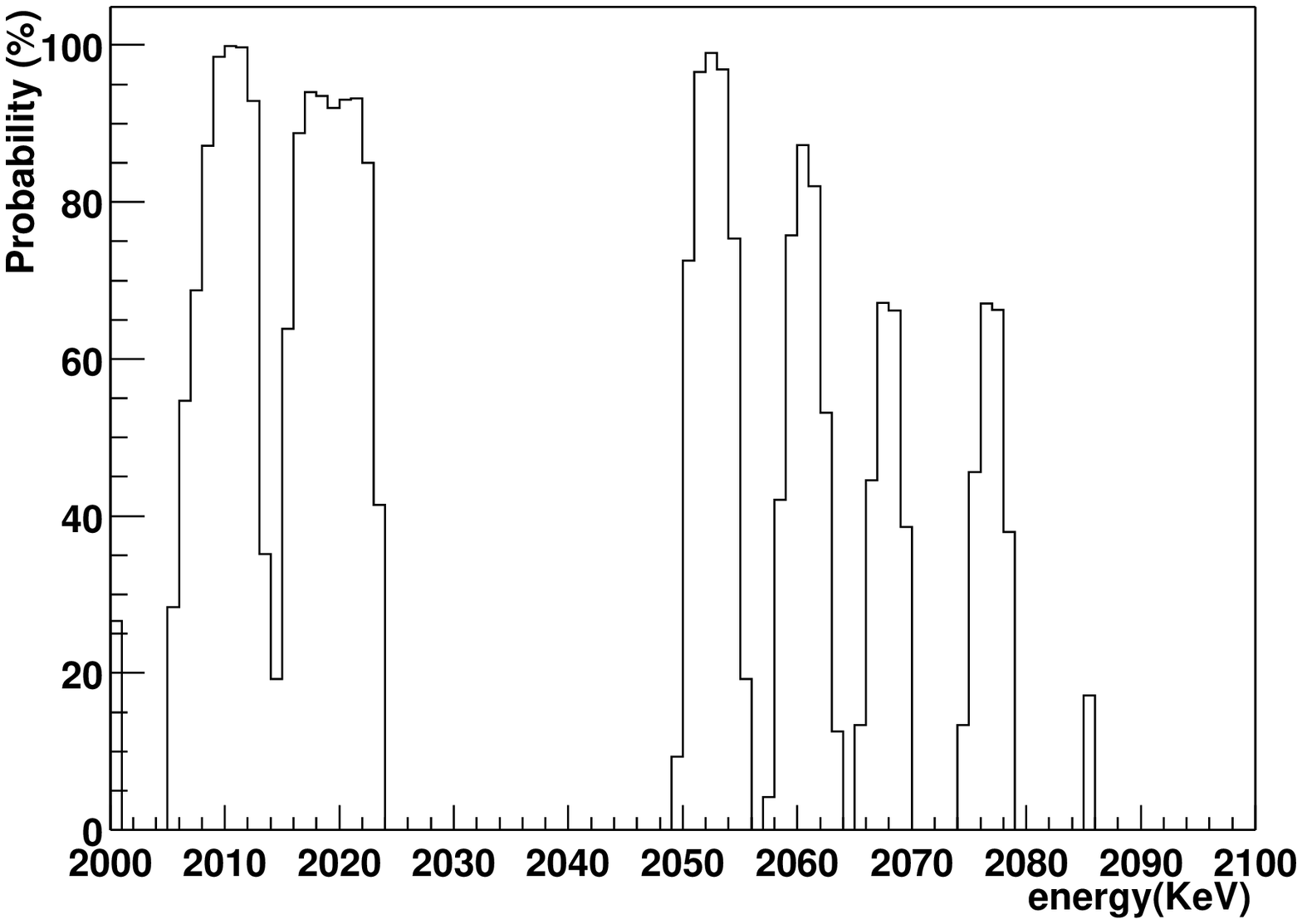}
\includegraphics
[height=4.3cm,width=10.5cm]{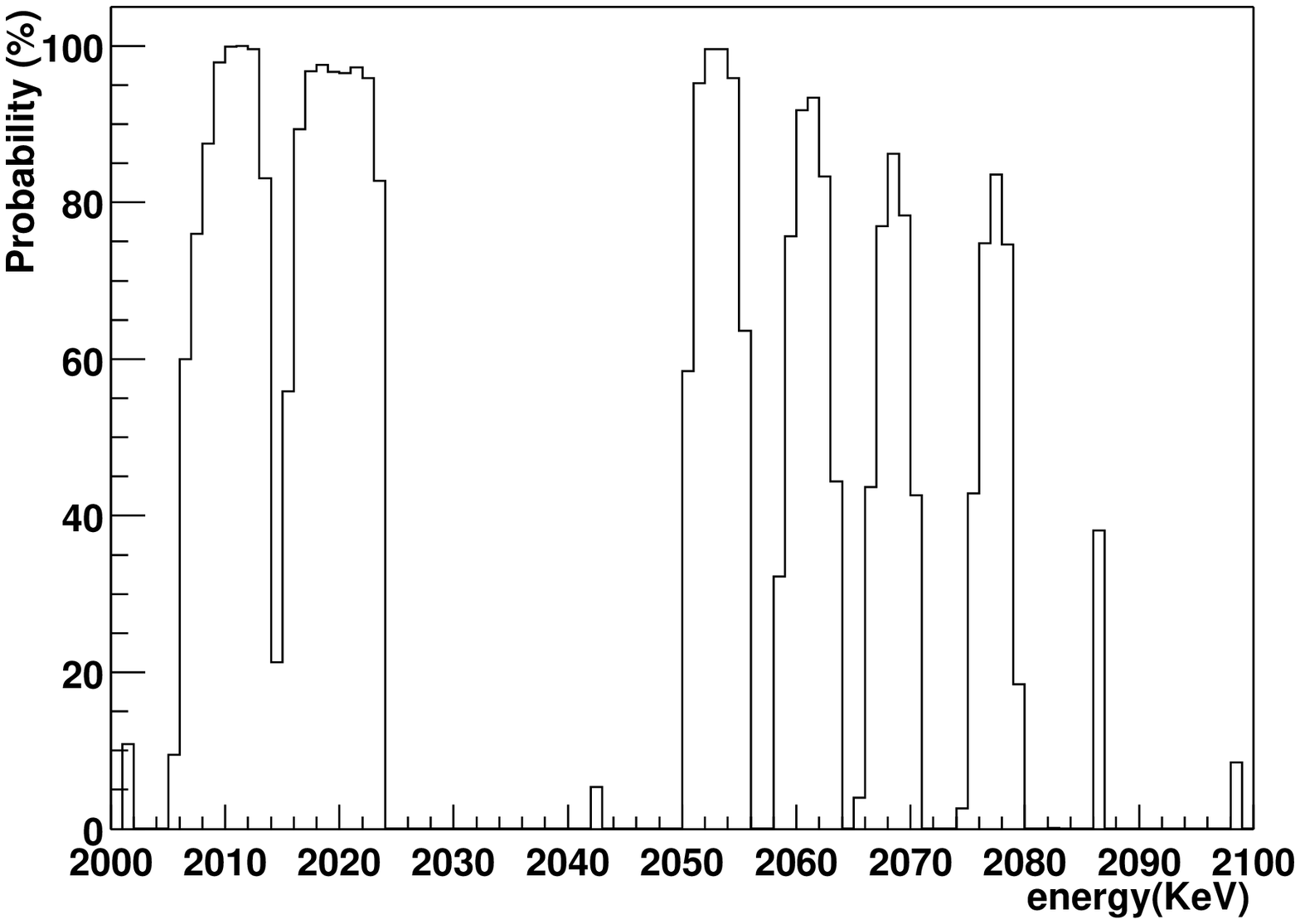}
\end{center}
\caption[]{Peak scanning of the spectrum measured by Caldwell et al. 
	\cite{Caldw91} with the Maximum Likelihood method (upper part), 
	and with the Bayesian method (lower part) (as in Fig. 
\ref{Bay-Alles-90-00-gr-kl-Bereich},\ref{Bay-Chi-all-90-00-gr},\ref{Bay-Hell-95-00-gr-kl-Bereich}) (see 
\cite{KK-new02}).} 
\label{Caldw-Max-Bayes}
\end{figure}

\begin{figure}[h]
\begin{center}
\includegraphics
[height=5.5cm,width=10.5cm]{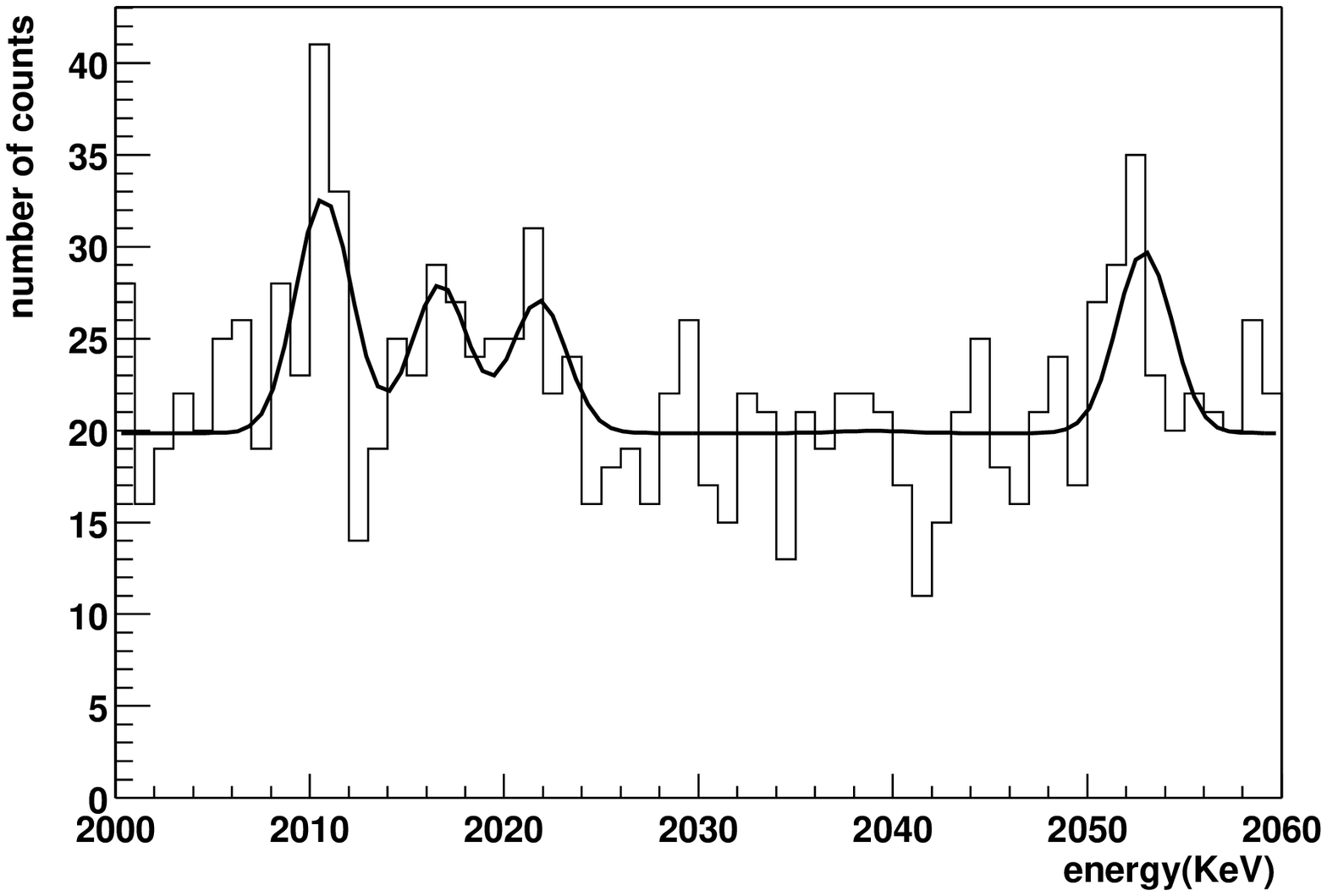}
\end{center}
\caption[]{Analysis of the spectrum measured by D. Caldwell et al. 
	\cite{Caldw91},
	with the Maximum Likelihood Method, in the energy 
	range 2000 - 2060\,keV (as in Fig. 
\ref{Bi-Lines-0_36keV}) 
	assuming lines at 2010.78, 2016.70, 2021.60, 2052.94, 2039.0\,keV.
	In contrast to Fig. 
\ref{Bi-Lines-0_36keV}
	no indication for a signal at 2039\,kev is observed (see text).} 
\label{Caldw-Sp-Q}
\end{figure}

	The peak scanning finds (Fig. 
\ref{Caldw-Max-Bayes})
	indications for peaks essentially at  the same energies as in Figs.
\ref{Bay-Alles-90-00-gr-kl-Bereich},\ref{Bay-Chi-all-90-00-gr},\ref{Bay-Hell-95-00-gr-kl-Bereich}.
	This shows that these peaks are not fluctuations. 
	In particular it sees the 2010.78 and 2052.94\,keV $^{214}{Bi}$ 
	lines with 3.6 and 2.8 $\sigma$ c.l., respectively. 
	It finds, however, {\it no line at Q$_{\beta\beta}$} (see also Fig. 
\ref{Caldw-Sp-Q}).
	This is consistent with the expectation from the rate 
	found from the HEIDELBERG-MOSCOW experiment. 
	About 17 observed events in the latter correspond to 0.6 
	{\it expected} events in the Caldwell experiment, 
	because of the use of non-enriched material 
	and the shorter measuring time.


%
%

%


	Another Ge experiment (IGEX) using 9\,kg of enriched $^{76}{Ge}$, 
	but collecting since beginning of the experiment 
	in the early nineties till shutdown in end of 1999 only 8.8\,kg\,y 
	of statistics 
\cite{DUM-RES-AVIGN-2000}, 
	because of this low statistics also naturally cannot 
	see any signal at 2039\,keV.


\subsection{Some Comments on the Bayesian, ${\chi}^2$ 
	and \protect\newline Maximum Likelihood Methods} 

	We probed the sensitivity of peak identification 
	for the three methods: Bayesian, ${\chi}^2$, 
	Maximum-Likelihood, for the latter two using codes from 
\cite{Kchi-kvadr}.

	The disadvantage of the latter two methods is, 
	that at low counting rates observation of lines 
	with negative counting rate is possible. 
	This is {\it excluded} in the Bayesian method.
	We find that the Bayesian method 
	tends to systematically give 
	too conservative confidence limits (see Table
\ref{Simul-Sp}
	in section 3.2). 
	We shall discuss technical details of the three methods 
	in a separate paper 
\cite{KK-new02}.


\begin{figure}[h]
\begin{center}
\includegraphics*
[height=5.cm,width=10cm]{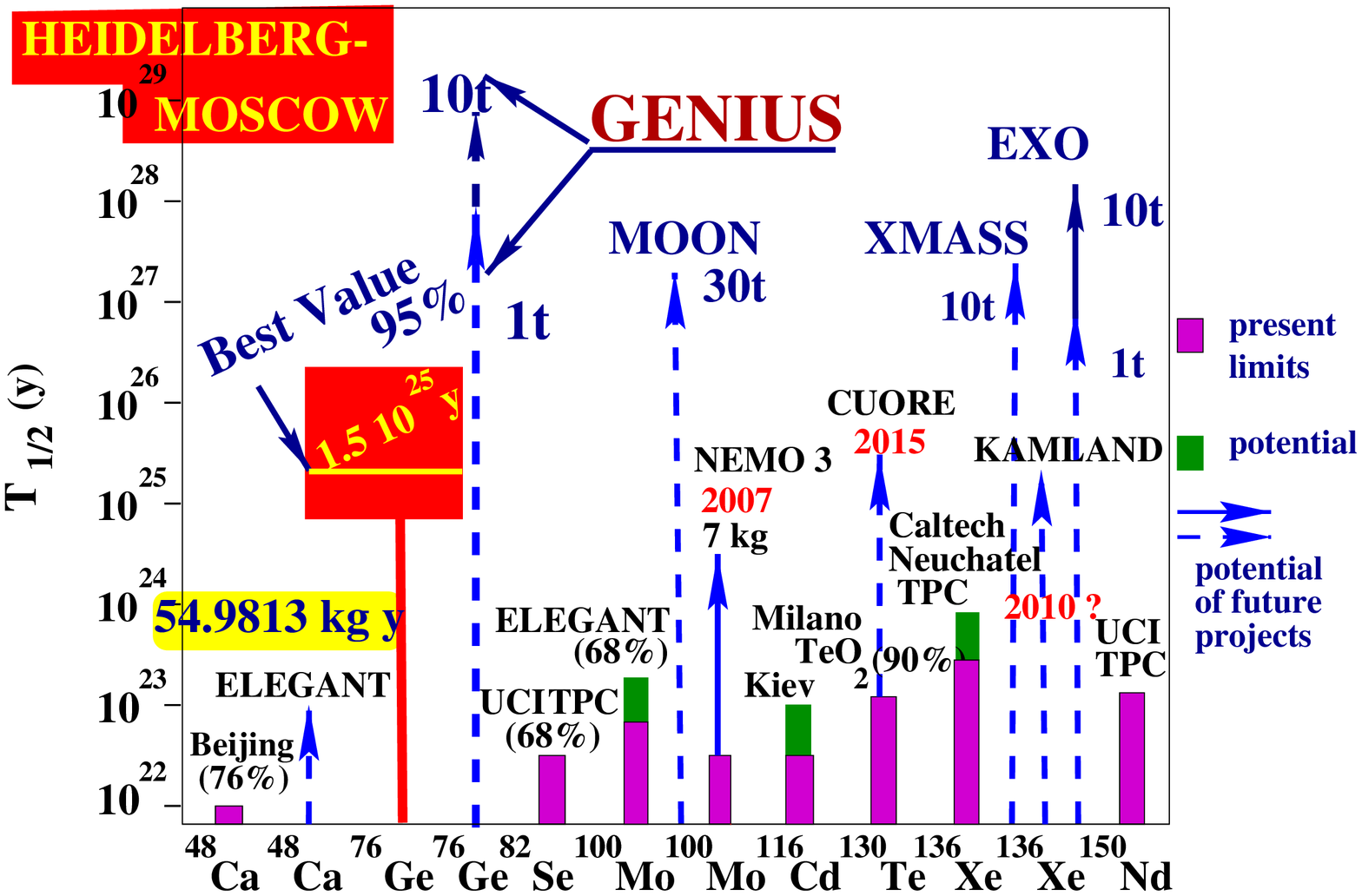}
\end{center}
\caption[]{
	Present evidence for \onbb decay from data 
	of the HEIDELBERG-MOSCOW experiment and the potential of present 
	and future $\beta\beta$ experiments. 
	Vertical axis - half life limits 
	(only HEIDELBERG-MOSCOW gives a value) in years, horizontal - 
	isotopes used in the various experiments.
}
\label{Diagr-s2}
\begin{center}
\includegraphics*
[height=5.cm,width=10cm]{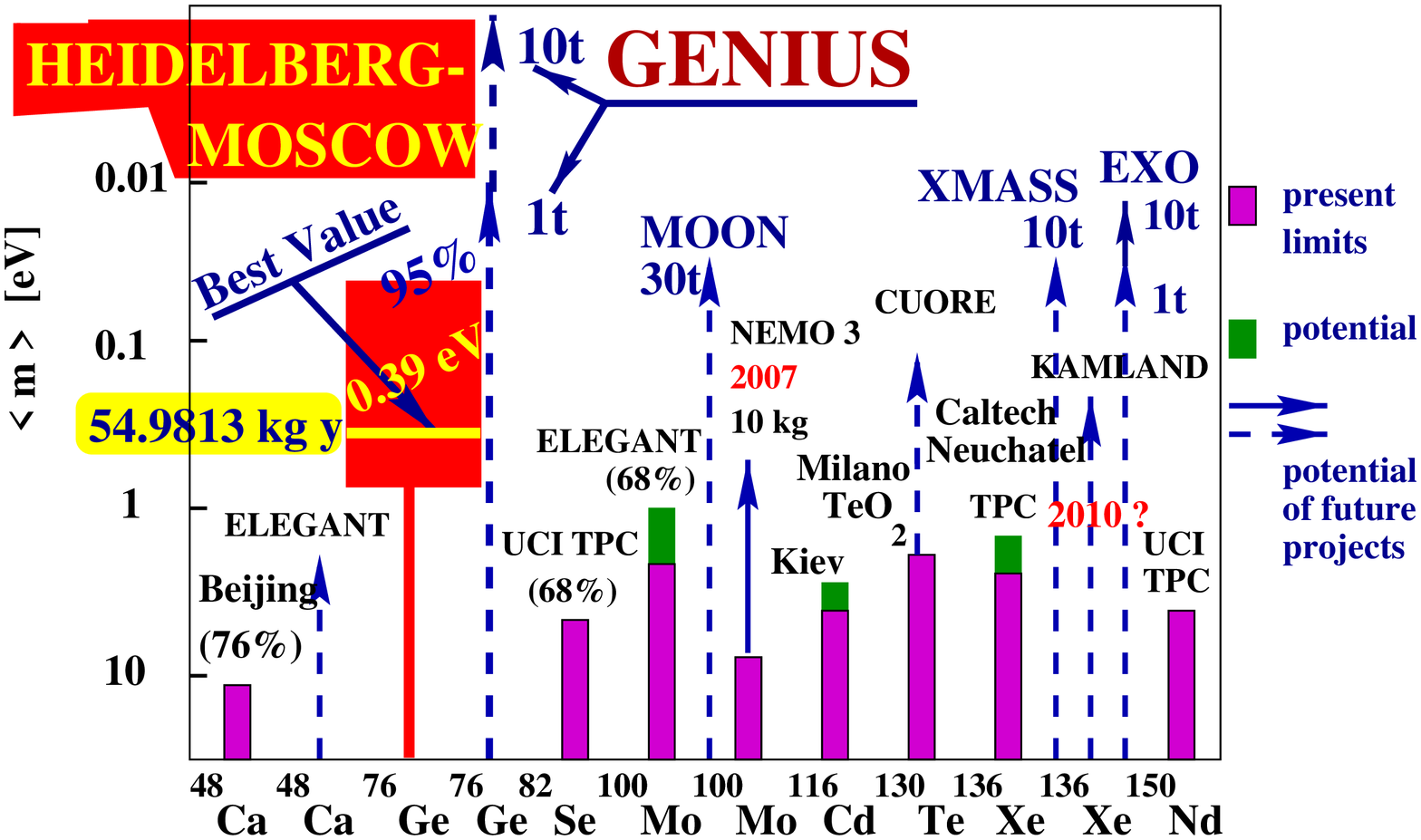}
\end{center}
\caption[]{
	Present evidence for \onbb decay from data 
	of the HEIDELBERG-MOSCOW experiment and the potential of present 
	and future $\beta\beta$ experiments. 
	Vertical axis - effective neutrino mass limits 
	(only HEIDELBERG-MOSCOW gives a value) in eV, horizontal - 
	isotopes used in the various experiments.
}
\label{Diagr-s1}
\end{figure}


\section{Half-life of the Neutrinoless Mode and 
\protect\newline Effective Neutrino Mass}

	We emphasize that we find in all analyses 
	of our spectra a line at the value of Q$_{\beta\beta}$. 
	We have shown that to our present knowledge 
	the signal at Q$_{\beta\beta}$  
	does not originate from a background 
	$\gamma$-line. 
	On this basis we 
	translate the observed number of events 
	into half-lives for the neutrinoless double beta decay. 
	We give in Table 
\ref{Results}
	 conservatively the values obtained with 
	the Bayesian method and not those obtained with 
	the Feldman-Cousins method. 
	Also given in Table 
\ref{Results} 
	are the effective neutrino masses    
	$\langle m \rangle $ deduced using the matrix elements of 
\cite{Sta90,Mut89}.

	We derive from the data taken with 46.502\,kg\,y 
	the half-life 
	${\rm T}_{1/2}^{0\nu} = (0.8 - 18.3) \times 10^{25}$ 
	${\rm y}$ (95$\%$ c.l.). 
	The analysis of the other data sets, shown in Table 
\ref{Results}  
	confirm this result.
	Of particular importance is that we see 
	the \onbb signal in the single site spectrum.

	The result obtained is consistent with the limits 
	given earlier by the HEI\-DELBERG-MOSCOW experiment 
\cite{HDM01}.  
	It is also consistent with all other double beta experiments -  
	which still reach less sensitivity (see Figs.
\ref{Diagr-s1},\ref{Diagr-s2}). 
	A second Ge-expe\-riment  
\cite{DUM-RES-AVIGN-2000}, 
	which has stopped operation in 1999 after 
	reaching a significance of 9\,kg\,y, 
	yields (if one believes their method of 'visual inspection' 
	in their data analysis) in a conservative analysis a limit of 
	${\rm T}_{1/2}^{0\nu} > 0.55 \times 10^{25}
	{\rm~ y}$  (90\% c.l.). 
	The $^{128}{Te}$ geochemical experiment 
\cite{Ber92}
	yields $\langle m_\nu \rangle < 1.1$ eV (68 $\%$ c.l.), 
	the $^{130}{Te}$ cryogenic experiment yields 
\cite{Ales00}
	$\langle m_\nu \rangle < 1.8$ eV 
	and the CdWO$_4$ experiment 
\cite{Dan2000}
	$\langle m_\nu \rangle < 2.6$ eV,  
	all derived with the matrix elements of 
\cite{Sta90} 
	to make the results comparable to the present value.

	Concluding we obtain, on the above basis, 
	with more than 95$\%$ probability, 
	first evidence for the neutrinoless 
	double beta decay mode.


\begin{table}[h]
\begin{center}
\newcommand{\m}{\hphantom{$-$}}
\renewcommand{\arraystretch}{1.}
\setlength\tabcolsep{6.2pt}
\begin{tabular}{c|c|c|c|c}
\hline
&&&&\\
Significan-	&	Detectors		
&	${\rm T}_{1/2}^{0\nu}	{\rm~ \,y}$	
& $\langle m \rangle $ eV	&	Conf. \\
	ce $[ kg\,y ]$	&&&& level\\
\hline
&&&&\\
	54.9813	
&	1,2,3,4,5	
&	$(0.80 - 35.07) \times 10^{25}$	
& (0.08 - 0.54)	
& 	$95\%  ~c.l.$	\\
	&	
&	$(1.04 - 3.46) \times 10^{25}$	
& (0.26 - 0.47)	
& $68\%  ~c.l.$	\\
 		&	
&	$1.61 \times 10^{25}$	
& 0.38 	
& Best Value	\\
\hline
 	46.502	
&	1,2,3,5
&	$(0.75 - 18.33) \times 10^{25}$	
& (0.11 - 0.56)
& $95\%  ~c.l.$	\\
&	
&	$(0.98 - 3.05) \times 10^{25}$	
& (0.28 - 0.49)
& $68\% ~c.l.$	\\
&	
&	$1.50 \times 10^{25}$	
& 0.39 
& Best Value		\\
\hline
	28.053	
&	2,3,5  SSE	
&	$(0.88 - 22.38) \times 10^{25}$	
& (0.10 - 0.51)	
& $90\% ~c.l.$		\\
&	
&	$(1.07 - 3.69) \times 10^{25}$	
& (0.25 - 0.47)	
& $68\% ~c.l.$		\\
&	
&	 $1.61 \times 10^{25}$	
& 0.38
& Best Value	\\
\hline
\end{tabular}
\end{center}
\caption[]{\label{Results}Half-life for the neutrinoless decay mode 
	and deduced effective neutrino mass 
	from the HEIDELBERG-MOSCOW experiment.
}
\end{table}

\begin{table}[h]
\begin{center}
\newcommand{\m}{\hphantom{$-$}}
\renewcommand{\arraystretch}{.5}
\setlength\tabcolsep{7.3pt}
\begin{tabular}{c|c|c|c|c|c|c|c|c}
\hline
\hline
&&&&&&&&\\
$M^{0\nu}$ from	& \cite{Mut89,Sta90}	& \cite{Tom91}	& \cite{Hax84}	
& \cite{Cau96} & \cite{FaesSimc97}	&	\cite{Sim99}	
& 	\cite{Faes01}	&	\cite{St-KK01-1,St-KK01-2}\\
&&&&&&&&\\
\hline
&&&&&&&&\\
$\langle m \rangle$\,eV	95$\%$&	0.39	&	0.37	&	0.34	
&	1.06	&	0.87	&	0.60	&	0.53	
&	0.44 - 0.52\\
&&&&&&&&\\
\hline
\hline
\end{tabular}
\end{center}
\caption{\label{Matr-Elem}The effect of nuclear matrix elements 
	on the deduced effective neutrino masses. 
	Shown are the neutrino masses deduced from the best 
	value of ${\rm T}_{1/2}^{0\nu} = 1.5 \times 10^{25} 
	{\rm~ y}$ determined 
	in this work with 97$\%$\,c.l. when using matrix elements 
	from various calculations and a phase factor 
	of $F^{0\nu}_1 = 6.31\times {10}^{-15}\,y^{-1}$.
}
\end{table}


	As a consequence, at this confidence level,  
	lepton number is not conserved. 
	Further our result implies that the neutrino is a Majorana particle.
	Both of these conclusions are {\it independent} 
	of any discussion of nuclear matrix elements.

	The matrix element enters when we derive 
	a {\it value} for the effective neutrino mass. 
	If using the nuclear matrix element from 
\cite{Sta90,Mut89}, 
	we conclude from the various 
	analyses given above the effective mass 
	$\langle m \rangle $ 
	to be $\langle m \rangle $ 
	= (0.11 - 0.56)\,eV (95$\%$ c.l.), 
	with best value of 0.39\,eV. 
	Allowing conservatively for an uncertainty of the nuclear 
	matrix elements of $\pm$ 50$\%$
	(for detailed discussions of the status 
	of nuclear matrix elements we refer to 
\cite{Mut88,Gro89/90,Tom91,FaesSimc,KK60Y,Vog2001,Faes01,St-KK01-1,St-KK01-2}) 
	this range may widen to 
	$\langle m \rangle $ 
	= (0.05 - 0.84)\,eV (95$\%$ c.l.). 
	In Table 
\ref{Matr-Elem}
	we demonstrate the situation of nuclear matrix 
	elements by showing the neutrino masses deduced 
	from different calculations. 
	It should be noted that the value obtained 
	in Large Scale Shell Model Calculations 
\cite{Cau96}
	is understood to be too large by almost a factor of 2 
	because of the two small configuration space, (see, e.g. 
\cite{FaesSimc}), 
	and that the second highest value given (from 
\cite{FaesSimc97}), 
	has now been reduced to 0.53\,eV 
\cite{Faes01}. 
	The recent studies by 
\cite{St-KK01-1,St-KK01-2} 
	yield an effective mass of (0.44 - 0.52)\,eV. 
	We see that the early calculations 
\cite{Sta90}
	done in 1989 agree within less than 25$\%$ 
	with the most recent values.

\begin{figure}[h]
\begin{center}
\includegraphics*
[height=7.cm,width=12cm]{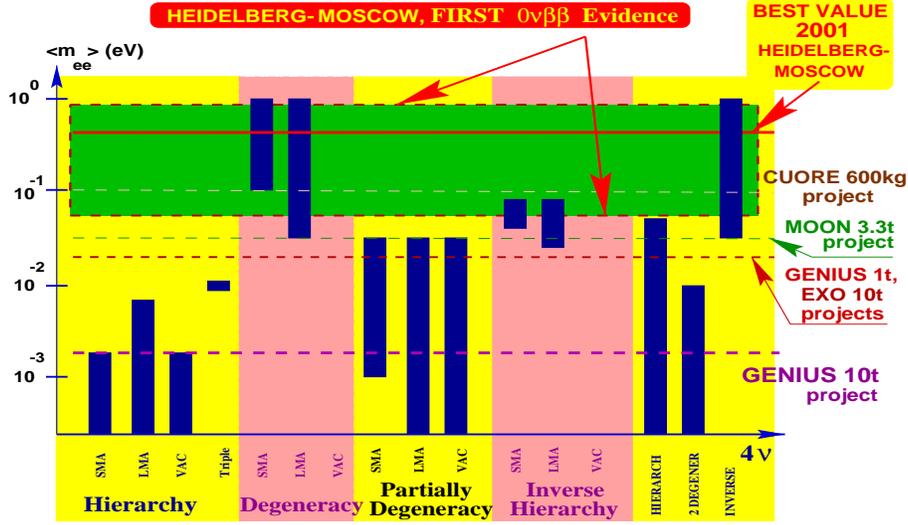}
\end{center}
\caption[]{
	The impact of the evidence obtained for neutrinoless 
	double beta decay in this paper (best value 
	of the effective neutrino mass 
	$\langle m \rangle$ = 0.39\,eV, 95$\%$ 
	confidence range (0.05 - 0.84)\,eV - 
	allowing already for an uncertainty of the nuclear 
	matrix element of a factor of $\pm$ 50$\%$)  
	on possible neutrino mass schemes. 
	The bars denote allowed ranges of $\langle m \rangle$ 
	in different neutrino mass scenarios, 
	still allowed by neutrino oscillation experiments.
	Hierarchical models are excluded by the 
	new \onbb decay result. Also shown are 
	the expected sensitivities 
	for the future potential double beta experiments 
	CUORE, MOON, EXO  
	and the 1 ton and 10 ton project of GENIUS 
\cite{KK-00-NOON-NOW-NANP-Bey97-GEN-prop} 
	(from  
\cite{KK-Sar_Evid01}).
}
\label{Jahr00-Sum-difSchemNeutr}
\end{figure}


	In the above conclusion for the effective neutrino mass, 
	it is assumed that contributions 
	to \onbb decay from processes other than  
	the exchange of 
	a Majorana neutrino (see, e.g. 
\cite{KK60Y,Moh91} 
	and section 1) are negligible.
	It has been discussed, however, recently 
\cite{Ueh02-Ev}
	that the possibility that \onbb decay is caused 
	by R-parity violation, may experimentally not be excluded, 
	although this would require making R-parity violating 
	couplings generation-dependent.


\section{Conclusions and Outlook}

	With the value deduced for the effective neutrino mass,  
	the HEIDELBERG-MOSCOW experiment excludes several 
	of the neutrino mass 
	scenarios 
	allowed from present neutrino oscillation experiments
	(see Fig.
\ref{Jahr00-Sum-difSchemNeutr}) 
	- allowing mainly only for degenerate 
	mass scenarios, and an inverse hierarchy 3$\nu$ and 4$\nu$- scenario
	(the former of these being, however, strongly disfavored 
	by a recent analysis
	of SN1987A
\cite{Minak-Nuno}). 
	For details we refer to 
\cite{KK-Sar_Evid01}. 
	In particular, hierarchical mass schemes 
	are excluded at the present level of accuracy.

	According to 
\cite{Barg02-Sol}
	a global analysis of all solar neutrino data including 
	the recent SNO neutral-current rate selects 
	the Large Mixing Angle (LMA) at the 90\% c.l., however, 
	the LOW solution is also viable, with 0.89 goodness of fit.

\begin{figure}[t]
\vspace{9pt}
\centering{
\includegraphics*[scale=0.40]
{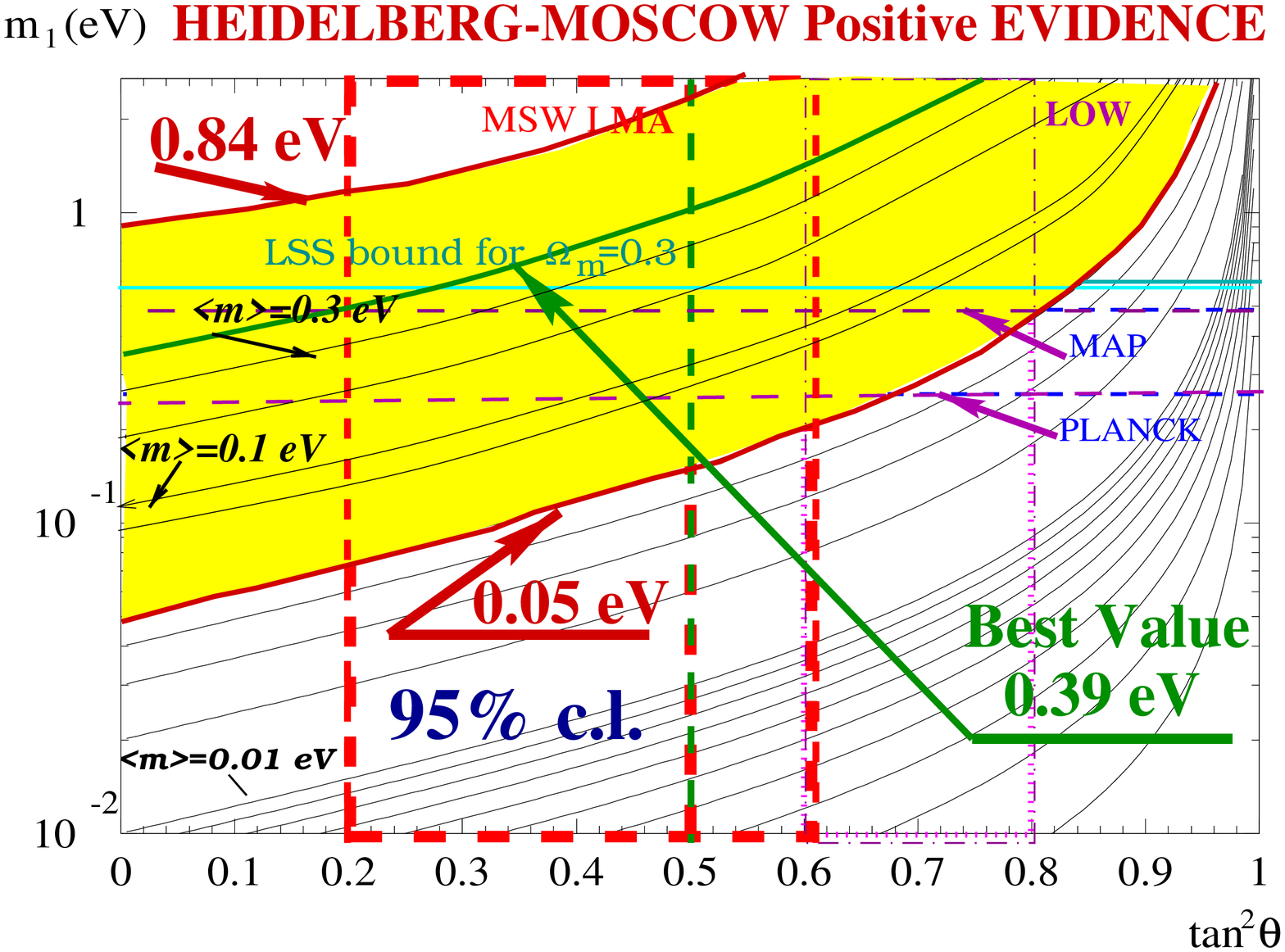}}

\vspace{-0.3cm}
\caption[]{
       Double beta decay observable 
	$\langle m\rangle$
	 and oscillation parameters: 
	 The case for degenerate neutrinos. 
	 Plotted on the axes are the overall scale of neutrino masses 
	 $m_0$ and mixing $\tan^2\, \theta^{}_{12}$. 
	 Also shown is a cosmological bound deduced from a fit of 
	 CMB and large scale structure 
\cite{Lop} 
	and the expected sensitivity of the satellite experiments 
	 MAP and PLANCK. 
	 The present limit from tritium $\beta$ decay of 2.2 eV 
\cite{Trit00} 
	would lie near the top of the figure. 
     The range of 
	$\langle m\rangle$
	 fixed by the HEIDELBERG-MOSCOW experiment 
	is, in the case of small solar neutrino mixing, already in the 
	 range to be explored by MAP and PLANCK. 
\label{fig:Map-Planck}}
\end{figure}

	Assuming the degenerate scenario to be realized in nature 
	we fix - according to the formulae derived in 
\cite{KKPS1-2} - 
	the common mass eigenvalue of the degenerate neutrinos 
	to m = (0.05 - 3.4)\,eV. 
	Part of the upper range is already excluded by 
	tritium experiments, which give 
	a limit of m $<$ 2.2\,eV, or 2.8\,eV (95$\%$ c.l.) 
\cite{Trit00}.
	The full range can only  partly 
	(down to $\sim$ 0.5 eV) be checked by future  
	tritium decay experiments,  
	but could be checked by some future $\beta\beta$ 
	experiments (see, e.g. 
\cite{KK60Y,KK-00-NOON-NOW-NANP-Bey97-GEN-prop}). 
	The deduced 95$\%$ interval for the sum of the degenerate 
	neutrino masses is consistent with the range 
	for $\Omega_\nu$ deduced from recent cosmic microwave 
	background measurements and large scale structure (redshift) 
	surveys, which still allow for a $\sum_{i} m_i \le 4.4$\,eV 
\cite{Teg00,Wang02}.
	The range of $\langle m \rangle $ fixed in this 
	work is, already now, in the range to be explored 
	by the satellite experiments MAP and PLANCK 
\cite{Lop,KK-China01}
	(see Fig. 
\ref{fig:Map-Planck}).
	It lies in a range of interest for Z-burst models recently 
	discussed as explanation for super-high energy cosmic ray events 
	beyond the GKZ-cutoff 
\cite{Farj00-04keV,PW01-Wail99,Fodor02-Evid}. 
	Finally, 
	the deduced best value for the mass 
	is consistent with expectations from experimental 
	$\mu ~\to~ e\gamma$
	branching limits in models assuming the generating 
	mechanism for the neutrino mass to be also responsible 
	for the recent indication for an anomalous magnetic moment 
	of the muon
\cite{MaRaid01}.
	A recent model with underlying A$_4$ symmetry 
	for the neutrino mixing matrix (and the quark mixing matrix) 
	also leads to degenerate neutrino masses consistent 
	with the present experiment 
\cite{BMV02-Evid}. 
	This model succeeds to consistently describe the large (small) 
	mixing in the neutrino (quark) sector.


	The neutrino mass deduced leads 
	to 0.002\,$\, \le\Omega_\nu h^2\le\,0.1$, 
	and thus may allow neutrinos to still play 
	an important role as hot dark matter in the Universe (see also 
\cite{Barg02-Ev}).

	With the HEIDELBERG-MOSCOW experiment, the era of the small smart 
	experiments is over. 
	New approaches and considerably enlarged experiments 
	(as discussed, e.g. in 
\cite{KK60Y,KK-LeptBar98,KK-00-NOON-NOW-NANP-Bey97-GEN-prop,KK-NANP01-TAUP01})
	will be required in future 
	to fix the neutrino mass with higher accuracy. 
	
	Since it was realized in the HEIDELBERG-MOSCOW experiment, 
	that the remaining small background is coming from the material 
	close to the detector (holder, copper cap, ...), 
	elimination of {\it any} material close to the detector 
	will be decisive. Experiments which do not take this 
	into account, like, e.g. CUORE 
\cite{Ales00},
	and MAJORANA 
\cite{MAJOR-WIPP00},
	will allow only rather limited steps in sensitivity.

	Another crucial point is - see eq. (6) - the energy resolution, 
	which can be optimized {\it only} in experiments 
	using Germanium detectors or bolometers. 
	It will be difficult to probe evidence for this rare decay 
	mode in experiments, which have to work - as result of their 
	limited resolution - with energy windows around 
	Q$_{\beta\beta}$ of up to several hundreds of keV, such as NEMO III 
\cite{ApPec02}, 
	EXO 
\cite{EXO-LowNu2}, 
	CAMEO
\cite{CAMEO-Taup01}.

	For example - according to eq. (6) - to compensate 
	for the projected energy resolution of only 130\,keV of EXO 
\cite{EXO-LowNu2},
	the mass of the EXO experiment 
	has to be increased to almost half a ton 
	of enriched material, to reach the sensitivity 
	of the HEIDELBERG-MOSCOW experiment. Only {\it after that} 
	one can think about improving the sensitivity, 
	by improving the background.
	Correspondingly, according to eq. (6), a potential 
	future 100\,kg $^{82}{Se}$ NEMO experiment would be because 
	of its low efficiency, equivalent only to a 10\,kg 
	experiment (not talking about the energy resolution).

	In the first proposal for a third generation double 
	beta experiment, the GENIUS proposal 
\cite{KK-Neutr98,KK-00-NOON-NOW-NANP-Bey97-GEN-prop},
	the idea is to use 'naked' Germanium detectors in a huge tank 
	of liquid nitrogen. It seems to be at present the {\it only} 
	proposal, which can fulfill {\it both} requirements mentioned above. 
	The potential of GENIUS is together with that of some later proposals 
	indicated in Fig. 
\ref{Jahr00-Sum-difSchemNeutr}. 
	GENIUS would - with only 100\,kg of enriched $^{76}{Ge}$ - 
	increase the confidence level of the present \onbb signal 
	to 5$\sigma$ within one year of measurement. 
	A GENIUS Test Facility is at present under construction 
	in the GRAN SASSO Underground Laboratory 
\cite{GENIUS-TF02}.


\section{Acknowledgments}

\vspace{.5cm}
\noindent
	The authors are indebted to C. Tomei and C. D\"orr 
	for their help in the analysis of the Bi lines,
	to H.L. Harney for cooperation on the Bayes method, 
	and to M. Zavertiaev for many inspiring discussions.


\end{document}